\documentclass[prd,twocolumn,preprintnumbers,showpacs,floatfix
]{revtex4}
\usepackage[colorlinks=true,linkcolor=blue,citecolor=blue,urlcolor=blue]{hyperref}
\usepackage{gensymb,comment}
\usepackage{amsmath,amssymb,epsfig,bm,mathrsfs,color}
\usepackage{slashed}

\newcommand{\be}{\begin{equation}}
\newcommand{\ee}{\end{equation}}
\newcommand{\bea}{\begin{eqnarray}}
\newcommand{\eea}{\end{eqnarray}}
\newcommand{\half}{\frac1 2}

\newcommand{\ep}{\epsilon}

\newcommand{\vecq}{{\bm q}}
\newcommand{\vecp}{{\bm p}}

\newcommand{\veck}{{\bm k}}

\newcommand{\vectau}{{\bm \tau}}
\newcommand{\vecrho}{{\bm \rho}}

\newcommand{\ie}{{\it i.e.}}
\newcommand{\eg}{{\it e.g.}}

\newcommand{\fromto}{\leftrightarrow}

\definecolor{red}{rgb}{0.8,0,0}
\definecolor{orange}{rgb}{0.8,0.2,0.0}
\definecolor{blue}{rgb}{0.3,0.0,0.8}
\definecolor{green}{rgb}{0,0.5,0.0}
\definecolor{darkred}{rgb}{0.7,.1,.2}
\definecolor{bgred}{rgb}{1.,.95,.95}
\definecolor{bgblue}{rgb}{.95,.95,1.}
\definecolor{bluegreen}{rgb}{0.,.5,.3}
\definecolor{darkred}{rgb}{0.7,.1,.2}
\definecolor{darkgreen}{rgb}{0.1,.6,.0}
\definecolor{lightyellow}{rgb}{1.,1.,.8}
\definecolor{darkcyan}{rgb}{0.,.7,.9}
\definecolor{lightblue}{rgb}{0.6,0.8,1}
\definecolor{lightgreen}{rgb}{0.7,1.,.9}
\definecolor{money}{rgb}{0.4,0.8,0.}
\definecolor{purple}{rgb}{0.9,0.0,0.8}
\definecolor{orange}{rgb}{0.9,0.5,0.0}
\definecolor{newgr}{rgb}{0.2,0.8,0.2}
\definecolor{newbl}{rgb}{0.3,0.6,0.8}
\definecolor{newor}{rgb}{1.0,0.6,0.}

\usepackage[normalem]{ulem}  

\begin{document}

\title{ Bulk viscosity from Urca processes: $npe\mu$ matter in the neutrino-trapped regime}

\author{Mark Alford} \email{alford@physics.wustl.edu}
\affiliation{Department of Physics, Washington University, St.~Louis,
  Missouri 63130, USA}

\author{ Arus Harutyunyan} \email{arus@bao.sci.am}
\affiliation{Byurakan Astrophysical Observatory,
  Byurakan 0213, Armenia\\
  Department of Physics, Yerevan State University, Yerevan 0025, Armenia}

\author{Armen Sedrakian}
\email{sedrakian@fias.uni-frankfurt.de}
\affiliation{Frankfurt Institute for Advanced Studies, D-60438
  Frankfurt am Main, Germany\\
 Institute of Theoretical Physics, University of Wroc\l{}aw,
50-204 Wroc\l{}aw, Poland}

\date{15 August 2021} 

\begin{abstract}

In this work, we extend our previous study of the bulk viscosity of hot and 
dense $npe$ matter induced by the Urca processes in the neutrino trapped regime 
to  $npe\mu$ matter by adding the  muonic Urca processes as well as the purely 
leptonic electroweak processes involving electron-muon transition. The nuclear 
matter is modeled in a relativistic density functional approach with two different 
parametrizations which predict neutrino dominated matter (DDME2 model) and 
antineutrino dominated matter (NL3 model) at temperatures for which 
neutrinos/antineutrinos are trapped. In the case of neutrino-dominated matter, 
the main equilibration mechanism is lepton capture, whereas in the case of 
antineutrino-dominated matter this is due to neutron decay. We find that the 
equilibration rates of Urca processes are higher than that of the pure leptonic 
processes, which implies that the Urca-process-driven bulk viscosity can be computed 
with the leptonic reactions assumed to be frozen. We find that the bulk viscosity 
decreases with temperature as $\zeta\sim T^{-2}$ at moderate temperatures. 
At high temperatures this scaling breaks down by sharp drops of the bulk viscosity 
close to the temperature where the proton fraction is density-independent and the 
matter becomes scale-invariant. This occurs also when the matter undergoes a 
transition from the antineutrino-dominated regime to the neutrino-dominated 
regime where the bulk viscosity attains a local maximum. We also estimate the 
bulk viscous dissipation timescales and find that these are in the range
$\gtrsim$ 1 s for temperatures above the neutrino trapping temperature. These 
timescales would be relevant only for long-lived objects formed in binary neutron 
star mergers and hot proto-neutron stars formed in core-collapse supernovas. 
\end{abstract}

\maketitle

\section{Introduction}
\label{sec:intro}

Binary neutron star mergers, 
which were observed in gravitational waves by the LIGO-Virgo collaboration, offer a new setting in which to study the properties of superdense, strongly interacting matter. These events are complementary 
to the studies of cold neutron stars, which probe the near zero-temperature limit 
and heavy-ion collisions which are covering less baryon-dense finite systems. 
Thus, they offer an opportunity to gain insight into the physics of hot, dense 
and highly isospin asymmetric matter by analyzing the premerger gravitational waves
(already observed in two merger events, GW170817 and GW190425~\cite{Abbott2017,Abbott2021}) 
and the postmerger signal which 
will be accessible to advanced  LIGO and the next-generation gravitational-wave 
observatories, such as  the Einstein Telescope~\cite{Maggiore2020JCAP} and the Cosmic
Explorer~\cite{Reitze2019}.
Furthermore, electromagnetic counterparts of the gravitational waves 
produced in neutron star mergers can be used to set bounds on the properties of compact stars. 

Numerical simulations of neutron star mergers using the nondissipative hydrodynamics~\cite{Perego:2019adq,Hanauske:2019qgs,Hanauske:2017oxo,Kastaun:2016elu,Bernuzzi:2015opx,Foucart:2015gaa,Kiuchi:2012mk,Sekiguchi:2011zd,Ruiz2016,East:2016,Most2019,Bauswein2019}  (for reviews see \cite{Baiotti:2016qnr,Baiotti2019,Faber2012:lrr}) show that the matter in the postmerger object undergoes oscillations which may be damped by dissipative processes. The initial estimates of the potential impact of dissipation on these oscillations based on cold-matter transport in neutron stars~\cite{Alford2018a} highlighted the potential importance of bulk viscosity in damping the modes. Subsequent studies computed the bulk viscosity of dense matter in various regimes~\cite{Alford2019a,Alford2019b,Haber2021}. In particular, our previous work~\cite{Alford2019b}, focused on the neutrino-trapped regime and computed the bulk viscosity of hot nuclear matter using the relativistic density functional method for the equation of state (EoS) and single-particle spectra of baryons consistent with the prevailing conditions in the postmerger object. It was found that in the regime where neutrinos are trapped the bulk viscosity is reduced compared to the neutrino free-streaming regime. Our estimates of the damping timescales~\cite{Alford2020} indicate that the bulk viscous damping would be most efficient close the temperatures $T_{\rm tr}\sim 5$~MeV~\cite{Roberts:2012um,Alford2018b}  at which the transition from trapped to the free-streaming neutrino regime occurs. The efficacy of the bulk viscosity was estimated by embedding it in the ideal hydrodynamics simulations~\cite{Most:2021zvc},  but this study was restricted to the free-streaming regime only. 
  
   The aim of this work is twofold. First, we extend our previous 
   study~\cite{Alford2019b} of neutron-proton-electron ($npe$) matter 
   to include muons. It is well established that muons appear in 
   significant amounts slightly above the nuclear saturation density, 
   which makes their proper treatment mandatory. Their appearance gives rise to new types of Urca processes and opens up the possibility of purely leptonic electroweak processes. Thus, it is the purpose of this work to assess the impact of these processes on the bulk viscosity of $npe\mu$ matter. 
   The second purpose of this work is to improve on the approximations used in Ref.~\cite{Alford2019b}, by computing the reactions rates of baryonic Urca processes in a fully relativistic manner. We show below that using relativistic rather than approximate nonrelativistic forms of baryon spectra produces a sizeable effect already above twice the nuclear saturation density. Below we focus again on the neutrino-trapped regime, in which neutrinos have a mean free path that is significantly shorter than the stellar size. The resulting nonzero lepton chemical potential affects both the composition of 
   matter and the reaction rates, and constitutes the main difference between this work and
   the extensively studied low-temperature limit of $npe$ and $npe\mu$ compositions  ~\cite{Sawyer1979ApJ,Sawyer1980ApJ,Sawyer1989,Haensel1992PhRvD,Haensel2000,Haensel2001,Dong2007,Alford2010JPhG,Kolomeitsev2015,Alford:2010jf}. We demonstrate explicitly how the low-temperature expressions
   are obtained from their more general counterparts derived here in Appendix~\ref{app:rates}.
    As in Ref.~\cite{Alford2019b} we will assume that thermal conduction is
efficient enough to keep matter isothermal as it undergoes oscillations. 
While such assumption is needed for the treatment of the oscillations, the 
rates of various weak processes we compute below are local quantities and do 
not require such an assumption. The background matter will be treated within the   covariant density functional models based on the DDME2 parametrization~\cite{Lalazissis2005} with density-dependent 
couplings and NL3 parametrization~\cite{Lalazissis1997} which features density-independent couplings but is supplemented with nonlinear self-interaction terms for scalar mesons. More details on these models are given in Ref.~\cite{Alford2019b}. 

Our study is focused on the bulk viscosity, but the methods and results 
are of more general interest, as they can be applied to obtain other 
microphysical characteristic of dense matter, for example, neutrino opacities.

The density-temperature regime studied here occurs in neutron star mergers and also in supernovae and proto-neutron stars, albeit in those cases the lepton fraction is 
  larger ($Y_e\simeq 0.4)$ than in the merger case $(Y_e \simeq 0.1)$~\cite{Prakash1997,Malfatti:2019tpg,Weber2019}. 
  It is worthwhile to note that the importance of muons has been  
  addressed recently in the supernova context as well~\cite{Guo:2020tgx,Fischer:2020vie}.
  
This paper is organized as follows. In Sec.~\ref{sec:urca_rates} we
discuss the rates of the nucleonic Urca and purely leptonic processes.  Section~\ref{sec:bulk} derives the corresponding expressions for the bulk viscosity.  In Sec.~\ref{sec:num_results} we present the results of numerical evaluation of the rates and bulk viscosity on the basis of two density functional theory models at a finite temperature which account for a neutrino component with nonzero chemical potential.
Our conclusions are given in Sec.~\ref{sec:conclusions}. Appendix~\ref{app:rates} 
details the computation of the phase-space integrals needed to evaluate the rates of the Urca processes. Finally, Appendix~\ref{app:A_coeff} details the computation of the susceptibilities of nucleonic matter, which are required for the evaluation of the bulk viscosity coefficient.

In this work we use natural (Gaussian) units with 
$\hbar = c = k_B = 1$,   
and the metric $g_{\mu\nu} = \textrm{diag}(1, -1, -1, -1)$.

\begin{widetext}

\section{Weak processes in neutron star matter}
\label{sec:urca_rates}

We consider neutron-star matter composed of neutrons,
protons, electrons, muons, and electron and muon neutrinos in the 
density range $0.5n_0\leq n_B\leq 5 n_0$, where $n_0\simeq 0.152$ 
fm$^{-3}$ is the nuclear saturation density, and temperature range
$T_{\rm tr}\leq T \simeq 100$ MeV with $T_{\rm tr}\simeq 5$ 
MeV being the temperature above which  neutrinos (or antineutrinos) 
are trapped in a neutron star \cite{Alford2018b}.

The $\beta$-equilibration processes among the baryons 
we consider below are the direct Urca processes
\bea\label{eq:e_decay}
&& n\rightleftarrows p + e^-+\bar{\nu}_e\quad{\rm (neutron~e\!-\!decay)},\\
\label{eq:e_capture}
&& p + e^-\rightleftarrows n+{\nu}_e\quad{\rm (electron~capture)},\\
\label{eq:mu_decay}
&& n\rightleftarrows p + \mu^-+\bar{\nu}_\mu\quad{\rm (neutron~\mu\!-\!decay)},\\
\label{eq:mu_capture}
&& p + \mu^-\rightleftarrows n+{\nu}_\mu\quad{\rm (muon~capture)}.
\eea
If muons are present in matter,  the following purely leptonic reactions are operative
\bea\label{eq:reaction_L1}
&& \mu^- \rightleftarrows e^- +\bar{\nu}_e +{\nu}_\mu \quad{\rm (muon~decay)},\\
\label{eq:reaction_L2}
&& \mu^- +{\nu}_e \rightleftarrows e^-+ {\nu}_\mu \quad{\rm (neutrino~scattering)},\\
\label{eq:reaction_L3}
&& \mu^- +\bar{\nu}_\mu  \rightleftarrows e^-+\bar{\nu}_e \quad{\rm (antineutrino~scattering)}.
\eea
Stellar matter is in approximate  $\beta$-equilibrium  which implies $\mu_n+\mu_{\nu_l}=\mu_p+\mu_l$, where $l=\{e,\mu\}$. 
We assume that neutrino flavor conversion can be neglected, so there are four exactly conserved quantities:
baryon number $n_B=n_n+n_p$, electric charge (the system remains charge neutral $n_p=n_e+n_\mu$), 
and lepton numbers $n_{L_l}=n_l+n_{\nu_l}=Y_{L_l}n_B$ for each flavor $l$ separately. 
Here $Y_{L_l}$ are the lepton fractions, which have  typical values 
$Y_{Le} = Y_{L\mu}=0.1$ in the BNS mergers~\cite{Baiotti2019} and $Y_{Le} = Y_{L\mu}=0.4$ in proto-neutron stars and supernovas~\cite{Prakash1997,Malfatti:2019tpg,Weber2019}.
Since there are 4 conserved quantities and 6 particle species, this leaves two chemical potentials \eqref{eq:delta_mu_1} and \eqref{eq:delta_mu_2} discussed below that are driven to zero by weak interactions on timescales that are potentially comparable to the density variations in a merger. In this paper, we calculate the resultant bulk viscosity.

\subsection{Urca process rates }

The  neutron decay processes \eqref{eq:e_decay} and \eqref{eq:mu_decay} 
can be written compactly as $n\rightarrow p + l^-+\bar{\nu}_l$, 
where $l^-$ is an electron or muon and $\bar{\nu}_l$ is the corresponding 
antineutrino. Then, the rate of each of these processes can be written as 
\bea \label{eq:Gamma1p_def1}
\Gamma_{n\to p l \bar\nu} &=& 
\int\!\! \frac{d^3p}{(2\pi)^32p_0} \int\!\!
\frac{d^3p'}{(2\pi)^32p'_0} \int\!\! \frac{d^3k}{(2\pi)^32k_0}
\int\!\! \frac{d^3k'}{(2\pi)^32k'_0}\sum \vert {\cal M}_{\text{Urca}}\vert^2 \nonumber\\
& \times & \bar{f}(k)\bar{f}(p) \bar{f}(k') f(p') (2\pi)^4\delta^{(4)}(k+p+k'-p'),
\eea
where $f(p)$ is the  Fermi distribution function,
 $\bar{f}(p)\equiv 1-f(p)$, and  the mapping between the
 particles and their momenta is as follows: 
$(l) \to k$, $(\nu_l/\bar{\nu}_l) \to k'$, $(p) \to p$,
 and $(n) \to p'$. Similarly, the lepton capture processes~\eqref{eq:e_capture} 
 and \eqref{eq:mu_capture} can be written 
  as $p+l^-\to n+\nu_l$ and the corresponding rate is given by 
\bea \label{eq:Gamma2n_def1}
\Gamma_{p l\to n\nu} &=& 
\int\!\! \frac{d^3p}{(2\pi)^32p_0} \int\!\!
\frac{d^3p'}{(2\pi)^32p'_0} \int\!\! \frac{d^3k}{(2\pi)^32k_0}
\int\!\! \frac{d^3k'}{(2\pi)^32k'_0}
\sum \vert {\cal M}_{\text{Urca}}\vert^2 \nonumber\\
& \times &
{f}(k) {f}(p) \bar{f}(k') \bar{f}(p')(2\pi)^4\delta(k+p-k'-p').
\eea
The matrix element of these processes is~\cite{Greiner2000gauge}
\be\label{eq:matrix_el_full}
\sum \vert {\cal M}_{\text{Urca}}\vert^2 = 32 G_F^2\cos^2
\theta_c \left[(1+g_A)^2(k\cdot p) (k'\cdot p')
+(1-g_A)^2(k\cdot p') (k'\cdot p)
+(g_A^2-1)m^*_n m^*_p(k\cdot k')\right],
\ee 
where $G_F=1.166\cdot 10 ^{-5}$ GeV$^{-2}$ is the Fermi coupling constant,
$\theta_c$ is the Cabibbo angle ($\cos\theta_c=0.974$), $g_A=1.26$
is the axial-vector coupling constant, and $m_n^*$ and $m_p^*$ are 
the effective masses of neutron and proton, respectively. Because $g_A\approx 1$,
the second and the third terms in Eq.~\eqref{eq:matrix_el_full} 
are suppressed as compared to the first one so we neglect them in our further computations. 


The equilibration rates given by Eq.~\eqref{eq:Gamma1p_def1}
and \eqref{eq:Gamma2n_def1} can be computed once we specify
the spectrum of strongly interacting nucleons. We apply the covariant
density functional theory (CDF) of nuclear matter which is
based on phenomenological baryon-meson Lagrangians introduced 
ba Walecka, Boguta-Bodmer and others~\cite{Glendenning_book,Weber_book}. 

The Lagrangian density of matter is given by 
\bea\label{eq:lagrangian}
{\cal L} & = &
\sum_N\bar\psi_N\bigg[\gamma^\mu \left(i\partial_\mu-g_{\omega
}\omega_\mu - \half g_{\rho }\vectau\cdot\vecrho_\mu\right)
- m^*_N\bigg]\psi_N +
 \sum_{\lambda}\bar\psi_\lambda(i\gamma^\mu\partial_\mu -
m_\lambda)\psi_\lambda,\nonumber\\
 & + & \half \partial^\mu\sigma\partial_\mu\sigma-\half
m_\sigma^2\sigma^2 -U(\sigma)- \frac{1}{4}\omega^{\mu\nu}\omega_{\mu\nu}
 + \half m_\omega^2\omega^\mu\omega_\mu -
\frac{1}{4}\vecrho^{\mu\nu}\cdot \vecrho_{\mu\nu} + \half
m_\rho^2\vecrho^\mu\cdot\vecrho_\mu, 
\eea 
where $N$ sums over nucleons, $\psi_N$ are the nucleonic Dirac 
fields with effective masses $m_N^*=m_N - g_{\sigma}\sigma$, with 
$m_N$ being the nucleon mass in vacuum;  $\sigma,\omega_\mu$, 
$\vecrho_\mu$ are, respectively, the scalar-isoscalar, vector-isoscalar 
and vector-isovector meson fields which mediate the interaction 
between baryons; $\omega_{\mu\nu}=\partial_\mu\omega_\nu-\partial_\nu
\omega_\mu$ and $\vecrho_{\mu\nu}=\partial_\mu \vecrho_{\nu}-\partial_\nu 
\vecrho_{\mu}$ are the field strength tensors of vector mesons; 
$m_{i}$ are the meson masses and $g_{i}$ are the baryon-meson 
couplings with $i=\sigma,\omega,\rho$; finally, $U(\sigma)$ is the 
self-interaction potential of scalar meson field. The leptonic part of
the Lagrangian is given by the second sum in Eq.~\eqref{eq:lagrangian}, 
where $\psi_\lambda$, $\lambda \in (e^-, \mu^-, \nu_e, \nu_{\mu})$,
 are the  free Dirac fields of leptons with masses  $m_{e^-} = 0.51$~MeV, 
 $m_{\mu^-} = 105.7$~MeV, and $m_{\nu_{e}} =m_{\nu_{\mu}} =0$.
In the following we will adopt two different parametrizations of 
Lagrangian~\eqref{eq:lagrangian}, specifically, the model 
DDME2~\cite{Lalazissis2005} in which the nucleon-meson couplings 
are density-dependent and $U(\sigma)=0$, and the model 
NL3~\cite{Lalazissis1997}, which has density-independent 
nucleon-meson couplings but nonzero self-interaction among 
$\sigma$-meson fields, which is contained in the potential 
$U(\sigma)=g_2\sigma^3/3+g_3\sigma^4/4$.

The spectrum of nucleonic excitations derived from Eq.~\eqref{eq:lagrangian} 
in the mean-field approximation is given by~\cite{Glendenning_book} 
\bea\label{eq:spectrum}
E_k = \sqrt{k^2+m^{* 2}_N} + g_{\omega}\omega_0 + I_3 g_{\rho}\rho_{03}+\Sigma_r,
\eea
where $I_3$ is the third component of the nucleon isospin, $\Sigma_r$ is so-called rearrangement self-energy~\cite{Typel1999} which should be introduced to maintain the thermodynamical consistency 
(specifically the energy conservation and fulfillment of the
Hugenholtz-van Hove theorem) of the system in the case of 
density-dependent couplings. 
Defining the nucleon effective chemical potentials as $\mu^*_N 
= \mu_N-g_{\omega }\omega_0 - I_3 g_{\rho }\rho^0_3-\Sigma_r$ we can write
the argument of nucleon Fermi-functions as $E_k-\mu_N=\sqrt{k^2
+m^{* 2}_N}-\mu_N^*$ which formally coincides with the spectrum of
free nucleons with effective masses and effective chemical potentials.


The details of computation of the phase-space integrals in 
Eqs.~\eqref{eq:Gamma1p_def1} and \eqref{eq:Gamma2n_def1} 
are given in Appendix~\ref{app:rates}. The final result reads
\bea\label{eq:Gamma1p_final1} 
\Gamma_{n\to p l \bar\nu} (\mu_{\Delta_l})
&=& -\frac{{G}^2T^4}{(2\pi)^5} 
\int_{-\infty}^\infty\!\!\! dy\,
\!\int_0^\infty\!\! dx\, 
\left[(\mu_{\nu_l} +\mu_n^*+yT)^2 
 -m_n^{*2}-x^2T^2\right]\nonumber\\
&&\times \left[(\mu_l +\mu_p^* +\bar{y}_lT)^2
-m_l^2-m_p^{*2} -x^2T^2\right]\nonumber\\
&&\times \int_{m_l/T-\alpha_l}^{\alpha_p +\bar{y}_l}\!
dz\, \bar{f}(z){f}(z-\bar{y}_l)\,\theta_x\!
\int_{\alpha_{\nu_l}}^\infty\! 
dz'\,f(z'+y)\bar{f}(z')\,\theta_y,\quad\\
\label{eq:Gamma2n_final1} 
\Gamma_{p l\to n\nu} (\mu_{\Delta_l}) 
&=& \frac{{G}^2T^4}{(2\pi)^5}\int_{-\infty}^\infty\!
dy\! \int_0^\infty\! 
dx\, \left[(\mu_{\nu_l} +\mu_n^*+yT)^2 
-m_n^{*2}-x^2T^2\right]\nonumber\\
&&\times \left[(\mu_l +\mu_p^* +\bar{y}_lT)^2
-m_l^2-m_p^{*2} -x^2T^2\right]\nonumber\\
&&\times \int_{m_l/T-\alpha_l}^{\alpha_p +\bar{y}_l}\!
dz\, f(z)f(\bar{y}_l-z)\,\theta_x\!
\int_{-\alpha_{\nu_l}}^{\alpha_n+y}\! 
dz'\, {f}(z'-y)\bar{f}(z')\,\theta_z,
\eea
where $G=G_F \cos\theta_c(1+g_A)$, $\alpha_j= \mu_j^*/T$ for baryons and 
$\alpha_j= \mu_j/T$ for leptons, $\bar{y}_l=y+\mu_{\Delta_l}/T$ with $\mu_{\Delta_l}=\mu_n+\mu_{\nu_l}-\mu_p-\mu_l$ being the chemical 
potential imbalances (see Sec.~\ref{sec:bulk}). The $\theta$-functions 
in Eqs.~\eqref{eq:Gamma1p_final1} and \eqref{eq:Gamma2n_final1} 
impose the constraints
\bea\label{eq:thetax2}
\theta_x &: &
(z_k-x)^2 \leq \left(z -\alpha_p 
-\bar{y}_l\right)^2 -m_p^{*2}/T^2\leq (z_k+x)^2,\\
\label{eq:thetay2}
\theta_y &: &
(z_k'-x)^2 \leq \left(z' +\alpha_n+ y\right)^2 
-m_n^{*2}/T^2\leq (z_k'+x)^2,\\
\label{eq:thetaz2}
\theta_z &: &
(z_k'-x)^2 \leq \left(z'-\alpha_n-y\right)^2 
-m_n^{*2}/T^2\leq (z_k'+x)^2.
\eea 
The integration variables $y$ and $x$ are the transferred energy 
and momentum, respectively, normalized by the temperature; the variables $z$ and $z'$ are the normalized-by-temperature lepton  and neutrino energies, respectively, computed from their chemical 
potentials, and $z_k=\sqrt{(z+\alpha_l)^2-m_l^2/T^2}$ and $z'_k=z'\mp \alpha_{\nu_l}$ are the normalized-by-temperature momenta of the lepton and the antineutrino/neutrino, respectively. The rates of the inverse processes can be obtained from 
Eqs.~\eqref{eq:Gamma1p_final1} and \eqref{eq:Gamma2n_final1} 
by interchanging $f(p_i)\leftrightarrow \bar{f}(p_i)$ for all particles.

In beta equilibrium we have $\mu_{\Delta_l}=0$ and the rates of the direct and inverse processes are equal for each lepton flavor: $\Gamma_{n\to p l \bar\nu}=\Gamma_{p l \bar\nu\to n}\equiv \Gamma_{n\fromto p l \bar\nu}$, $\Gamma_{p l\to n\nu}=\Gamma_{n\nu\to pl}\equiv \Gamma_{p l\fromto n\nu}$. For small departures 
from $\beta$-equilibrium $\mu_{\Delta_l} \ll T$, we can assume linear response
where the net proton production rate due to the neutron decay and its inverse processes is $\Gamma_{n\to p l \bar\nu}-\Gamma_{p l \bar\nu\to n} = \lambda_{n\fromto p l \bar\nu}\,\mu_{\Delta_l}$. Similarly, the net proton production rate due to the inverse and direct lepton capture processes is $\Gamma_{n\nu\to pl}-\Gamma_{p l\to n\nu} = \lambda_{p l\fromto n\nu}\,\mu_{\Delta_l}$.
Pushing the system out of beta equilibrium by a chemical potential $\mu_{\Delta_l}$ just replaces one power of $T$ in the rate with a power of $\mu_{\Delta_l}$, so
the expansion coefficients $\lambda_{n\fromto p l \bar\nu}$ and $\lambda_{p l\fromto n\nu}$ (see~Appendix~\ref{app:rates}) are given by
\bea \label{eq:lambda1}
\lambda_{n\fromto p l \bar\nu}& = & \left(\frac{\partial\Gamma_{n\to p l \bar\nu}}
{\partial\mu_{\Delta_l}}-
\frac{\partial\Gamma_{p l \bar\nu\to n}}
{\partial\mu_{\Delta_l}}\right)\bigg\vert_{\mu_{\Delta_l}=0}
=\frac{\Gamma_{n\fromto p l \bar\nu}}{T},\\
\label{eq:lambda2}
\lambda_{p l\fromto n\nu}& = & \left(\frac{\partial\Gamma_{n\nu \to pl}}
{\partial\mu_{\Delta_l}}-
\frac{\partial\Gamma_{p l\to n\nu}}
{\partial\mu_{\Delta_l}}\right)\bigg\vert_{\mu_{\Delta_l}=0}
=\frac{\Gamma_{p l\fromto n\nu}}{T}.
\eea

\subsection{Lepton process rates}
\label{app:rates_leptons}

The general form of the lepton reaction rates~\eqref{eq:reaction_L1},
\eqref{eq:reaction_L2} and \eqref{eq:reaction_L3} reads
\bea \label{eq:Gamma_lep1}
\Gamma_{\mu\to e\bar{\nu}\nu}  &=& 
\int d\Omega_k \sum \vert {\cal M}_{\rm lep}\vert^2 
f(k_\mu) \bar{f}(k_e)\bar{f}(k_{\bar{\nu}_e}) \bar{f}(k_{\nu_\mu}) 
(2\pi)^4\delta^{(4)}(k_e+k_{\bar{\nu}_e}+k_{\nu_\mu}-k_\mu),\\
\label{eq:Gamma_lep2}
\Gamma_{\mu\nu\to e{\nu}} &=& 
\int d\Omega_k \sum \vert {\cal M}_{\rm lep}\vert^2 
f(k_\mu) {f}(k_{\nu_e})  \bar{f}(k_e)\bar{f}(k_{\nu_\mu}) 
(2\pi)^4\delta^{(4)}(k_e+k_{\nu_\mu}-k_{\nu_e}-k_\mu),\\
\label{eq:Gamma_lep3}
\Gamma_{\mu\bar{\nu}\to e\bar{\nu}}  &=& 
\int d\Omega_k \sum \vert {\cal M}_{\rm lep}\vert^2 
f(k_\mu){f}(k_{\bar{\nu}_\mu}) \bar{f}(k_e) \bar{f}(k_{\bar{\nu}_e})  
(2\pi)^4\delta^{(4)}(k_e+k_{\bar{\nu}_e}-k_{\bar{\nu}_\mu}-k_\mu),
\eea
where the short-hand notation $d\Omega_k$ is the Lorentz-invariant
momentum phase-space element, \ie,
\bea 
\int d\Omega_k = \int\! \frac{d^3k_e}{(2\pi)^3\, 2k_{0e}} \int\!
\frac{d^3k_\mu}{(2\pi)^3\,2k_{0\mu}} \int\! \frac{d^3k_{\nu_e/\bar{\nu}_e}}{(2\pi)^3\, 2k_{0\,\nu_e/\bar{\nu}_e}}\int\!
\frac{d^3k_{\nu_\mu/\bar{\nu}_\mu}}{(2\pi)^3\, 2k_{0\,\nu_\mu/\bar{\nu}_\mu}}.
\eea
The spin-averaged relativistic matrix 
element of lepton reactions reads~\cite{Guo:2020tgx}
\be\label{eq:matrix_lep}
\sum \vert {\cal M}_{\rm lep}\vert^2 = 128 
G_F^2 \left(k_e\cdot k_{\nu_\mu/\bar{\nu}_\mu}\right) 
\left(k_{\nu_e/\bar{\nu}_e}\cdot k_\mu\right).
\ee 
Computation of lepton process rates can be performed analogously to the Urca process rates. The final expressions suitable for numerical evaluation are 
\bea\label{eq:Gamma1e_final} 
\Gamma_{\mu\to e\bar{\nu}\nu}\, (\mu_\Delta^L)
&=& -\frac{4{G}^2_FT^4}{(2\pi)^5} 
\int_{-\infty}^\infty\!\!\! dy\,
\!\int_0^\infty\!\! dx\, \left[(\mu_e +\mu_{\nu_\mu} 
+\tilde{y}T)^2-m_e^2-x^2T^2\right]\nonumber\\
&&\times \left[(\mu_{\nu_e} +\mu_\mu+yT)^2 
 -m_\mu^{2}-x^2T^2\right]\nonumber\\
&&\times \int_{m_e/T-\alpha_e}^{\alpha_{\nu_\mu}
+\tilde{y}}\! dz\, \bar{f}(z){f}(z-\tilde{y})\,\theta_x^L\!
\int_{\alpha_{\nu_e}}^\infty\!
dz'\,f(z'+y)\bar{f}(z')\,\theta_y^L,\\
\label{eq:Gamma2e_final} 
\Gamma_{\mu\nu\to e{\nu}}\, (\mu_\Delta^L) 
&=& \frac{4{G}^2_FT^4}{(2\pi)^5}\int_{-\infty}^\infty\!
dy\! \int_0^\infty\! dx\, \left[(\mu_e +\mu_{\nu_\mu} 
+\tilde{y}T)^2-m_e^2-x^2T^2\right]\nonumber\\
&&\times \left[(\mu_{\nu_e} +\mu_\mu+yT)^2 
-m_\mu^{2}-x^2T^2\right]\nonumber\\
&&\times \int_{m_e/T-\alpha_e}^{\alpha_{\nu_\mu} 
+\tilde{y}}\! dz\, \bar{f}(z)\bar{f}(\tilde{y}-z)\,
\theta_x^L\!\int_{-\alpha_{\nu_e}}^{\alpha_\mu+y}\! 
dz'\, \bar{f}(z'-y) {f}(z')\,\theta_z^L,\\
\label{eq:Gamma3e_final} 
\Gamma_{\mu\bar{\nu}\to e\bar{\nu}}\, (\mu_\Delta^L) 
&=& \frac{4{G}^2_FT^4}{(2\pi)^5} 
\int_{-\infty}^\infty\!\!\! dy\,
\!\int_0^\infty\!\! dx\, 
\left[(\mu_e +\mu_{\nu_\mu} +\tilde{y}T)^2-m_e^2
-x^2T^2\right]\nonumber\\
&&\times \left[(\mu_{\nu_e} +\mu_\mu+yT)^2 
 -m_\mu^{2}-x^2T^2\right]\nonumber\\
&&\times \int_{z_{\rm min}}^{\infty}\! 
dz\, \bar{f}(z){f}(z-\tilde{y})\,\theta_x^L\!
\int_{\alpha_{\nu_e}}^\infty\!
dz'\,f(z'+y)\bar{f}(z')\,\theta_y^L,
\eea
where $\mu_{\Delta}^L\equiv \mu_\mu+\mu_{\nu_e}-\mu_e-\mu_{\nu_\mu}=\mu_{\Delta_e}-\mu_{\Delta_\mu}$ is the chemical imbalance for leptons,
$\tilde{y}=y+\mu_{\Delta}^L/T$, $z_{\rm min} = {\rm max}\{m_e/T-
\alpha_e;\alpha_{\nu_\mu} +\tilde{y}\}$, and the $\theta$-functions impose the constraints
\bea\label{eq:thetaxL}
\theta_x^L &: &
(z_k-x)^2 \leq \left(z -\alpha_{\nu_\mu} 
-\tilde{y}\right)^2 \leq (z_k+x)^2,\\
\label{eq:thetayL}
\theta_y^L &: &
(z_k'-x)^2 \leq \left(z' +\alpha_\mu+ y\right)^2 
-m_\mu^{2}/T^2\leq (z_k'+x)^2,\\
\label{eq:thetazL}
\theta_z^L &: &
(z_k'-x)^2 \leq \left(z'-\alpha_\mu-y\right)^2 
-m_\mu^{2}/T^2\leq (z_k'+x)^2,
\eea 
with $z_k=\sqrt{(z+\alpha_e)^2-m_e^2/T^2}$ and 
$z'_k=z'\mp \alpha_{\nu_e}$ for $\theta_y/\theta_z$.

\section{Bulk viscosity of $npe\mu$ matter}
\label{sec:bulk}

In this section, we derive a microscopic formula for the bulk 
viscosity of neutrino-trapped $npe\mu$ matter arising from the
Urca processes~\eqref{eq:e_decay}--\eqref{eq:mu_capture}. 
In the temperature and density range where the neutrinos are 
trapped the $\beta$-equilibration rates are much 
higher than the characteristic frequency of density oscillations; 
this corresponds to the  {\it fast equilibration regime}~\cite{Alford2019a}. 
Then the analysis can be restricted to the ``subthermal'' case, 
where the matter is only slightly perturbed from equilibrium. 
 
Consider now small-amplitude density oscillations in nuclear matter
with a given frequency $\omega$ for which 
we can write $n_B(t)=n_{B0}+\delta n_B(t)$ and $n_{L_l}(t)=n_{L_l0}
+\delta n_{L_l}(t)$, where $\delta n_B(t),\delta n_{L_l}(t) \sim 
e^{i\omega t}$. The baryon and lepton conservation laws in the comoving frame imply  
\bea\label{eq:cont_j}
\delta n_B =-n_{B0}\frac{\theta}{i\omega},\qquad
\delta n_{L_l} =-n_{L_l0}\frac{\theta}{i\omega},\qquad l=\{e,\mu\},
\eea 
where $\theta=\partial_i v^i$ is the fluid velocity divergence.

The perturbations of particle densities can be separated into 
local equilibrium and nonequilibrium parts
\bea\label{eq:dens_j}
n_j(t)=n_{j0}+\delta n_j(t), && \delta n_j(t)=
\delta n^{\rm eq}_j(t)+\delta n'_j(t),
\eea 
where $j=\{n,p,e,\mu,\nu_e,\nu_\mu\}$ labels the particles. The variations
$\delta n^{\rm eq}_j(t)$ denote the shift of the equilibrium state for
the instantaneous values of the baryon and lepton densities $n_{B}(t)$
and $n_{L_l}(t)$, whereas $\delta n'_j(t)$ denote the deviations of the
corresponding densities from their equilibrium values. 

The compression and rarefaction drives the system out of chemical
equilibrium leading to nonzero $\delta n'_j(t)$, and, subsequently,
to chemical imbalances $\mu_{\Delta_l}=\delta\mu_n+\delta\mu_{\nu_l}-\delta\mu_p-
\delta\mu_l$, which can be written as
\bea\label{eq:delta_mu_1} 
\mu_{\Delta_e} &=& A_n \delta n_n +A_{\nu_e} \delta n_{\nu_e}
 -A_p \delta n_p -A_e \delta n_e,\\
\label{eq:delta_mu_2}
\mu_{\Delta_\mu} &=& A_n \delta n_n +A_{\nu_\mu} \delta n_{\nu_\mu}
-A_p \delta n_p -A_\mu \delta n_\mu,
\eea 
where $A_n=A_{nn}-A_{pn}$, $A_p=A_{pp}-A_{np}$, and 
$A_l=A_{ll}$, $A_{\nu_l}=A_{\nu_l \nu_l}$ with
\bea\label{eq:A_f}
A_{ij} = \left(\frac{\partial \mu_i}{\partial n_j}\right)_0,
\eea 
and index 0 refers to the equilibrium state.
The nuclear off-diagonal elements $A_{np}$ and $A_{pn}$ are nonzero
because of the cross-species strong interaction between neutrons and
protons. The computation of susceptibilities $A_{ij}$ is
performed in Appendix~\ref{app:A_coeff}.

To proceed further we need to determine how the lepton
reactions~\eqref{eq:reaction_L1}--\eqref{eq:reaction_L3} affect 
 the bulk viscosity from the Urca 
processes~\eqref{eq:e_decay}--\eqref{eq:mu_capture}.
Typically, we deal with two limiting cases: (a) fast 
lepton-equilibration limit, \ie, the lepton process rates 
are much higher than Urca process rates; (b) slow 
lepton-equilibration limit, where the lepton process 
rates are much lower than Urca process rates. We derive
analytic expressions for the bulk viscosity in terms of 
equilibration rates and particle susceptibilities in 
these two limiting cases in the next two subsections.

\subsection{Fast lepton-equilibration limit}

In this case, the chemical equilibration among leptons 
(processes \eqref{eq:reaction_L1}, \eqref{eq:reaction_L2}, \eqref{eq:reaction_L3})
takes
place much faster than the equilibration between baryons and 
leptons, therefore the condition $\mu_e-\mu_{\nu_e}=\mu_\mu-
\mu_{\nu_\mu}$ can be assumed to be satisfied while studying 
the bulk viscosity from the Urca processes. This implies 
$\mu_{\Delta_e}=\mu_{\Delta_\mu}\equiv \mu_\Delta$, \ie, the 
electronic and muonic Urca processes are described by a single 
chemical potential shift from equilibrium. The rate equations 
which take into account the loss and gain of particles read
\bea\label{eq:cont_n_fast}
\frac{\partial}{\partial t}\delta n_n(t) &=& 
-\theta  n_{n0} -(\lambda_e+\lambda_\mu) \mu_\Delta(t),\\
\label{eq:cont_p_fast}
\frac{\partial}{\partial t}\delta n_p(t) &=& 
-\theta n_{p 0} +(\lambda_e+\lambda_\mu)\mu_\Delta(t),\\
\label{eq:cont_e_fast}
\frac{\partial}{\partial t}\delta n_e(t) &=& 
-\theta n_{e 0} +\lambda_e\mu_\Delta(t)+I_{L},\\
\label{eq:cont_mu_fast}
\frac{\partial}{\partial t}\delta n_\mu(t) &=& 
-\theta n_{\mu 0} +\lambda_\mu\mu_\Delta(t)-I_{L},
\eea 
where $\lambda_e=\lambda_{n\fromto p e \bar\nu}+\lambda_{p e\fromto n\nu}$
and $\lambda_\mu=\lambda_{n\fromto p \mu \bar\nu}+\lambda_{p \mu\fromto n\nu}$ are the summed equilibration rates
of the electron and muon Urca reactions, respectively.
The quantity $I_L$ is the summed rate of the lepton
reactions~\eqref{eq:reaction_L1}, \eqref{eq:reaction_L2}, \eqref{eq:reaction_L3},
which arises as a result of an almost vanishing shift 
$\mu_{\Delta_e}-\mu_{\Delta_\mu}\ll \mu_\Delta$
but cannot be neglected 
because the relevant $\lambda$-coefficient can be very large, 
as already discussed in Ref.~\cite{Jones2001PhRvD}.

Only two of the balance equations are independent (one for
a baryon and one for a lepton) as the others can be obtained 
from them via exploiting the conditions of charge neutrality 
$\delta n_p=\delta n_e+\delta n_\mu$ and baryon conservation 
$\delta n_{B}=\delta n_n+\delta n_p$. The balance equations 
for neutrinos are obtained from Eqs.~\eqref{eq:cont_e_fast} 
and \eqref{eq:cont_mu_fast} and the constraints $\delta n_{L_l}
=\delta n_l+\delta n_{\nu_l}.$
 
The equilibrium with respect to lepton reactions implies
\bea\label{eq:delta_mu_L}
\delta\mu_\mu +\delta\mu_{\nu_e} -\delta\mu_e- \delta\mu_{\nu_\mu}
=A_{\mu}\delta n_\mu +A_{\nu_e}\delta n_{\nu_e} -A_e \delta n_e
- A_{\nu_\mu}\delta n_{\nu_\mu}  =0,
\eea 
which gives the constraints
\bea\label{eq:n_e_const}
\delta n_{e} &=& \frac{(A_{\mu}+A_{\nu_\mu})\delta n_p
+A_{\nu_e}\delta n_{L_e}- A_{\nu_\mu}\delta n_{L_\mu}}{A_L},\\
\label{eq:n_mu_const}
\delta n_\mu &=& \frac{(A_e+A_{\nu_e})\delta n_p
-A_{\nu_e}\delta n_{L_e}+ A_{\nu_\mu}\delta n_{L_\mu}}{A_L},
\eea 
with $A_L=A_e+A_{\nu_e}+A_{\mu}+A_{\nu_\mu}$.
Substituting these expressions into Eq.~\eqref{eq:delta_mu_1} we find
\bea\label{eq:mu_delta_vs_n}
\mu_\Delta &=& A_L^{-1}
\left[(N_n-N_p)\delta n_n +N_p\delta n_B+N_e\delta n_{L_e}+
N_\mu \delta n_{L_\mu}\right],
\eea 
where
\bea
\label{eq:N_n_def}
N_n &=& A_LA_n,\\
\label{eq:N_p_def}
N_p &=& -A_p A_L-(A_e+A_{\nu_e})(A_{\mu}+A_{\nu_\mu} ),\\
\label{eq:N_B_def}
N_B &=&N_n-N_p= A_L(A_n +A_p)+(A_e+A_{\nu_e})(A_{\mu}+A_{\nu_\mu} ),\\
\label{eq:N_e_def}
N_e &=& A_{\nu_e}(A_{\mu}+A_{\nu_\mu}),\\
\label{eq:N_mu_def}
N_\mu &=& A_{\nu_\mu}(A_e+A_{\nu_e}).
\eea 
Next we substitute $\mu_\Delta$ in Eq.~\eqref{eq:cont_n_fast},
assume that the time-dependence of density perturbations 
is given by $\delta n_j(t)\sim e^{i\omega t}$ and take into
account Eq.~\eqref{eq:cont_j} to obtain 
\bea
i\omega\delta n_n + \theta n_{n0}+
{\lambda} A_L^{-1}\left[N_B\delta n_n +N_p\delta n_B+N_e\delta n_{L_e}+
N_\mu \delta n_{L_\mu}\right]\nonumber\\
= A_L^{-1}\left[(i\omega A_L+{\lambda} N_B) \delta n_n
+{\lambda} N_p \delta n_B+ \theta n_{n0}A_L +{\lambda} 
(N_e\delta n_{L_e}+ N_\mu \delta n_{L_\mu})\right]=0,
\eea 
with ${\lambda}=\lambda_e+\lambda_\mu$. Solving for $\delta n_n$ 
gives
\bea\label{eq:delta_n}
\delta n_n 
&=&
\frac{\theta}{i\omega}\frac{
{\lambda} N_p n_{B0} +{\lambda} (N_e n_{L_e0}+N_\mu n_{L_\mu 0})-i\omega n_{n0} A_L}
{i\omega A_L+{\lambda} N_B}.
\eea 
Under similar assumptions Eq.~\eqref{eq:n_e_const} gives
\bea\label{eq:delta_e}
\delta n_{e} &=&  -\frac{\theta}{i\omega}
\frac{1} {(i\omega {A}_L+{\lambda} N_B)}
 \bigg\{ i\omega (A_{\mu}+A_{\nu_\mu})n_{p0}
 +{\lambda}A_n (A_{\mu}+A_{\nu_\mu})n_{B0} \nonumber\\
&&+(i\omega +{\lambda} A_2)
A_{\nu_e} n_{L_e0}-\left[i\omega +{\lambda}(A_n+A_p)\right]
A_{\nu_\mu} n_{L_\mu 0}\bigg\},
\eea 
where we exploited the relations 
\bea 
N_B A_{\nu_e} +N_e (A_{\mu}+A_{\nu_\mu}) 
&=& A_{\nu_e}A_L A_2,\\
N_B A_{\nu_\mu} - N_\mu (A_{\mu}+A_{\nu_\mu})&=&
 A_{\nu_\mu}A_L(A_n+A_p),
\eea 
 and 
\bea
\label{eq:def_A1}
A_1 &\equiv& A_n+A_p+A_{e}+A_{\nu_e},\\
\label{eq:def_A2}
A_2 &\equiv& A_n+A_p+A_{\mu}+A_{\nu_\mu}.
\eea

The equilibrium shifts of neutron and electron densities 
can be found from the $\lambda\to \infty$ limit of
Eqs.~\eqref{eq:delta_n} and \eqref{eq:delta_e},
respectively (see also Ref.~\cite{Alford2020})
\bea\label{eq:delta_n_eq}
\delta n_n^{\rm eq} 
&=&
\frac{\theta}{i\omega N_B }\Big\{\left[-A_p A_L-(A_e+A_{\nu_e})
(A_{\mu}+A_{\nu_\mu} )\right] n_{B0} +A_{\nu_e}(A_{\mu}+A_{\nu_\mu}) 
n_{L_e0}+A_{\nu_\mu}(A_e+A_{\nu_e}) n_{L_\mu 0}\Big\},\\
\label{eq:delta_e_eq}
\delta n_{e}^{\rm eq} &=&  -\frac{\theta}{i\omega N_B} 
\left[A_n (A_{\mu}+A_{\nu_\mu})n_{B0} +A_{\nu_e}A_2\,
n_{L_e0}-A_{\nu_\mu}(A_n+A_p) n_{L_\mu 0}\right].
\eea 
Finally, for the nonequilibrium shifts, we find
\bea\label{eq:delta_n_non_eq}
\delta n'_n 
&=& 
-\theta A_L\frac{N_B n_{n0}+ N_p n_{B0} +N_e n_{L_e0}+
N_\mu n_{L_\mu 0}}{N_B(i\omega A_L+{\lambda} N_B)},\\
\label{eq:delta_e_non_eq}
\delta n'_{e} &=& \frac{\theta(A_{\mu}+A_{\nu_\mu})} {N_B (i\omega {A}_L
 +{\lambda} N_B)} \left(N_n n_{n0}+N_p n_{p0} +
 N_{e} n_{L_e0}+N_{\mu} n_{L_\mu 0} \right),
\eea 
which can be written in a compact form 
\bea\label{eq:delta_n_non_eq2}
\delta n'_n &=& - \frac{\theta C}
{B(i\omega +{\lambda} B)},\\
\label{eq:delta_e_non_eq2}
\delta n'_{e} &=& \frac{A_{\mu}+A_{\nu_\mu}}{A_L}
\frac{\theta C} {B (i\omega  +{\lambda} B)},
\eea 
where $B=N_B/A_L$, and $C=(N_n n_{n0}+ N_p n_{p0} 
+N_e n_{L_e0}+N_\mu n_{L_\mu 0})/A_L$.

The full expression for the out-of-equilibrium pressure is given by 
\bea
p(t)=p(n_j(t))=p\left[n_{j0}+\delta n_j^{\rm eq}(t)\right]
+\delta p'(t)=p^{\rm eq}(t)+\delta p'(t),
\eea 
where the nonequilibrium part of the pressure, 
referred to as bulk viscous pressure, is given by
\bea\label{eq:Pi}
\Pi \equiv \delta p' =\sum_j
\left(\frac{\partial p}{\partial n_j}\right)_0\delta n'_j.
\eea 
Using the Gibbs-Duhem relation $dp=sdT+\sum_i n_i d \mu_i$, which is
valid also out of equilibrium, we can write \footnote{Note that the
temperature is assumed to be constant because, as argued in the
Sec.~\ref{sec:intro}, we assume that the thermal equilibration rate
is much larger than the chemical equilibration rate.}
\bea\label{eq:c_j_def}
c_j &\equiv &
\left(\frac{\partial p}{\partial n_j}\right)_0
=\sum_i n_{i0} \left(\frac{\partial \mu_i}{\partial n_j}\right)_0
=\sum_i n_{i0}A_{ij}.
\eea
Then, using also the relations $\delta n'_p = -\delta n'_n$,
$\delta n'_{\mu} = \delta n'_p-\delta n'_e$,
$\delta n'_{\nu_l} = -\delta n'_l$, we obtain 
\bea\label{eq:Pi1}
\Pi = \frac{\theta C}{A_LB(i\omega +{\lambda} B)}
\bigg[-(c_n-c_p-c_\mu +c_{\nu_\mu})A_L+(c_e-c_{\nu_e}-c_\mu +c_{\nu_\mu})
(A_{\mu}+A_{\nu_\mu})\bigg].
\eea 
Writing out  Eq.~\eqref{eq:c_j_def}  for each particle species 
and recalling the definitions of relations
$A_{pn}=A_{np}=A_{nn}-A_n =A_{pp}-A_{p}$ we find 
\bea
c_l=n_{l0}A_l,\quad
c_{\nu_l}=n_{\nu_l0}A_{\nu_l},\hspace{1.8cm} \\
c_n=n_{n0}A_{nn}+n_{p0}A_{pn}=n_{B0}A_{nn}-n_{p0}A_{n},\hspace{0.5cm} \\ 
c_p=n_{p0}A_{pp}+n_{n0}A_{np}=n_{B0}(A_{nn}-A_n)+n_{p0}A_{p},
\eea
which allows us to write \eqref{eq:Pi1} as 
\bea\label{eq:Pi2}
\Pi =-\frac{\theta C^2}{B(i\omega +{\lambda} B)}.
\eea 
The bulk viscosity is defined as the real part of $-\Pi/\theta$, \ie,
\bea\label{eq:zeta}
\zeta =\frac{C^2}{B}\frac{\gamma}{\omega^2+\gamma^2}, \qquad \gamma={\lambda}B.
\eea

Bulk viscosity given by Eq.~\eqref{eq:zeta} has the classic 
resonant form which depends on two quantities:
the thermodynamic prefactor $C^2/B$ which depends 
only on the EoS, and the relaxation rate $\gamma$
which depends on the weak interaction rates of electron 
and muon Urca processes. The limit of the absence of 
 muons is obtained from the above equations  
by setting $n_\mu=n_{\nu_\mu}=0$ and taking the limit
$A_\mu, A_{\nu_\mu}\to \infty$. Then $A_L= A_{\mu}+A_{\nu_\mu}$, 
and  the previous expressions \eqref{eq:N_p_def} and \eqref{eq:N_e_def} reduce to 
\bea\label{eq:N_n_nomu}
N_p = - (A_p+A_e+A_{\nu_e})A_L,\quad
N_e = A_{\nu_e}A_L,
\eea 
and 
\bea\label{eq:coeff_B1_C1}
B =A_1,\quad 
C 
= A_n n_{n0}- A_p n_{p0} -A_e n_{e0} +A_{\nu_e}n_{\nu_e0}.
\eea 
The coefficients $B$ and $C $ coincide with those given 
in our previous work~\cite{Alford2019b}.

\subsection{Slow lepton-equilibration limit}

When the lepton equilibration processes \eqref{eq:reaction_L1}, \eqref{eq:reaction_L2}, \eqref{eq:reaction_L3}
are slow compared to the Urca processes, \ie, 
$\mu_{\Delta_e}\neq \mu_{\Delta_\mu}$, there are two 
independent shifts in this case. Now $I_L\simeq0$, 
and rate equations take the form
\bea\label{eq:cont_n_slow}
\frac{\partial}{\partial t}\delta n_n(t) &=& 
-\theta  n_{n0} -\lambda_e\mu_{\Delta_e}(t)
-\lambda_{\mu} \mu_{\Delta_\mu}(t),\\
\label{eq:cont_p_slow}
\frac{\partial}{\partial t}\delta n_p(t) &=& 
-\theta n_{p 0} +\lambda_e\mu_{\Delta_e}(t)
+\lambda_{\mu} \mu_{\Delta_\mu}(t),\\
\label{eq:cont_e_slow}
\frac{\partial}{\partial t}\delta n_e(t) &=& 
-\theta n_{e 0} +\lambda_e\mu_{\Delta_e}(t),\\
\label{eq:cont_mu_slow}
\frac{\partial}{\partial t}\delta n_\mu(t) &=& 
-\theta n_{\mu 0} +\lambda_{\mu}\mu_{\Delta_\mu}(t).
\eea 
Substituting Eqs.~\eqref{eq:delta_mu_1} and \eqref{eq:delta_mu_2}
in the rate equations and assuming the same time-dependence of perturbation as 
above we find 
\bea
i\omega\delta n_n &=& 
-n_{n0}\theta -(\lambda_e+\lambda_{\mu}) A_n\delta n_n
+(\lambda_e+\lambda_{\mu}) A_p\delta n_p
-\lambda_e A_{\nu_e} \delta n_{\nu_e}+\lambda_e A_e\delta n_e
-\lambda_{\mu} A_{\nu_\mu} \delta n_{\nu_\mu}+\lambda_{\mu} A_\mu\delta n_\mu,\\
i\omega\delta n_e &=& -n_{e0}\theta +\lambda_e A_n\delta n_n
- \lambda_e A_p\delta n_p+\lambda_e 
A_{\nu_e} \delta n_{\nu_e}-\lambda_e A_e\delta n_e.
\eea 
This system of equations is closed upon using the relations 
$\delta n_p+\delta n_n=\delta n_B$,
$\delta n_e+\delta n_\mu=\delta n_p$, $\delta n_{L_e} 
=\delta n_e+\delta n_{\nu_e}$,
and $\delta n_{L_\mu}=\delta n_\mu+\delta n_{\nu_\mu}$, which leads us to  ($\lambda\equiv\lambda_e+\lambda_{\mu}$)
\bea
\label{eq:delta_ne}
\delta n_e &=& \frac{-n_{e0}\theta +\lambda_e (A_n+A_p)\delta n_n
- \lambda_e A_p \delta n_B +\lambda_e A_{\nu_e} \delta n_{L_e}}
{i\omega+\lambda_e (A_e+A_{\nu_e})},\\
\label{eq:delta_nn}
i\omega\delta n_n 
&=&
-n_{n0}\theta -(\lambda A_n+\lambda A_p+\lambda_{\mu} A_\mu+
\lambda_{\mu} A_{\nu_\mu})\delta n_n+(\lambda_e A_e+
\lambda_e A_{\nu_e}-\lambda_{\mu} A_\mu-\lambda_{\mu} A_{\nu_\mu})\delta n_e \\
&& +(\lambda A_p+\lambda_{\mu} A_\mu+\lambda_{\mu} A_{\nu_\mu})\delta n_B 
-\lambda_e A_{\nu_e}\delta n_{L_e} -\lambda_{\mu} A_{\nu_\mu}\delta n_{L_\mu}.
\eea
The coupled Eqs.~\eqref{eq:delta_ne} and \eqref{eq:delta_nn} 
can be solved to find 
\bea\label{eq:delta_nn1}
D\delta n_n
 &=&-\frac{\theta}{i\omega}\bigg\{i\omega \left[n_{n0}(i\omega+\lambda_e A_e
 +\lambda_e A_{\nu_e}) +n_{e0}(\lambda_e A_e+\lambda_e A_{\nu_e}-
 \lambda_{\mu} A_\mu -\lambda_{\mu} A_{\nu_\mu})\right]\nonumber\\
 &+&\left[i\omega(\lambda A_p+ \lambda_{\mu} A_\mu+\lambda_{\mu} A_{\nu_\mu})
 +\lambda_e \lambda_{\mu} (( A_1- A_n)( A_2-A_n)- A_p^2)\right]n_{B0}\nonumber\\
 &-& \lambda_e A_{\nu_e}\left[i\omega +\lambda_{\mu} (A_\mu+ A_{\nu_\mu}) 
 \right] n_{L_e0} -\lambda_{\mu} A_{\nu_\mu}\left[i\omega+\lambda_e 
 (A_e+ A_{\nu_e})\right] n_{L_\mu 0}\bigg\},\\
\label{eq:delta_ne1}
D\delta n_e 
&=&-\frac{\theta}{i\omega}\bigg\{i\omega n_{e0} 
\Big[i\omega +\lambda_{\mu} A_2+\lambda_e (A_n+A_p)\Big]
- \lambda_e  n_{B0} \Big[A_p
(i\omega+\lambda_{\mu} A_2)-\lambda_{\mu}(A_n+A_p)(A_2-A_n)\Big]\nonumber\\
&+&\lambda_e  n_{L_e0}A_{\nu_e}
(i\omega+\lambda_{\mu} A_2)
+\lambda_e (A_n+A_p) i\omega n_{n0}  
-\lambda_e \lambda_{\mu}(A_n+A_p) A_{\nu_\mu} n_{L_\mu 0}\bigg\},
\eea 
where 
\bea\label{eq:D}
D
=(i\omega+\lambda_e A_1)(i\omega+\lambda_{\mu} A_2)
-\lambda_e \lambda_{\mu}(A_n+A_p)^2. 
\eea 
The equilibrium shifts $\delta n_f^{\rm eq}$ are found as the 
limit $\lambda_i\to \infty$ of Eqs.~\eqref{eq:delta_nn1} and 
\eqref{eq:delta_ne1}. 
However, as we showed in Ref.~\cite{Alford2020}, one can use 
the quasiequilibrium solutions $\delta n_f^{0}=-{\theta}
 n_{f0}/{i\omega}$ instead, which arise in the $\lambda_{e,\mu}\to 0$ limit of 
Eqs.~\eqref{eq:delta_nn1} and \eqref{eq:delta_ne1}. We then find
\bea\label{eq:delta_nn5}
\delta n'_n 
 &=&\frac{\theta}{i\omega}
\frac{i\omega (\lambda_e {C_1} +\lambda_{\mu}{C_2})
+\lambda_e \lambda_{\mu} \big[{C_2}  (A_e+A_{\nu_e}) 
+{C_1} (A_\mu+A_{\nu_\mu})\big]}
{(i\omega+\lambda_e A_1)(i\omega+\lambda_{\mu} A_2)
-\lambda_e \lambda_{\mu}(A_n+A_p)^2},\\
\label{eq:delta_ne2}
\delta n'_e 
&=&-\frac{\theta}{i\omega}\frac{i\omega\lambda_e C_1
+\lambda_e \lambda_{\mu} \big[A_2 C_1 - (A_n+A_p)C_2\big]}
{(i\omega+\lambda_e A_1)(i\omega+\lambda_{\mu} A_2)
-\lambda_e \lambda_{\mu}(A_n+A_p)^2}.
\eea 
The bulk viscous pressure then reads
\bea\label{eq:Pi_slow1}
\Pi 
&=& \frac{\theta}{i\omega}
\frac{i\omega (\lambda_e C_1^2+\lambda_{\mu} C_2^2)
+\lambda_e \lambda_{\mu} \big[A_1 C_2^2 +A_2 C_1^2-2(A_n+A_p)C_1C_2\big]}
{(i\omega+\lambda_e A_1)(i\omega+\lambda_{\mu} A_2)
-\lambda_e \lambda_{\mu}(A_n+A_p)^2},
\eea 
where we used the relations 
\bea
\label{eq:def_C1}
c_n-c_p-c_e +c_{\nu_e}=n_{n0}A_n-n_{p0}A_{p}-
n_{e0}A_e+n_{\nu_e0}A_{\nu_e}\equiv C_1,\\ 
\label{eq:def_C2}
c_n-c_p-c_\mu +c_{\nu_\mu}=n_{n0}A_n-n_{p0}A_{p}-
n_{\mu0}A_\mu+n_{\nu_\mu0}A_{\nu_\mu}\equiv C_2.
\eea
Extracting the real part of Eq.~\eqref{eq:Pi_slow1} 
leads to the  final expression of the bulk-viscosity
\bea\label{eq:zeta_slow}
\zeta = \frac{\lambda_e \lambda_{\mu}\Big\{\lambda_e 
\left[(A_n+A_p) C_1- A_1 C_2\right]^2+\lambda_{\mu} \left[(A_n+A_p) C_2- A_2 C_1\right]^2 \Big\}+\omega^2(\lambda_e C_1^2+\lambda_{\mu} C_2^2)}
{\Big\{\lambda_e \lambda_{\mu}\left[A_1A_2-(A_n+A_p)^2\right]-\omega^2\Big\}^2
+\omega^2(\lambda_e A_1+\lambda_{\mu} A_2)^2}.
\eea
If the muon contribution is neglected ($\lambda_{\mu} = 0$) 
Eq.~\eqref{eq:zeta_slow} reduces to
\bea\label{eq:zeta_slow1}
\zeta_e = \frac{C_1^2}{A_1}\frac{\gamma_e}
{\omega^2+\gamma_e^2},
\eea
with $\gamma_e=\lambda_e A_1$, which coincides with 
the result of our previous work~\cite{Alford2019b}.

In the limit of high frequencies $\omega\gg \lambda A$ we find
from Eq.~\eqref{eq:zeta_slow}
\bea\label{eq:zeta_slow2}
\zeta = \frac{\lambda_e C_1^2+\lambda_{\mu} C_2^2}
{\omega^2}=\zeta_e+\zeta_\mu,
\eea
where $\zeta_e$ and $\zeta_\mu$ are the contributions
by electrons and muons, respectively.

In the low-frequency limit
\bea\label{eq:zeta_slow3}
\zeta 
=\frac{\lambda_e(C_1- a_1 C_2)^2
+\lambda_{\mu} (C_2- a_2 C_1)^2}
{\lambda_e \lambda_{\mu}(A_n+A_p)^2 (a_1 a_2-1)^2},
\eea
with $a_1=A_1/(A_n+A_p)$ and $a_2=A_2/(A_n+A_p)$.

\section{Numerical results}
\label{sec:num_results}

\subsection{$\beta$-equilibration rates}
\label{sec:rates}

\begin{figure}[t] 
\begin{center}
\includegraphics[width=0.45\columnwidth, keepaspectratio]{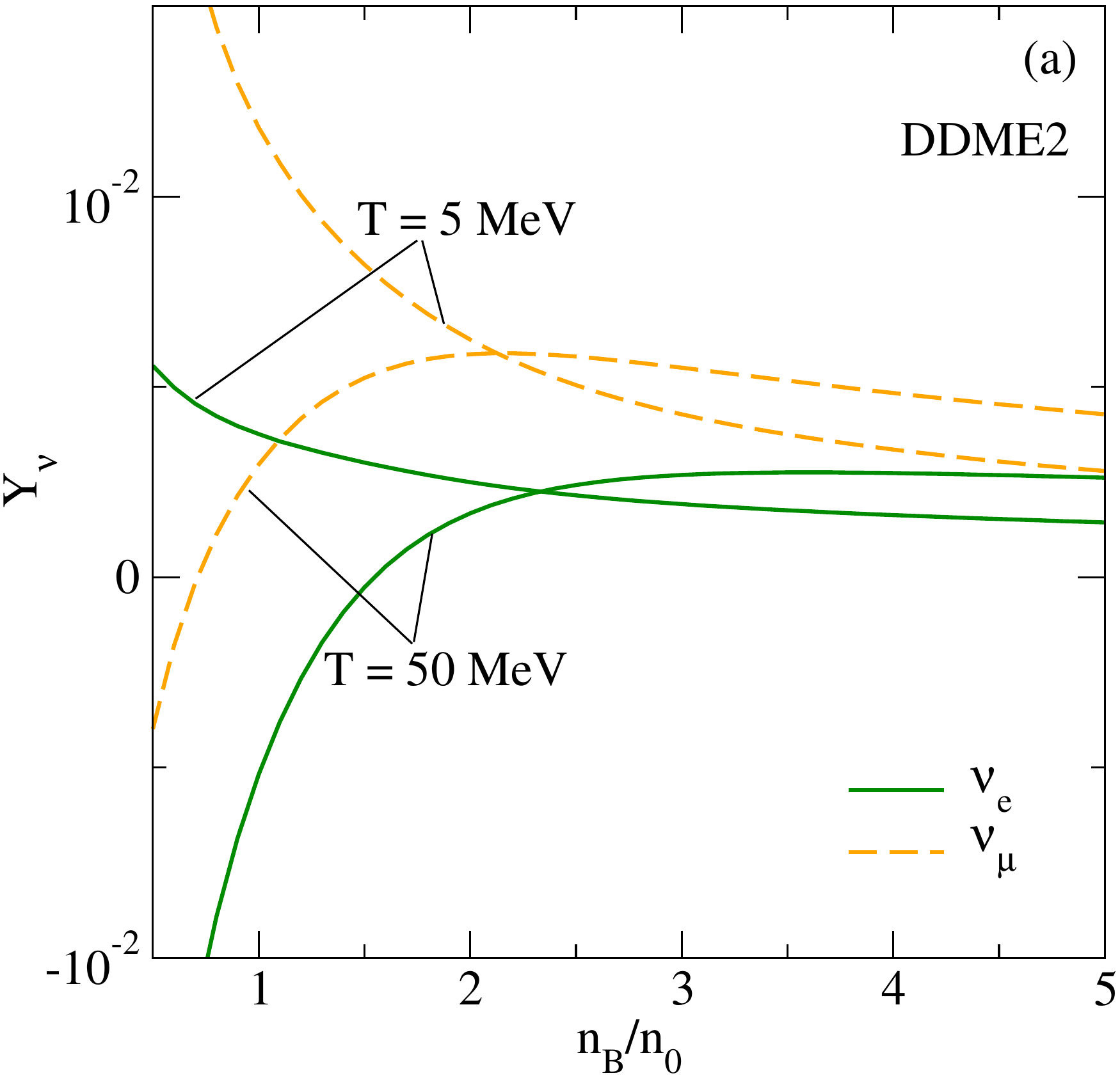}
\hspace{0.5cm}
\includegraphics[width=0.45\columnwidth, keepaspectratio]{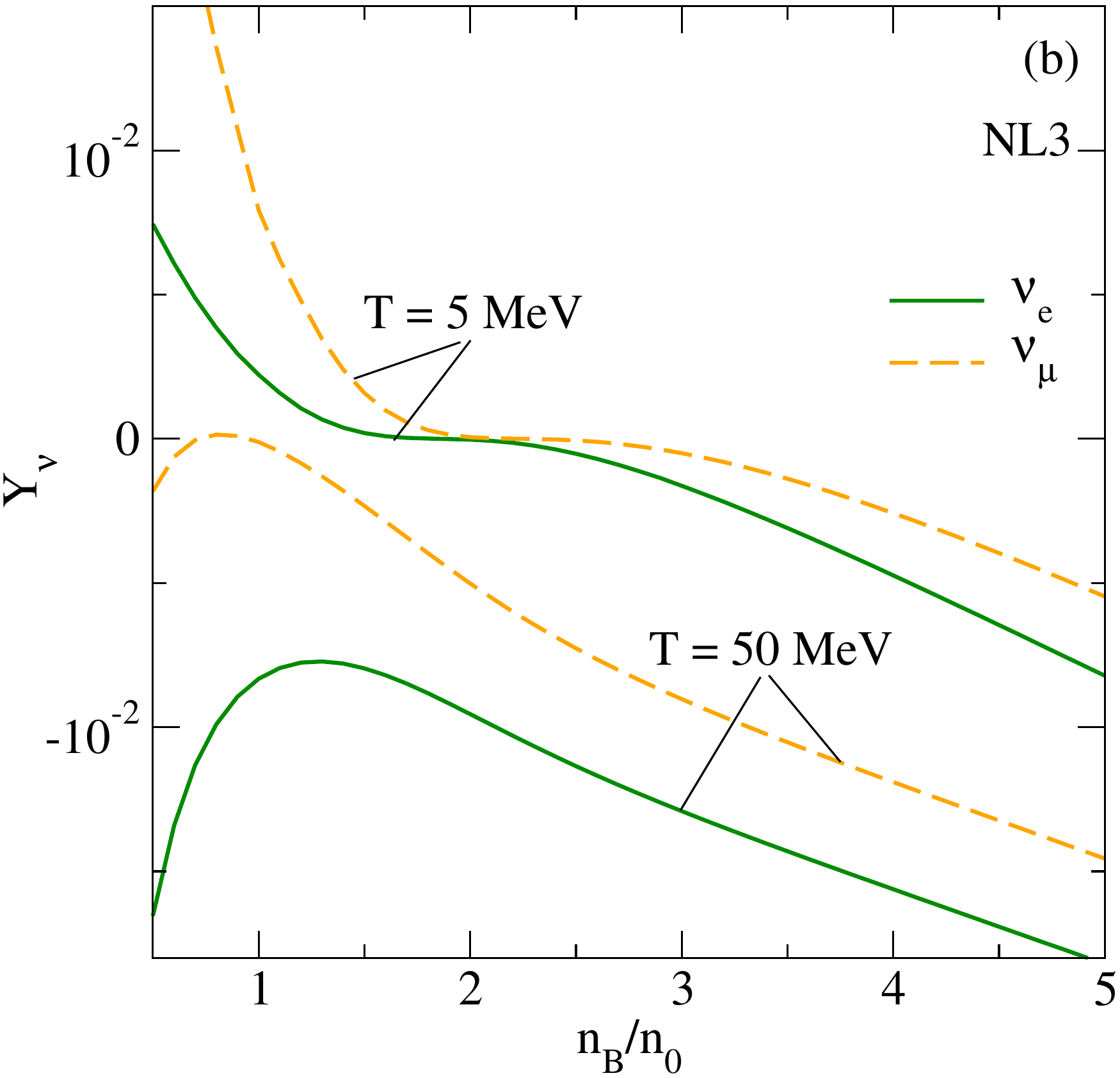}
\caption{ Neutrino fractions in neutron-star-merger matter with $Y_{L_e}=Y_{L_\mu}=0.1$ as 
  functions of the baryon density $n_B$ (in units of nuclear 
  saturation density $n_0$) for two values of the temperatures 
  for models DDME2 (a) and NL3 (b). 
  At high temperatures and 
  densities model NL3 becomes antineutrino-dominated. 
}
\label{fig:fractions_nu} 
\end{center}
\end{figure}

We start the discussion by presenting the relevant thermodynamics 
of $\beta$-equilibrated, neutrino-trapped $npe\mu$ matter for two 
parametrizations of the density functional theory -- the model 
DDME2 and the model NL3. The fractions of massive particles 
(\ie, nucleons, electrons and muons) are rather insensitive 
to the density and temperature. The particles abundances for 
$Y_{L_e}=Y_{L_\mu}=0.1$
are as follows: neutron fraction -- 80\%-82\%, proton 
fraction -- 18\%-20\%, electron fraction -- 9.5\%-10.5\%, muon fraction 
-- 9\%-10\% for model DDME2; and neutron fraction -- 77\%-81\%, 
proton fraction -- 19\%-23\%, electron fraction -- 10\%-12\%, muon 
fraction -- 9\%-11.5\% for model NL3 in the range $5\leq T\leq 50$ 
MeV and $1\leq n_B/n_0\leq 5$ with $n_0$ being the nuclear 
saturation density with the values $n_0=0.152$ fm$^{-3}$ 
for model DDME2 and $n_0=0.153$ fm$^{-3}$ for model NL3. 

In contrast to the massive particles, the fractions of 
neutrinos are rather sensitive both to the density and 
temperature, see Fig.~\ref{fig:fractions_nu}. At high 
temperatures and very low densities the net neutrino 
densities become negative in the DDME2 model, indicating 
that there are more antineutrinos than neutrinos in that 
regime. In the NL3 model instead only the low-temperature 
and the low-density regime is neutrino-dominated; the 
antineutrino population increases with the increase of 
both density and temperature. The reason for this behavior 
is the larger symmetry energy in the case of NL3 model
which favors larger proton fractions as compared to
the DDME2 model. Charge neutrality then requires larger electron
and muon fractions and, therefore, smaller neutrino 
fractions for the given values of $Y_{L_l}=Y_l+Y_{\nu_l}$.
Thus, we have an important difference in the composition of 
high-density and low-temperature, \ie, the degenerate regime of 
neutrino-trapped matter for these two models: while the trapped 
species are neutrinos in the DDEM2 matter, these are antineutrinos
in the case of the NL3 model. This feature leads to 
qualitatively different behavior of $\beta$-equilibration
rates and the bulk viscosity for these two models, see below.

\begin{figure}[t] 
\begin{center}
\includegraphics[width=0.45\columnwidth,keepaspectratio]{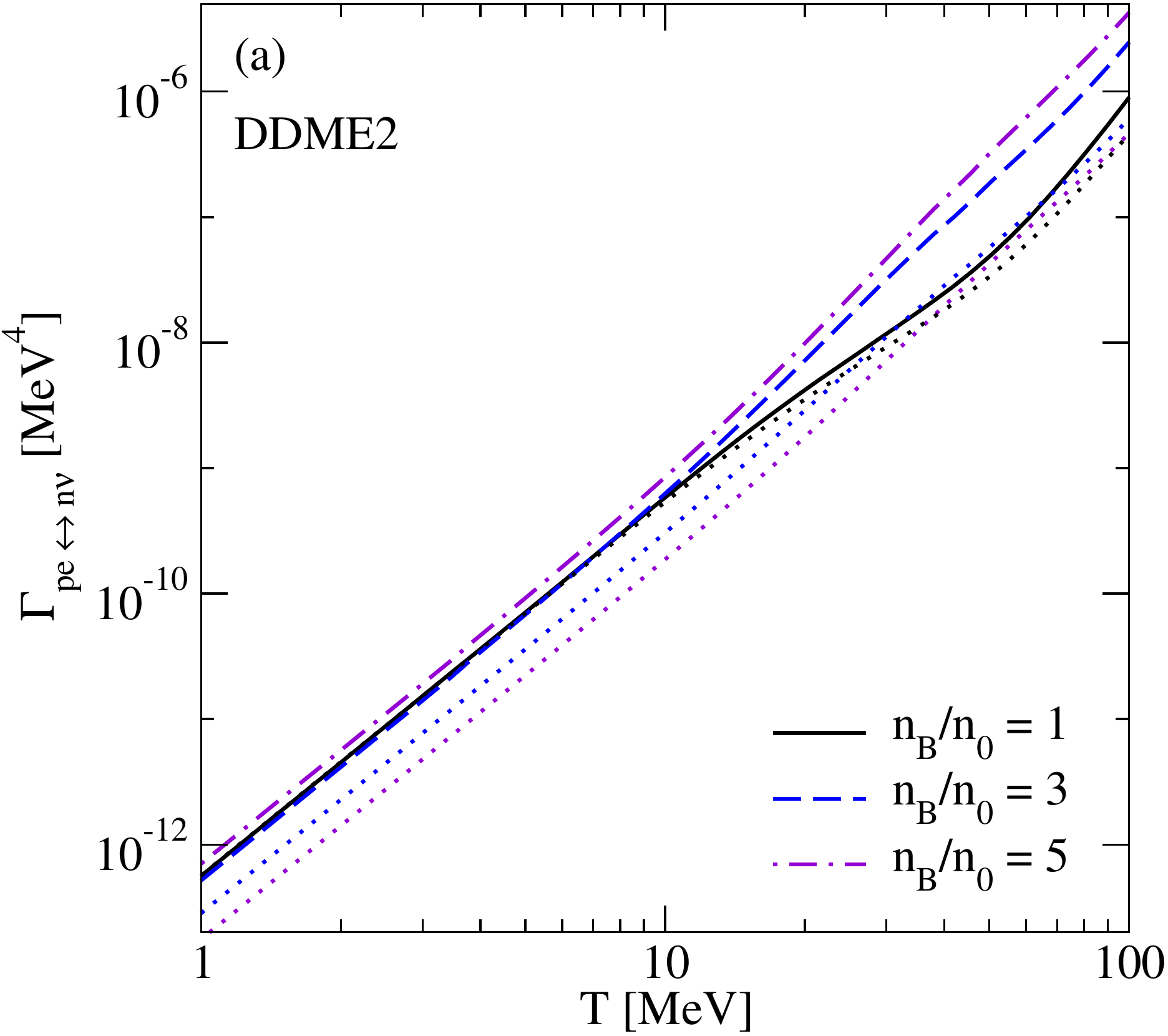}
\hspace{0.5cm}
\includegraphics[width=0.45\columnwidth,keepaspectratio]{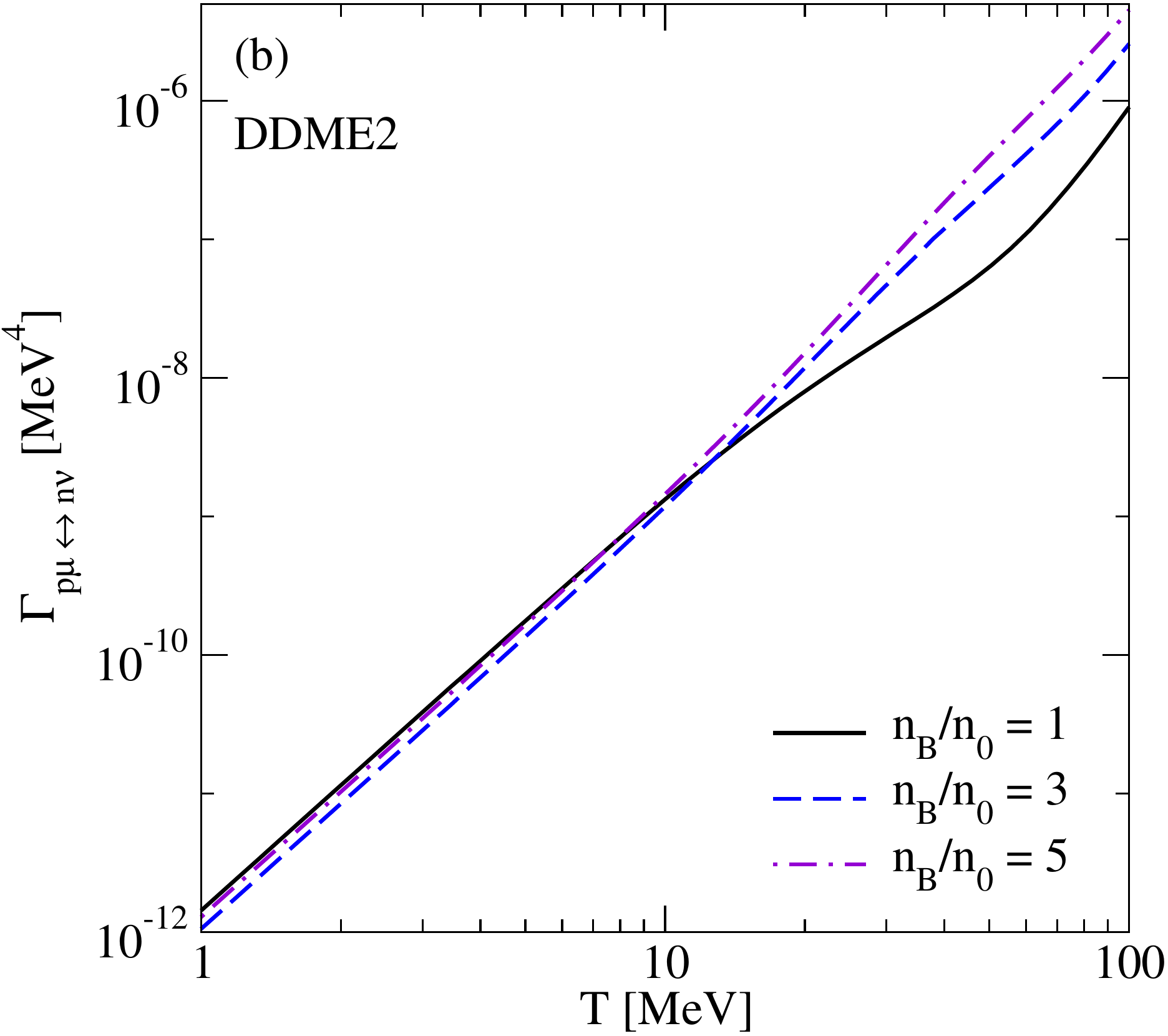}
\caption{ The electron capture rate $\Gamma_{pe\fromto n\nu}$ (a) and the muon capture rate $\Gamma_{p\mu\fromto n\nu}$ (b) 
as functions of the temperature for various densities 
  for the model DDME2. The neutron decay rates $\Gamma_{n\fromto p e \bar\nu}$ and $\Gamma_{n\fromto p \mu\bar\nu}$ are negligible compared to the lepton capture rates in the whole temperature-density range. 
   The dotted lines in panel (a) show the electron capture rates computed
  in Ref.~\cite{Alford2019b} within the approximation of nonrelativistic
  nucleons.}
\label{fig:Gamma_DDME2} 
\end{center}
\end{figure}

\subsubsection{Rates of Urca processes}

Next, we turn to the discussion of the Urca process rates.
As the neutron decay processes~\eqref{eq:e_decay} and 
\eqref{eq:mu_decay} involve antineutrinos, their rates
are expected to be much smaller than the lepton capture 
rates if the matter is neutrino-dominated, as discussed in Ref.~\cite{Alford2019b}. Our numerical calculations show that the neutron decay rate is negligibly small if 
the neutrino chemical potential (for the given lepton species) satisfies the condition $\alpha_{\nu_l}=\mu_{\nu_l}/T\geq -3$. This condition is satisfied for 
DDME2 model in the whole temperature-density range of 
interest, therefore the dominant equilibration processes
are the lepton capture processes. The rates of the electron
and muon capture processes for model DDME2 are shown, respectively, 
in panels (a) and (b) of Fig.~\ref{fig:Gamma_DDME2}. At  moderate temperatures, $T\leq 10$ MeV the lepton decay 
rates follow their low-temperature scaling given by
Eq.~\eqref{eq:Gamma2p_lowT_trap}, \ie, increase cubically 
with the temperature. At higher temperatures, this scaling 
breaks down. However, the deviation of the exact equilibration 
rates from their low-temperature limit is within a few 
factors (see Appendix~\ref{app:rates}). 
A comparison between the left and right panels in 
Fig.~\ref{fig:Gamma_DDME2} shows that the electron and
the muon capture rates are quite similar both qualitatively
and quantitatively. In panel (a) we show also the 
electron capture rates which were
computed in Ref.~\cite{Alford2019b} in the approximation
of nonrelativistic nucleons. As expected, the importance
of relativistic corrections to the nucleon spectrum 
rises with the density, and at the density $n_B/n_0=5$
the full relativistic rate is around one order of magnitude
larger than its nonrelativistic approximation. 

Figure~\ref{fig:Gamma_NL3} shows the summed $\beta$-equilibration (Urca) rates $\Gamma_l=\Gamma_{n\fromto p l \bar\nu}+\Gamma_{pl\fromto n\nu}$
for the model NL3. 
In contrast to the model DDME2, the model 
NL3 features two different regimes of equilibration -- the 
antineutrino-dominated regime in the low-temperature, high-density 
sector, where the dominant equilibration process is the neutron decay; 
and the neutrino-dominated regime in the high-temperature, low-density 
sector, where the dominant equilibration process is the lepton capture. 
As the antineutrino-dominated regime is realized at low temperatures 
and high densities where the matter is degenerate, the neutron decay 
rates follow the scaling $\propto T^3$ given by Eq.~\eqref{eq:Gamma1p_lowT_trap}. 
Numerically we find that the lepton capture rates are suppressed as long 
as the scaled-to-temperature neutrino chemical potential $\alpha_{\nu_l}\leq -6$. Although the net neutrino densities drop with the 
increase of temperature (see Fig.~\ref{fig:fractions_nu}), their scaled 
chemical potentials increase (remaining negative) thus allowing the 
neutrinos to come into the game already at $\alpha_{\nu_l}\simeq -6$. 
At higher temperatures, the neutron decay rates become suppressed exponentially, and the lepton capture processes become dominant at 
$\alpha_{\nu_l} \simeq -3$. As a consequence, there is always a sharp 
minimum in the net equilibration rate which arises in the transition 
region between these two regimes. The transition point moves to  higher 
temperatures with increasing density as the matter becomes more saturated 
with antineutrinos at higher densities. Note that there are no transitions 
at the density $n_B=n_0$; in this case, the lepton decay is the dominant 
process in the whole range of the temperature $1\leq T\leq 100$ MeV shown 
in Fig.~\ref{fig:Gamma_NL3}.

\begin{figure}[t] 
\begin{center}
\includegraphics[width=0.45\columnwidth,keepaspectratio]{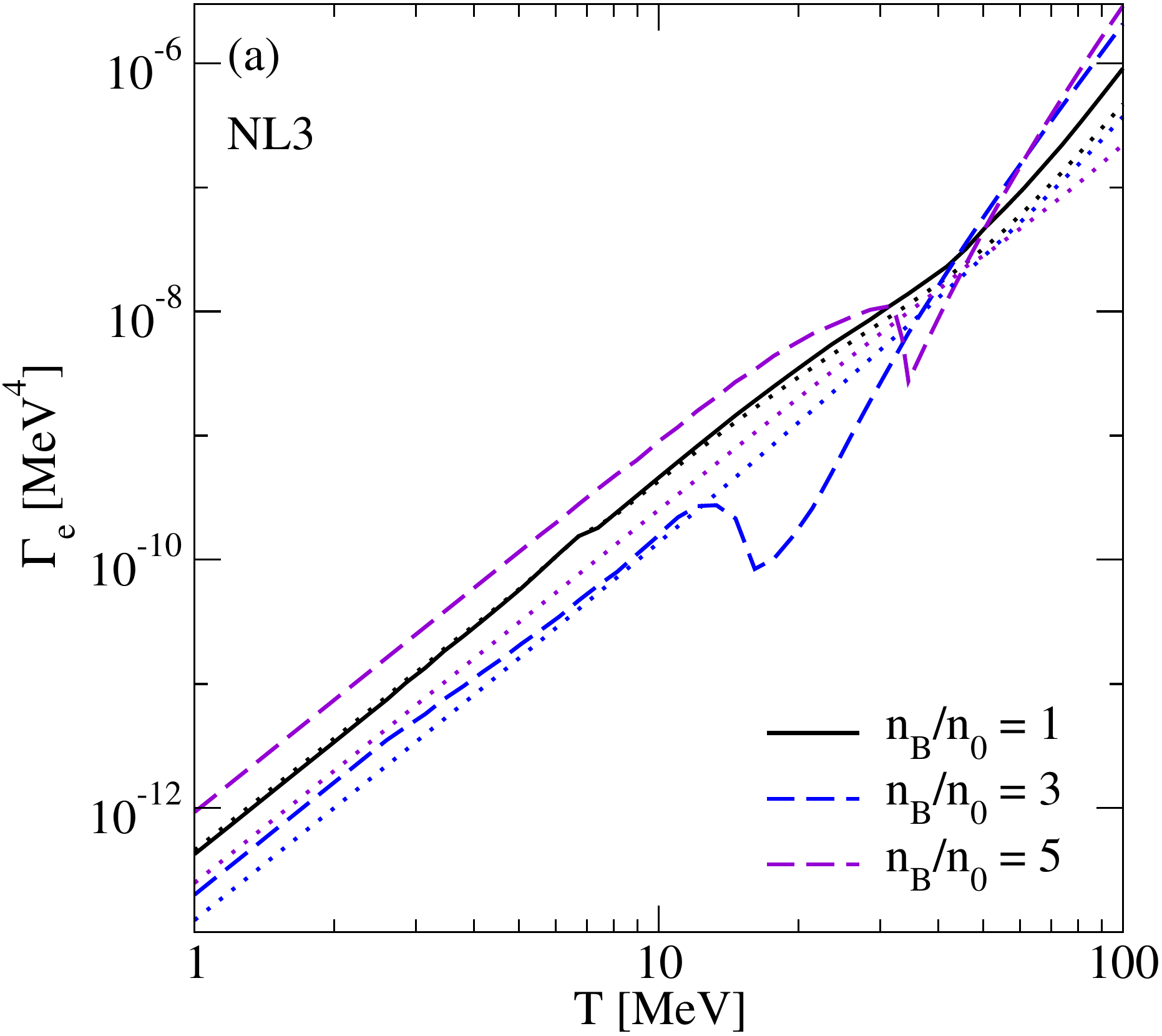}
\hspace{0.5cm}
\includegraphics[width=0.45\columnwidth,keepaspectratio]{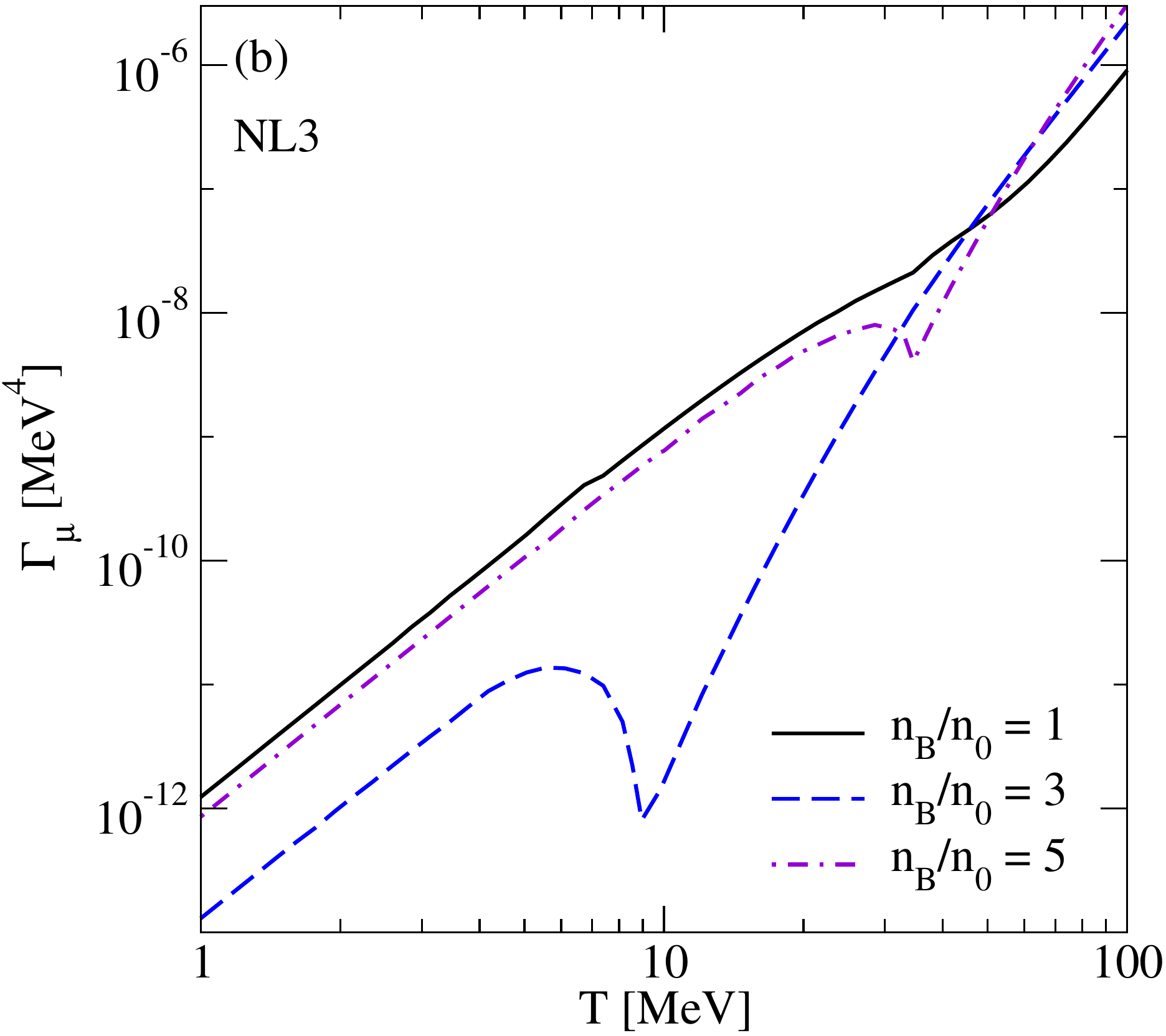}
\caption{ The summed $\beta$-equilibration rates $\Gamma_l=\Gamma_{n\fromto p l \bar\nu}+\Gamma_{pl\fromto n\nu}$ for electronic (a) and muonic (b)
  Urca processes as functions of the temperature for various densities 
  for the model NL3. In this case the dominant process is the neutron 
  decay at low temperatures and the lepton capture at high temperatures.
  The dotted lines in panel (a) show the electron capture rates computed
  in Ref.~\cite{Alford2019b} within the approximation of nonrelativistic
  nucleons.}
\label{fig:Gamma_NL3} 
\end{center}
\end{figure}
\begin{figure}[!] 
\begin{center}
\includegraphics[width=0.45\columnwidth,keepaspectratio]{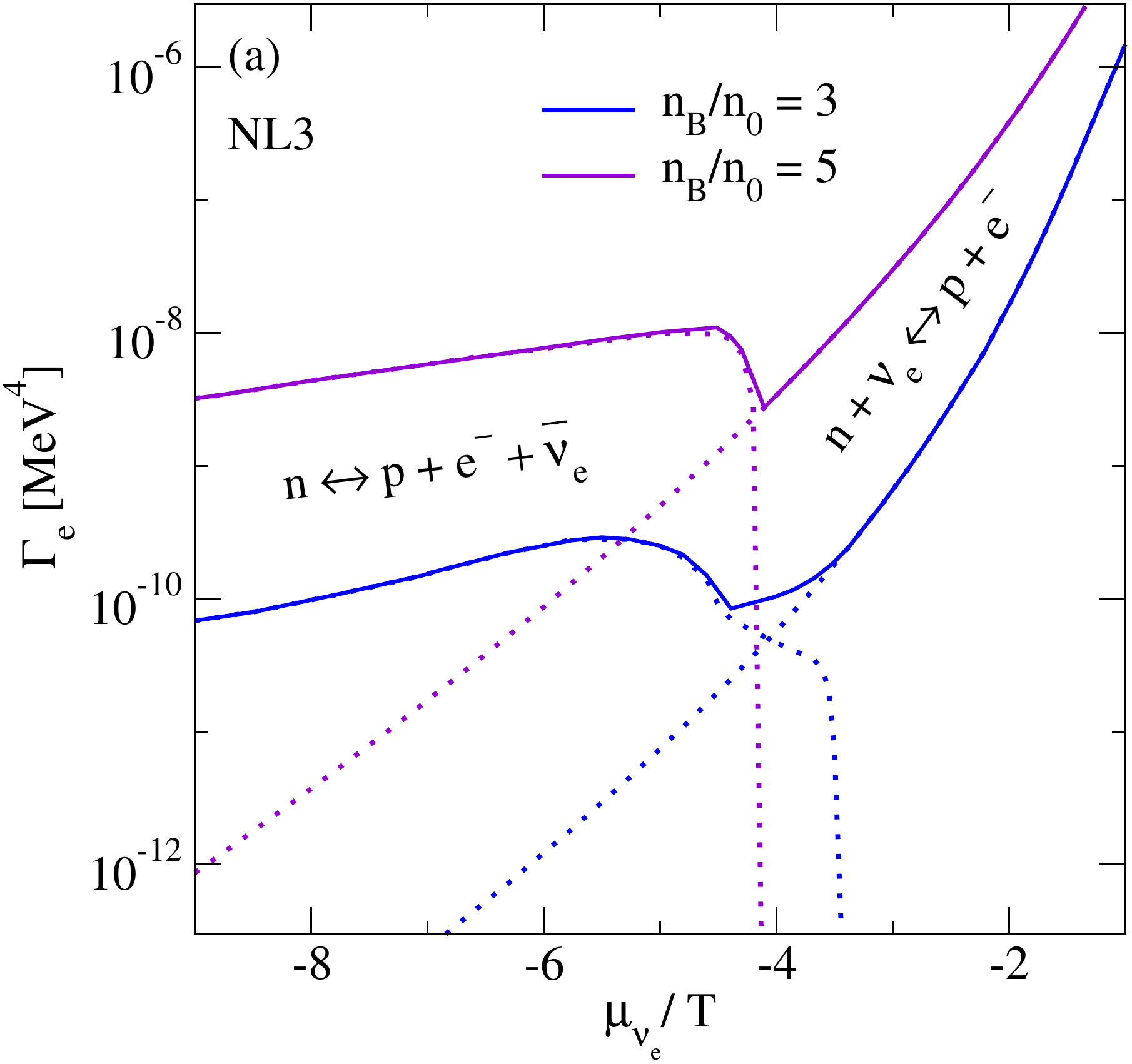}
\hspace{0.5cm}
\includegraphics[width=0.45\columnwidth,keepaspectratio]{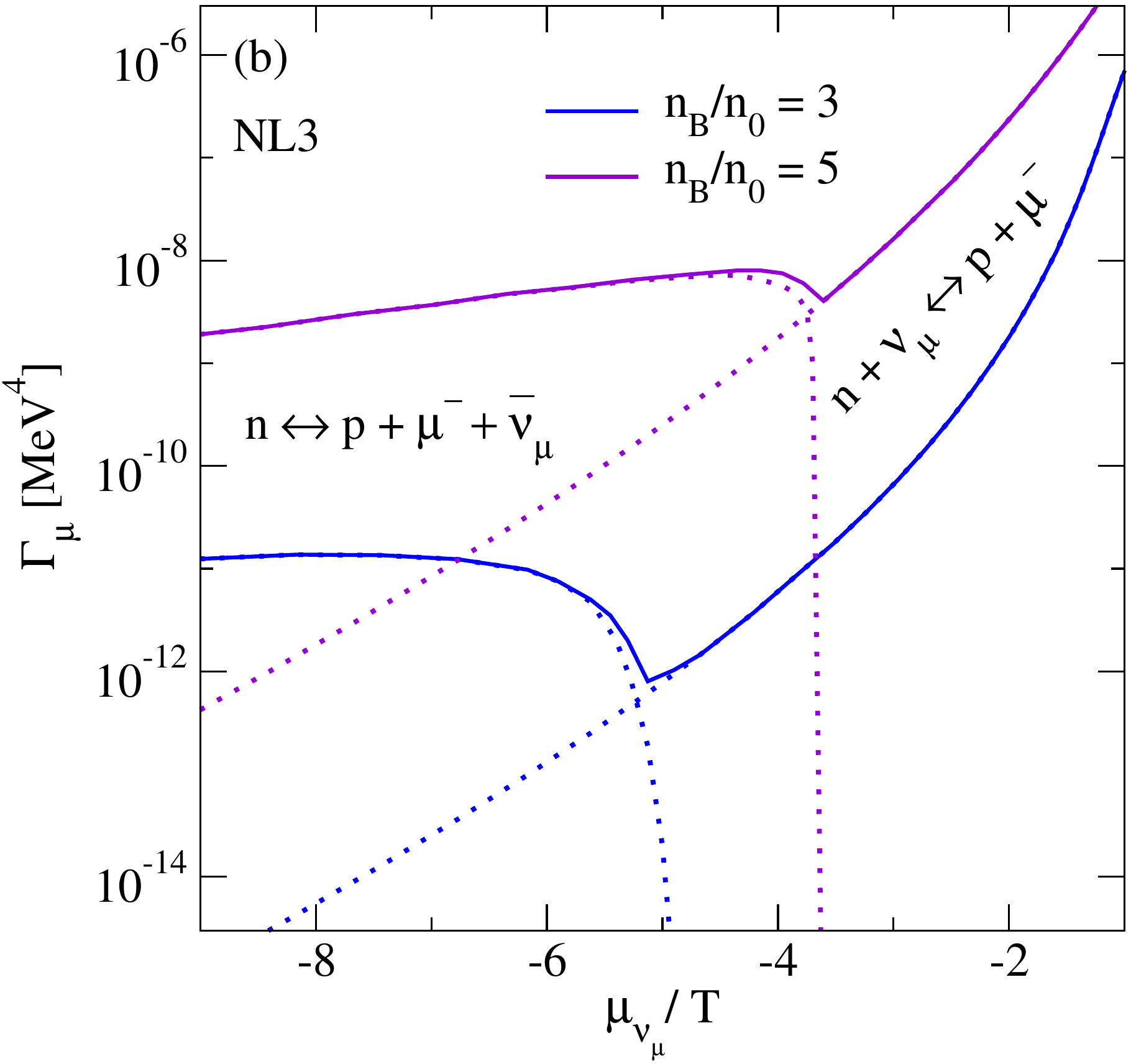}
\caption{ The relative rates of neutron decay and lepton capture 
 processes as functions of the scaled-to-temperature neutrino chemical potentials for electrons (a) and muons (b) for two values of the 
 density for the model NL3. }
\label{fig:Gamma12_NL3} 
\end{center}
\end{figure}

\begin{figure}[t]  
\begin{center}
\includegraphics[width=0.45\columnwidth,keepaspectratio]{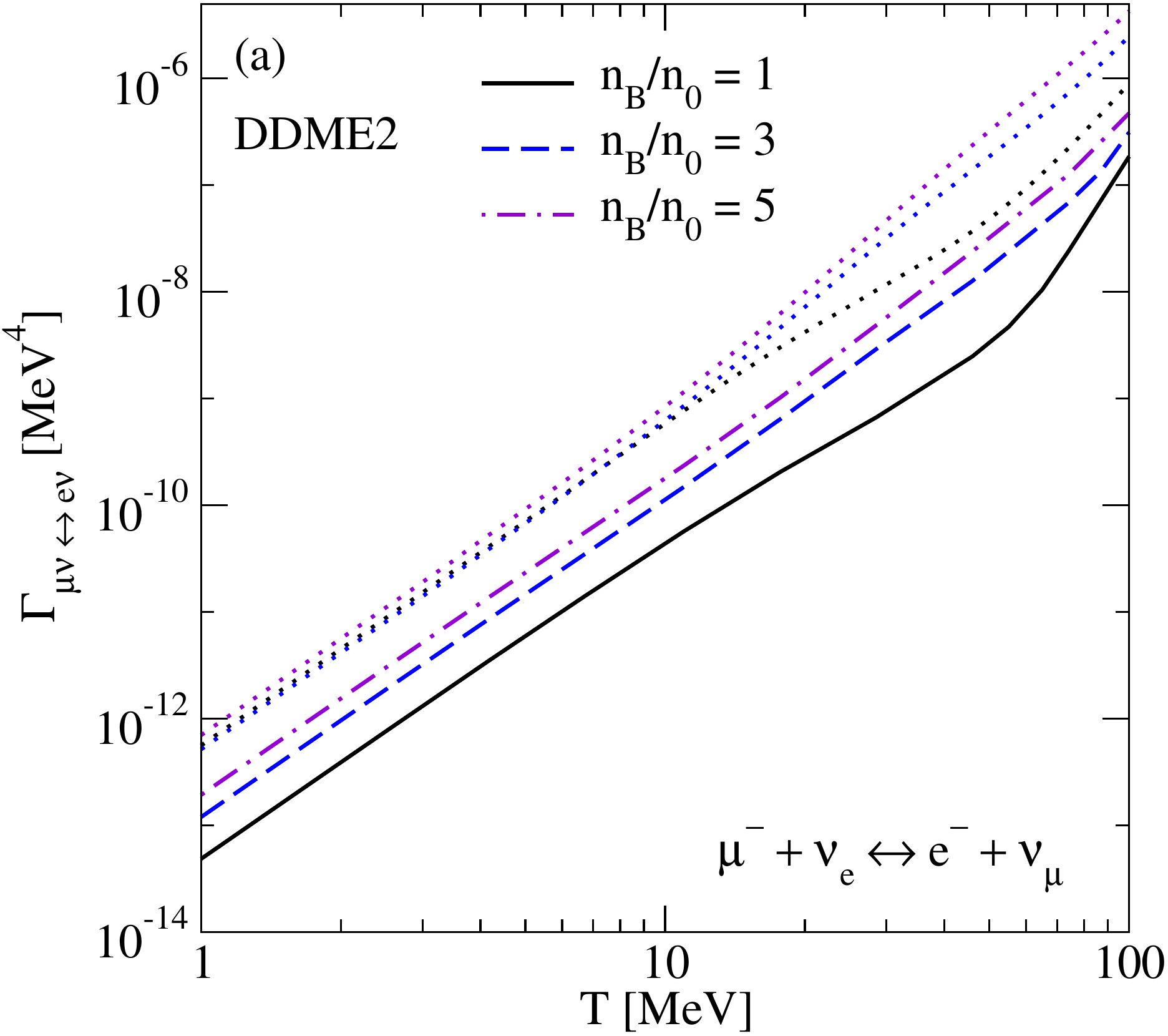}
\hspace{0.5cm}
\includegraphics[width=0.45\columnwidth,keepaspectratio]{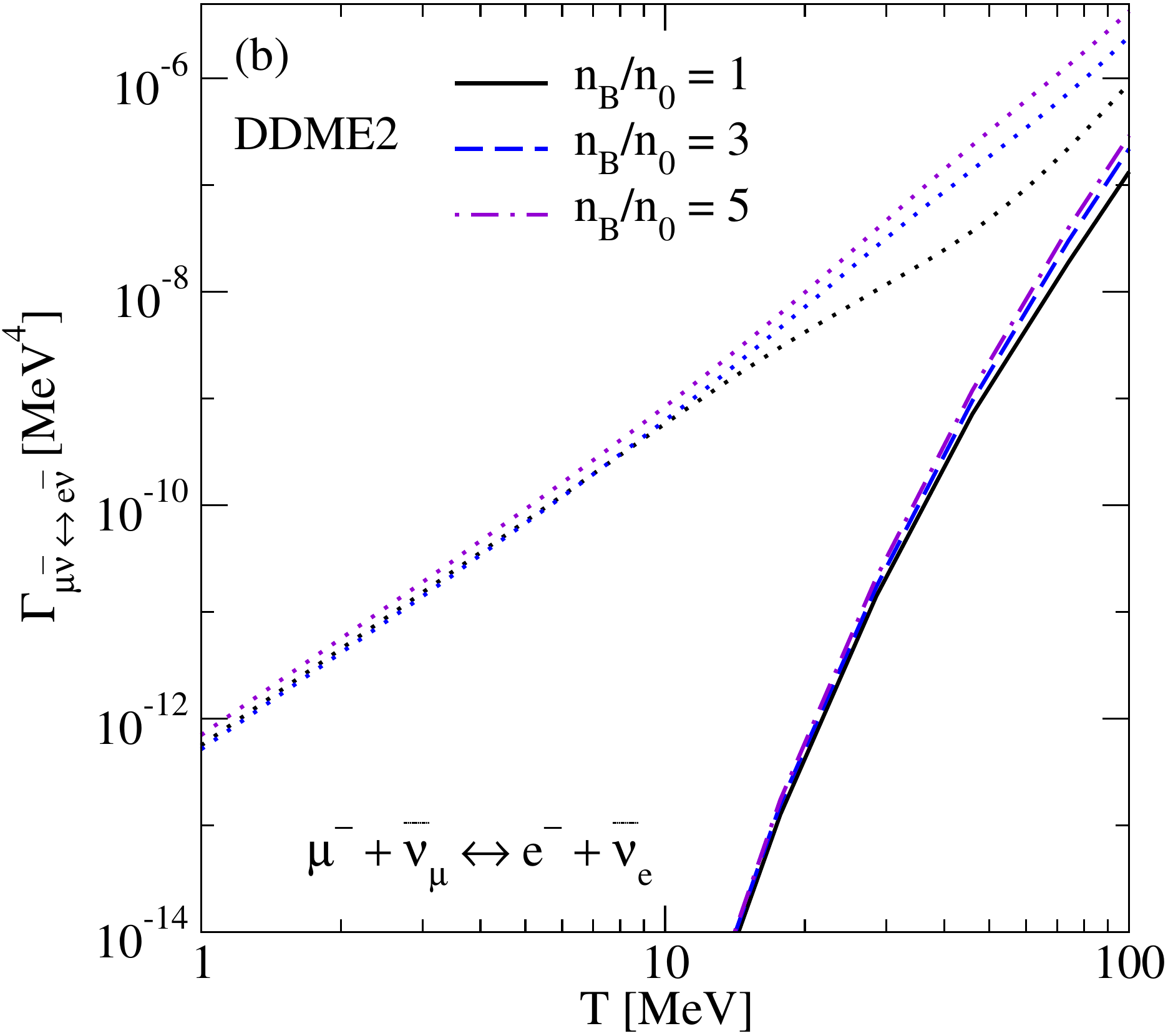}
\caption{Rates of leptonic $\beta$-equilibration processes 
as functions of the temperature for different values of the density 
for the model DDME2. The panel (a) refers to the neutrino absorption, 
and the panel (b) to the antineutrino absorption processes. We see that for DDME2 the leptonic rates are always at least an order of magnitude slower than the Urca electron
capture rates (shown by the dotted lines for comparison; the muon capture rates are slightly higher than the electron capture rates and are not shown.).}
\label{fig:Gamma_lep_DDME2}  
\end{center}
\end{figure}
\begin{figure}[!]  
\begin{center}
\includegraphics[width=0.45\columnwidth,keepaspectratio]{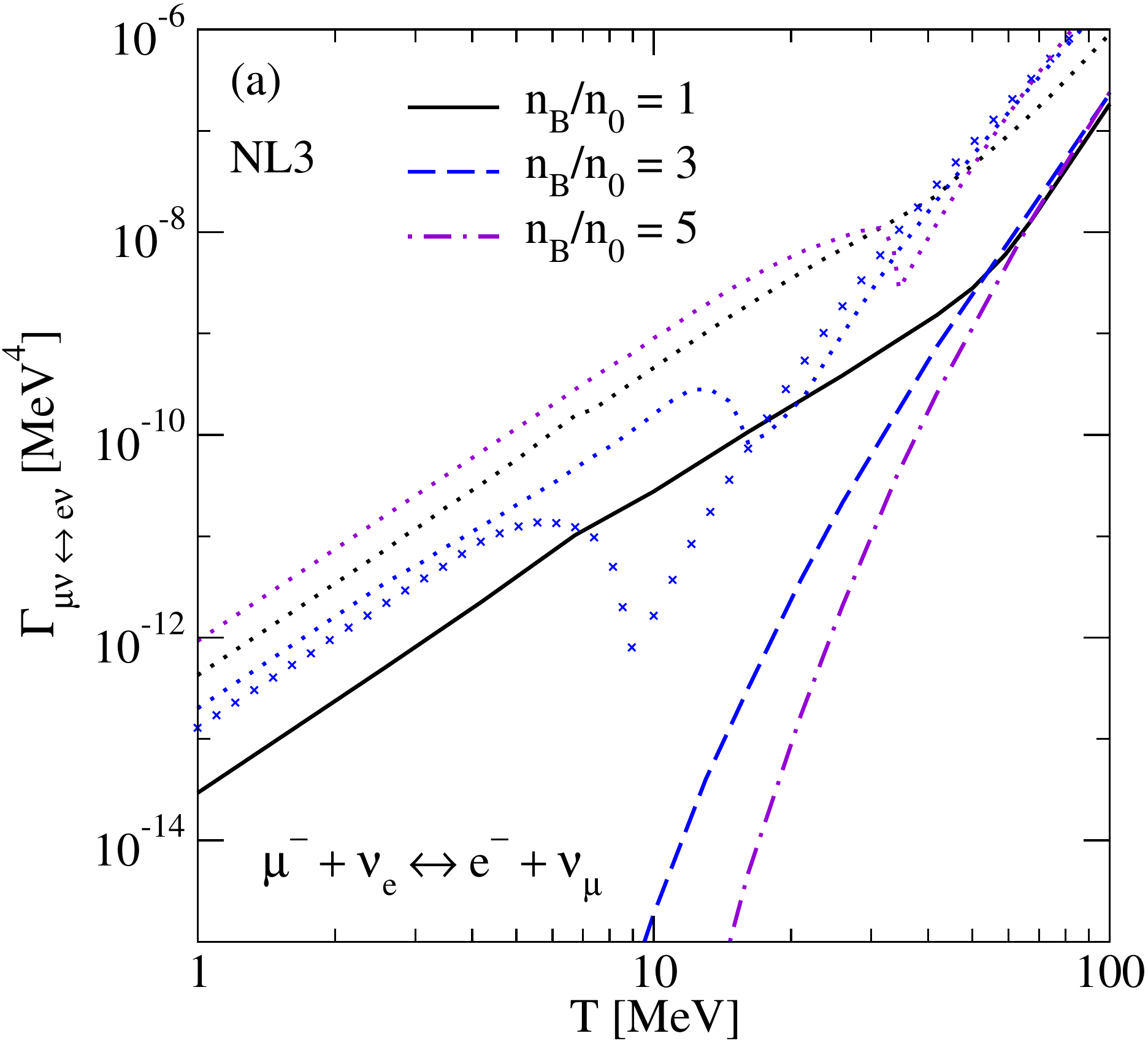}
\hspace{0.5cm}
\includegraphics[width=0.45\columnwidth,keepaspectratio]{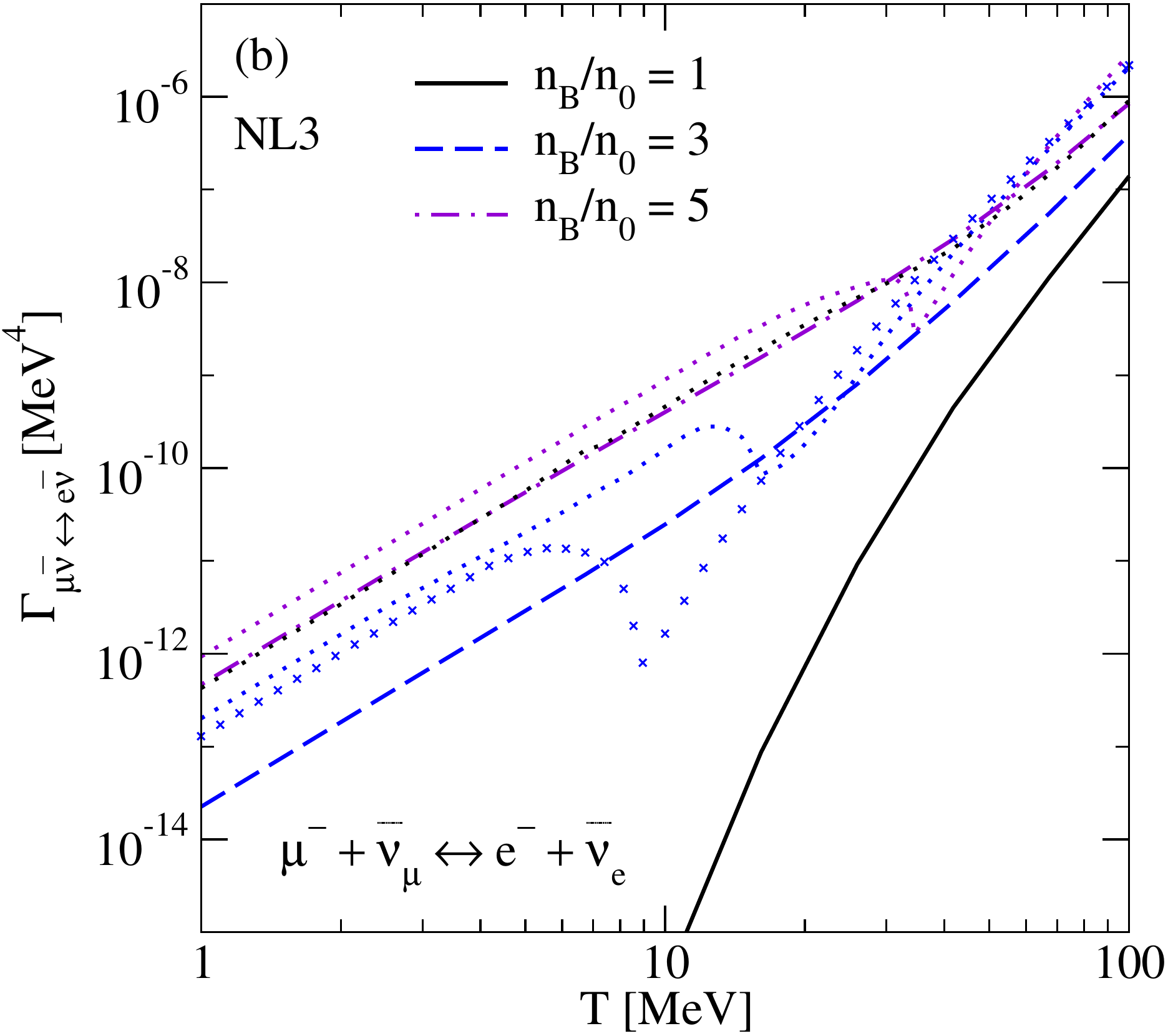}
\caption{Rates of leptonic $\beta$-equilibration processes 
as functions of the temperature for different values of the density 
for the model NL3. The panel (a) refers to the neutrino absorption, 
and the panel (b) to the antineutrino absorption processes. We see that for NL3 the leptonic rates are generally much slower than the summed Urca process rates $\Gamma_l\equiv\Gamma_{n\fromto p l \bar\nu}+\Gamma_{pl\fromto n\nu}$  (shown by the dotted lines for comparison for electrons; the muonic Urca process rates $\Gamma_\mu$ differ from $\Gamma_e$ significantly only at the density $n_B=3n_0$ and are shown by blue crosses) except near the transition point where the Urca rate goes through a minimum.
}
\label{fig:Gamma_lep_NL3} 
\end{center}
\end{figure}

To show the transition between the two regimes
we plot the equilibration rates for neutron decay and lepton 
capture processes separately as functions of the scaled chemical 
potentials $\alpha_{\nu_l}$ in Fig.~\ref{fig:Gamma12_NL3}.
As seen from the figure, the curves representing the rates
of the neutron decay and the lepton capture processes
intersect at a value of the scaled chemical potential within
the range $-5 \leq\alpha_{\nu_l}\leq -3$. Note that the regime 
of neutrino-dominated equilibration starts already around
$\alpha_{\nu_l}\simeq -3$, where the antineutrino density is 
still higher than the neutrino density. The reason for this 
is the difference in the available phase space for the 
neutron decay and lepton capture processes. Indeed, the 
lepton capture process has a larger kinematic phase space
than the neutron decay, therefore for equal densities of
neutrinos and antineutrinos (\ie, at vanishing neutrino 
chemical potential) the neutron decay rates are suppressed
as compared to the lepton decay rates. 

As in the case of DDME2 model, we show also the nonrelativistic 
electron capture rates in panel (a) of Fig.~\ref{fig:Gamma_NL3}.
The nonrelativistic approximation underestimates the exact rates
by factors from 1 to 10 in the regions away from the minimum, but
close to the minimum, we have the opposite behavior: the exact
relativistic rates are lower as there is no minimum in the 
nonrelativistic approximation (the transition between the 
antineutrino and neutrino-dominated regimes is smooth in the 
nonrelativistic approximation). We thus conclude that the 
sharp drop of the neutron decay rate and the minimum at the 
transition point is a purely relativistic effect and does not
appear in the nonrelativistic treatment.

\subsubsection{Rates of leptonic processes}

Next we discuss the results of the leptonic process rates 
given by Eqs.~\eqref{eq:Gamma1e_final}--\eqref{eq:Gamma3e_final}.
Figure~\ref{fig:Gamma_lep_DDME2} shows the neutrino (a) and
the antineutrino (b) absorption rates for the model DDME2. As 
seen from panel (a), the neutrino absorption rates
show similar temperature dependence to
the lepton capture rates (shown by dotted lines), but are smaller on average by an order of magnitude. The antineutrino absorption rates are always many orders of magnitude smaller than the neutrino absorption rates except in the very high-temperature domain. The rate of the muon decay process is negligible as compared to the neutrino and antineutrino absorption processes because of the very small scattering phase space. These rates are related to the rate coefficients $\lambda_X$ in the rate equations in a simple way, $\lambda_X = \Gamma_X/T$
(See Eqs.~\eqref{eq:lambda1} and \eqref{eq:lambda2}; similar relations hold for the leptonic reactions since they have exactly the same kinematics.)
We can therefore conclude
that within the DDME2 model the leptonic processes are always much slower than the Urca processes, putting the material in the ``slow lepton equilibration'' regime.

In the NL3 model, the neutrino absorption is more efficient at low densities but is suppressed at high densities
and low or moderate temperatures, see Fig.~\ref{fig:Gamma_lep_NL3}.
The antineutrino absorption rates show the opposite behavior:
they dominate the leptonic processes at high densities and are damped at low densities. However, the summed rate of leptonic processes in the case of NL3 model is qualitatively similar to those of the model DDME2. Consequently, as we see in
Fig.~\ref{fig:Gamma_lep_NL3},  
the material described by the NL3 model is almost always in the ``slow lepton equilibration'' regime.
The only exception is
the region around the transition point where the Urca process rate has a minimum. 
Note that the ``fast lepton equilibration'' regime is realized only around the minimum of the {\it muonic} Urca rates.

\subsection{Bulk viscosity of relativistic $npe$ matter}
\label{sec:bulk_electrons}

\begin{figure}[t] 
\begin{center}
\includegraphics[width=0.45\columnwidth,keepaspectratio]{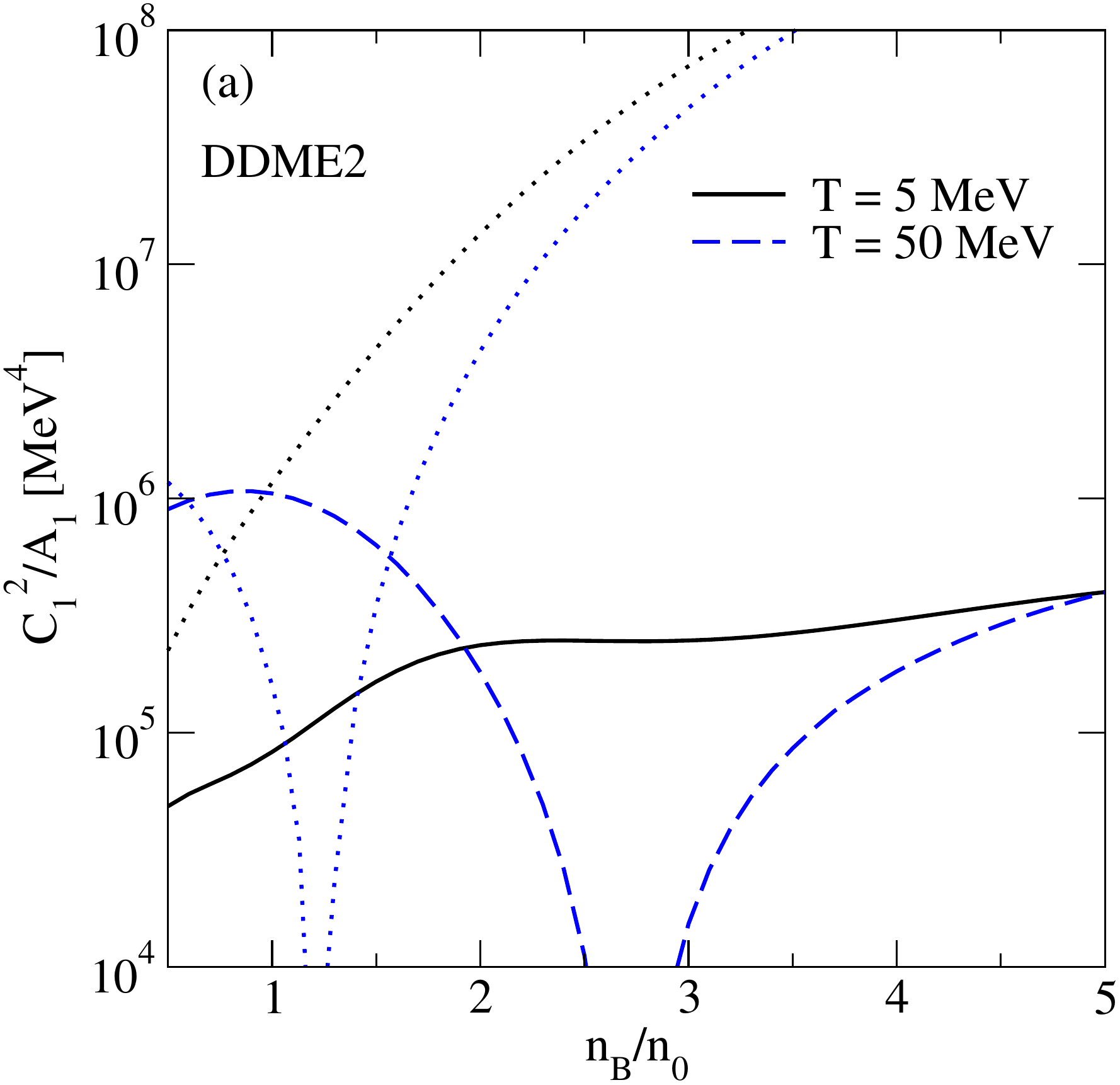}
\hspace{0.5cm}
\includegraphics[width=0.45\columnwidth,keepaspectratio]{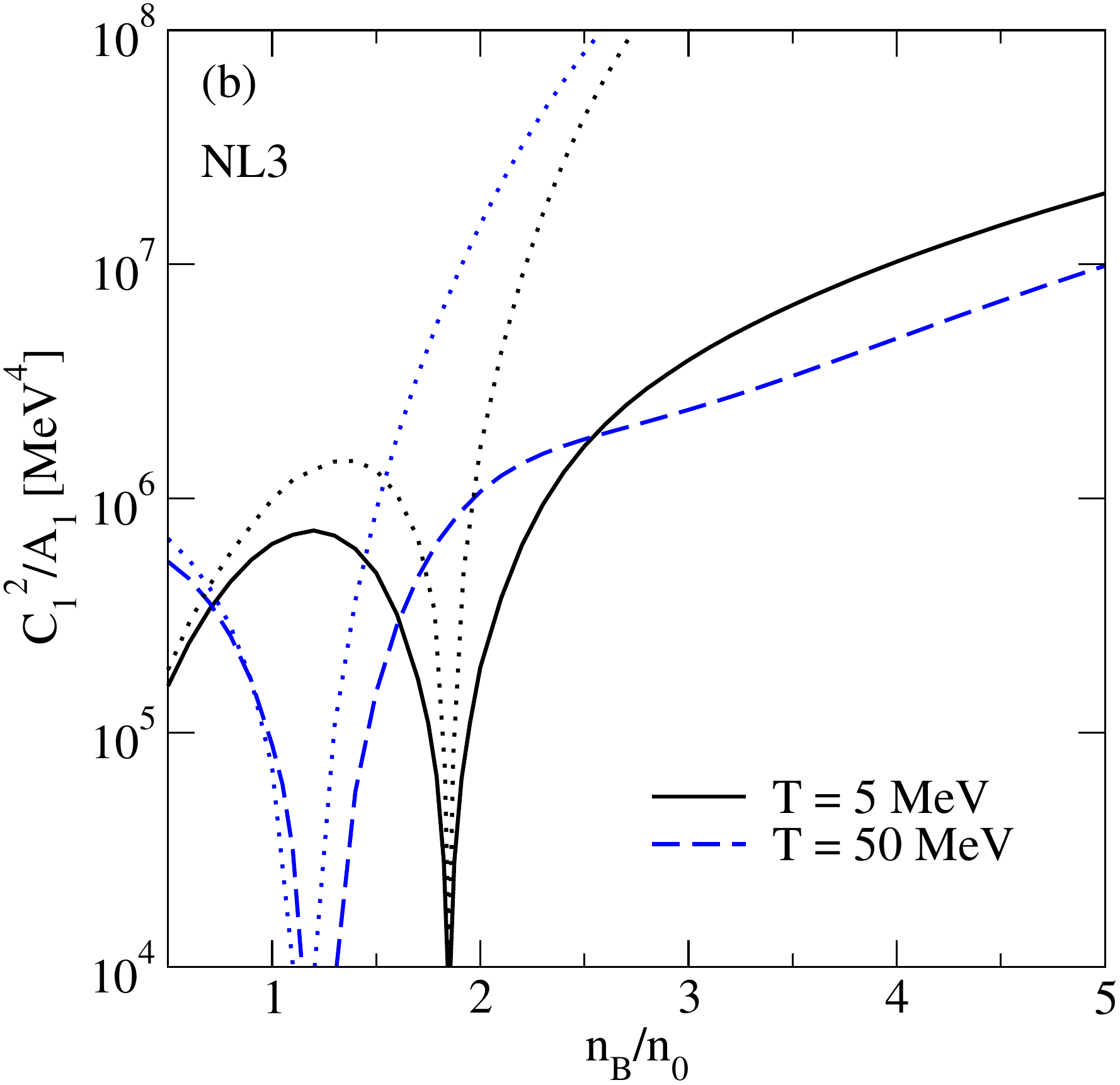}
\caption{ The susceptibility prefactor $C^2_1/A_1$ as a function
  of the baryon density for two values of the temperature for 
  (a) model DDME2 and (b) model NL3. The dotted lines show the 
  nonrelativistic results used in Ref.~\cite{Alford2019b}. }
\label{fig:C2A_dens} 
\end{center}
\end{figure}

In this subsection we will neglect muons and discuss the bulk 
viscosity arising only from electronic Urca processes. We include 
relativistic corrections to the nucleon spectrum both in the equilibration rates and the nucleon 
susceptibilities. The bulk viscosity of $npe\nu_e$ matter is 
given by Eq.~\eqref{eq:zeta_slow1} with the susceptibilities 
$C_1$ and $A_1$ defined by Eqs.~\eqref{eq:def_C1} and \eqref{eq:def_A1}. 

The susceptibility $A_1$ is not sensitive to the temperature
and the density, whereas $C_1$ increases with density and
typically crosses zero at a temperature-dependent value of 
the density where the proton fraction in $\beta$-equilibrated 
matter has a minimum as a function of the density. At this critical 
density, the system becomes scale-invariant, so compression
does not drive the system out of equilibrium. This implies 
vanishing bulk viscosity at critical densities. 

\begin{figure}[!] 
\begin{center}
\includegraphics[width=0.45\columnwidth,keepaspectratio]{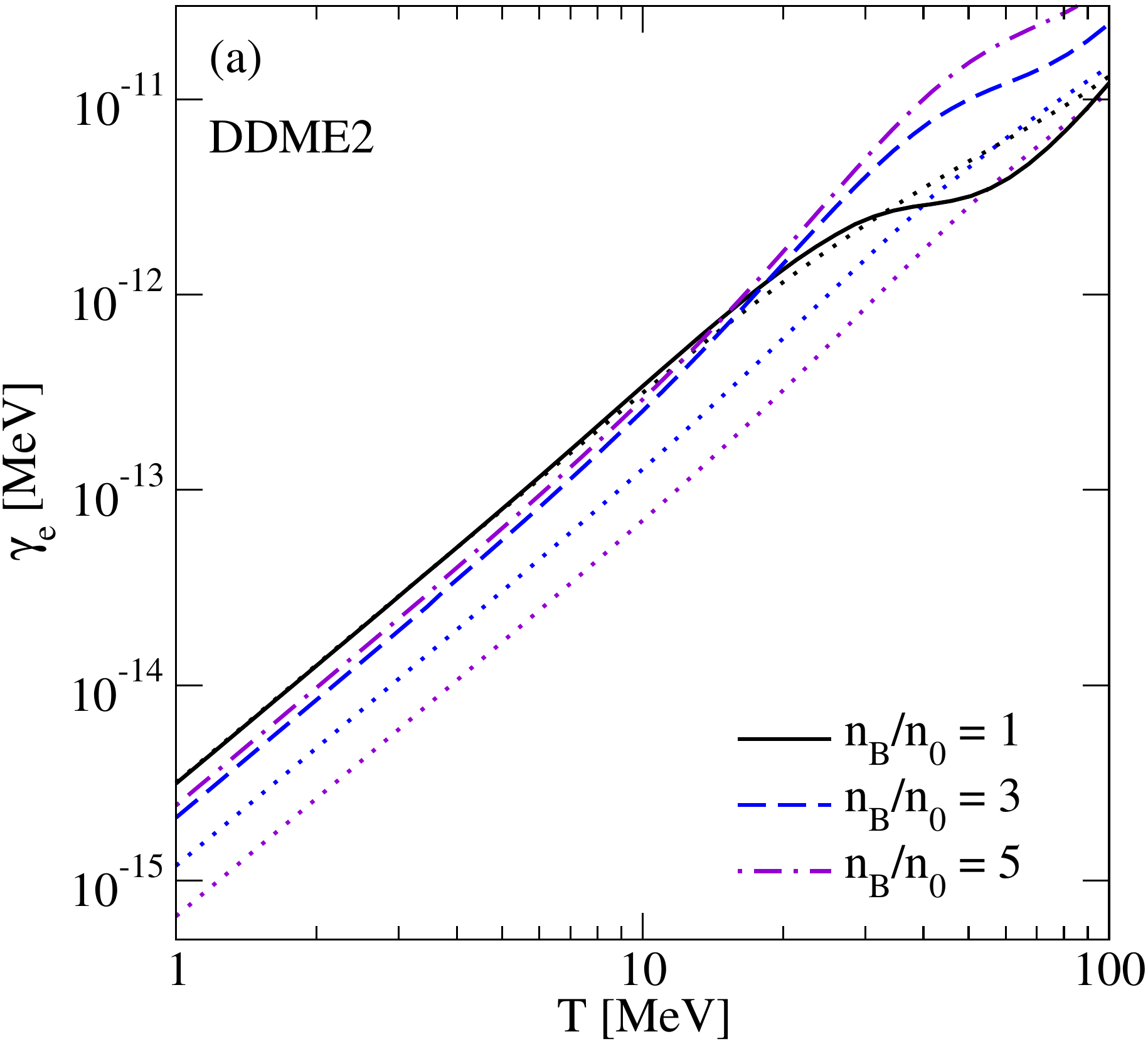}
\hspace{0.5cm}
\includegraphics[width=0.45\columnwidth,keepaspectratio]{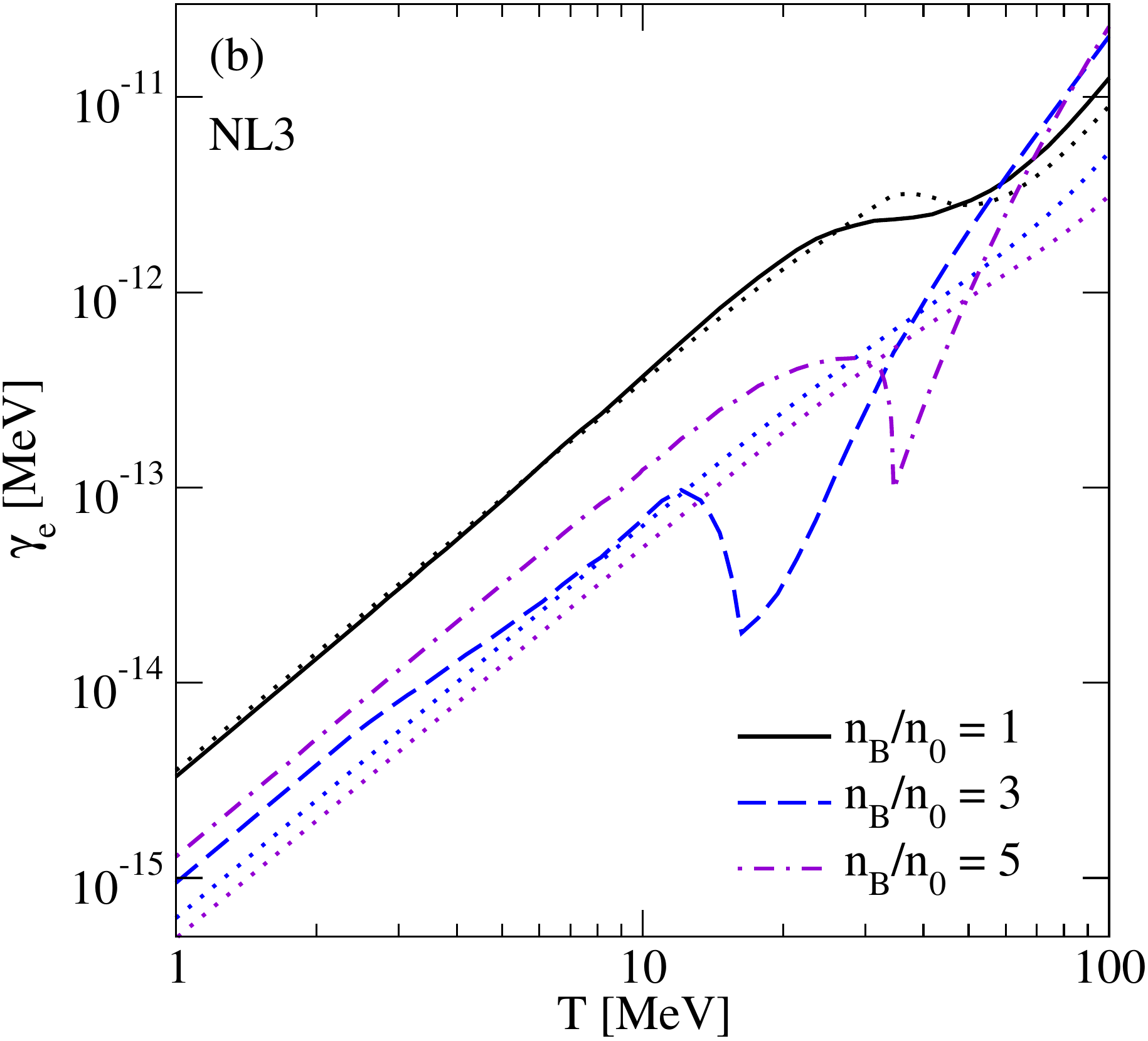}
\caption{The relaxation rate $\gamma_e$ as a function of the 
  temperature for fixed values of the density for (a) model DDME2; 
  (b) model NL3. The dotted lines show the relaxation rates computed
  in Ref.~\cite{Alford2019b} using the approximation of 
  nonrelativistic nucleons. Typical density oscillations in mergers are 
  at $\omega\sim 1\,\text{kHz} \approx 4\times 10 ^{-18}$\,MeV, so 
  neutrino-trapped matter is always in the fast equilibration regime.
  }
\label{fig:gamma_e_temp} 
\end{center}
\end{figure}
\begin{figure}[t] 
\begin{center}
\includegraphics[width=0.45\columnwidth, keepaspectratio]{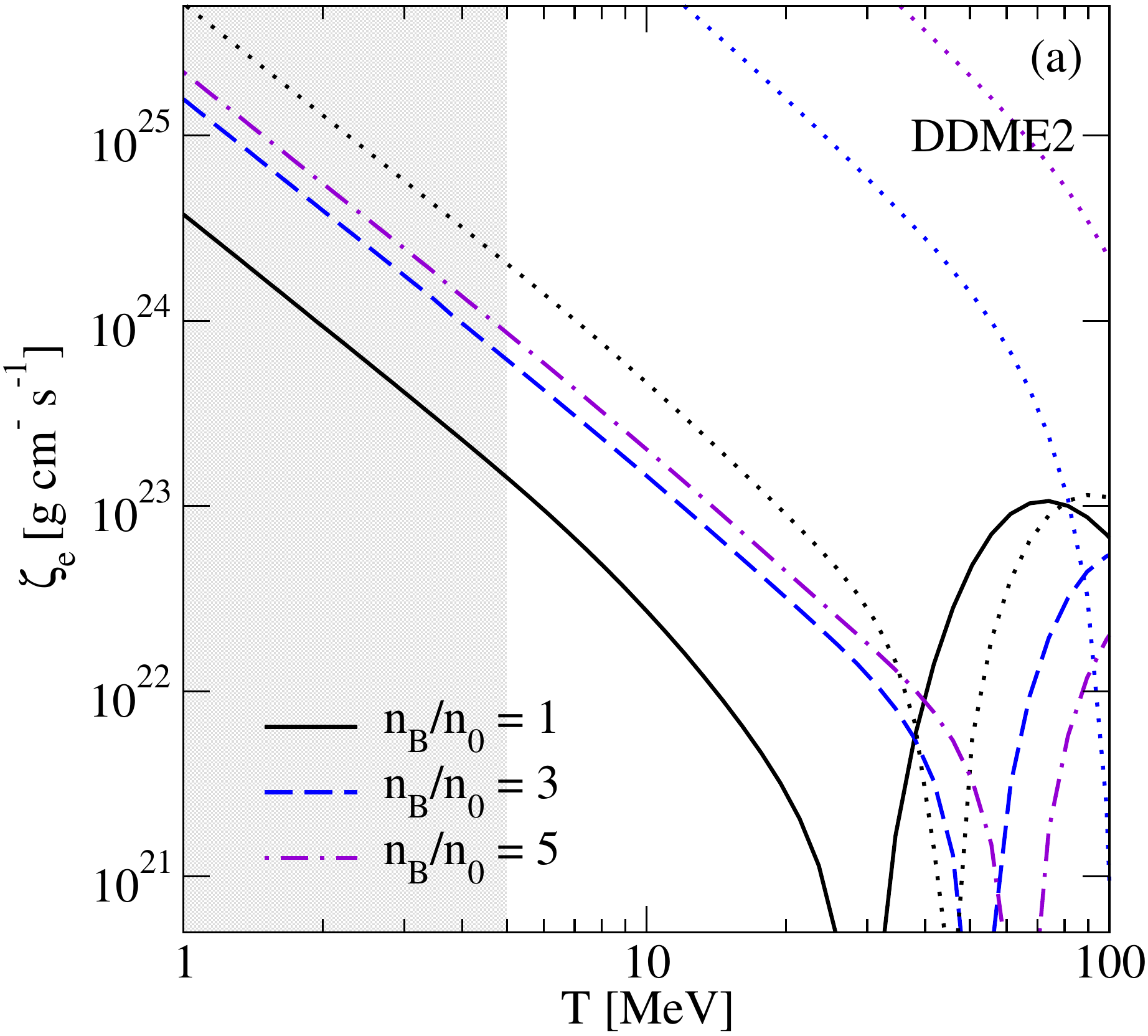}
\hspace{0.5cm} 
\includegraphics[width=0.45\columnwidth, keepaspectratio]{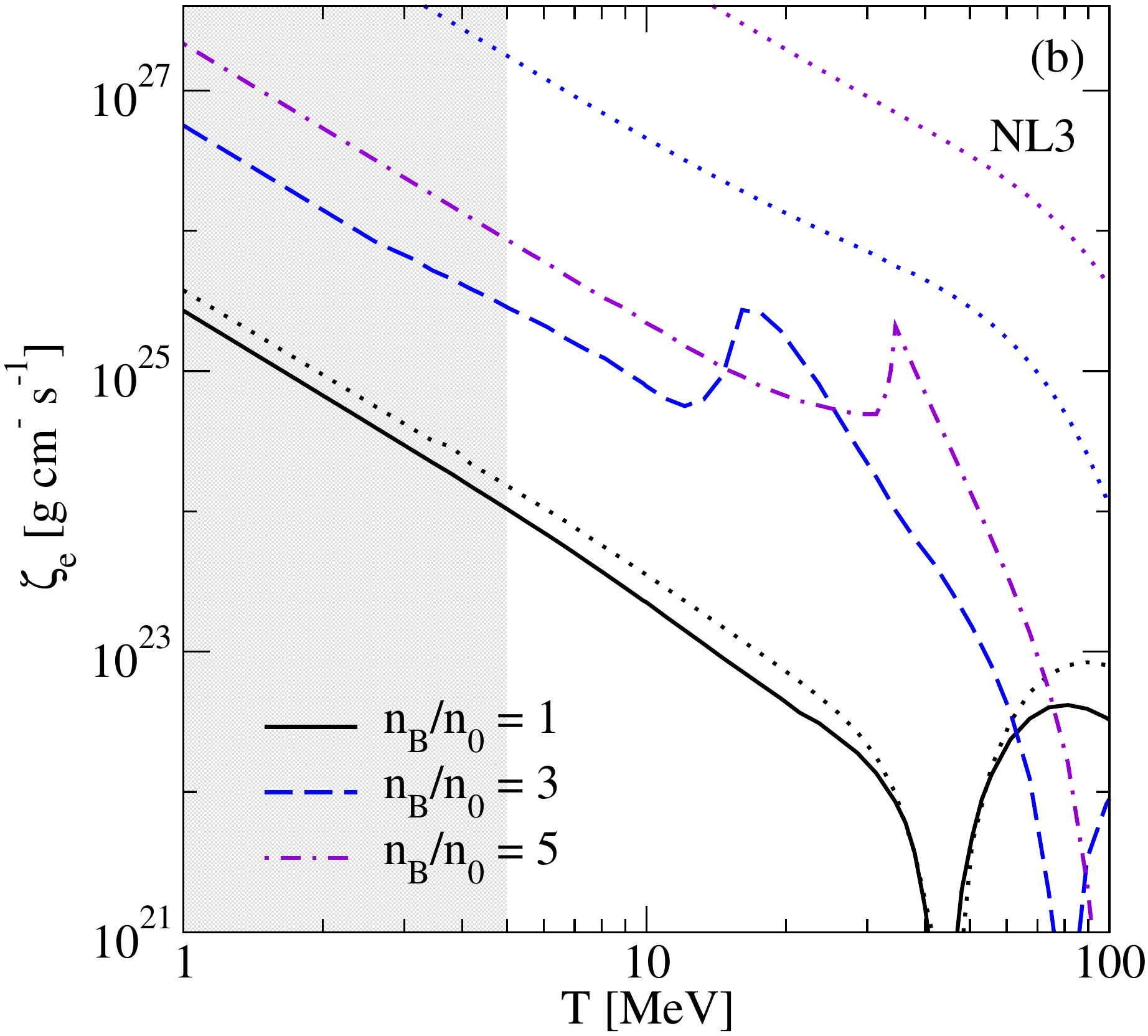}
\caption{ The bulk viscosity due to electron Urca processes as a
function of the temperature for (a) model DDME2; (b) model NL3.
 The region $T\le 5$~MeV is shaded because neutrinos are no longer 
 trapped at those temperatures. The dotted lines show the results
  of Ref.~\cite{Alford2019b} using the approximation of nonrelativistic nucleons. }
\label{fig:zeta_e_temp} 
\end{center}
\end{figure}

Figure~\ref{fig:C2A_dens} shows the susceptibility prefactor 
$C_1^2/A_1$ as a function of density for two values of the temperature. 
At the critical density, it drops to zero and slowly increases with the 
density above that point. For comparison we show also the results of 
our previous work~\cite{Alford2019b} with the dotted lines, which were 
obtained with the nonrelativistic spectrum for nucleons. We see that 
the nonrelativistic approximation strongly overestimates the 
susceptibility even at low densities $n_B\leq 2n_0$ where the 
relativistic corrections to the nucleonic spectrum are relatively small. 

The beta relaxation rates $\gamma_e=\lambda_e A_1$ of electronic 
Urca processes which determine the location of the resonant maximum 
of the bulk viscosity are shown in Fig.~\ref{fig:gamma_e_temp}. 
Qualitatively $\gamma_e$ closely follows the behavior of $\Gamma_e$. 
As the typical frequencies of density oscillations in neutron 
star mergers are several kHz, the relaxation is always fast, $\gamma_e\gg 
\omega$ (1 kHz corresponds to $4.14\cdot 10 ^{-18}$\,MeV). Thus, 
the neutrino-trapped matter is in the fast equilibration regime, 
and from \eqref{eq:zeta_slow1} the bulk viscosity is independent of the oscillation
frequency and is given by $\zeta = C_1^2/(A_1\gamma_e)$.

The results of the bulk viscosity arising from electronic Urca
processes are shown in Fig.~\ref{fig:zeta_e_temp}. At low temperatures, $T\leq 10$ MeV the bulk viscosity decreases according to the scaling $\zeta_e\sim T^{-2}$, which breaks down at higher temperatures
where the system approaches the point of scale-invariance. In the 
case of NL3 model, the bulk viscosity has a local maximum at high
densities due to the transition from the antineutrino-dominated
regime to the neutrino-dominated regime. At that maximum, the bulk 
viscosity jumps nearly by an order of magnitude. Comparing
these results with ones obtained within the nonrelativistic 
approximation for nucleons we see that the bulk viscosity 
decreases by orders of magnitude when the relativistic corrections
are properly taken into account. The main reason for this is much
lower susceptibility $C_1$ as compared to the nonrelativistic 
case, and also the higher $\beta$-equilibration rates. We also
observe that the local maxima in the case of NL3 model appear
only in full relativistic computation as was mentioned above.

\subsection{Bulk viscosity of relativistic $npe\mu$ matter}
\label{sec:bulk_muons}

\begin{figure}[t] 
\begin{center}
\includegraphics[width=0.45\columnwidth, keepaspectratio]{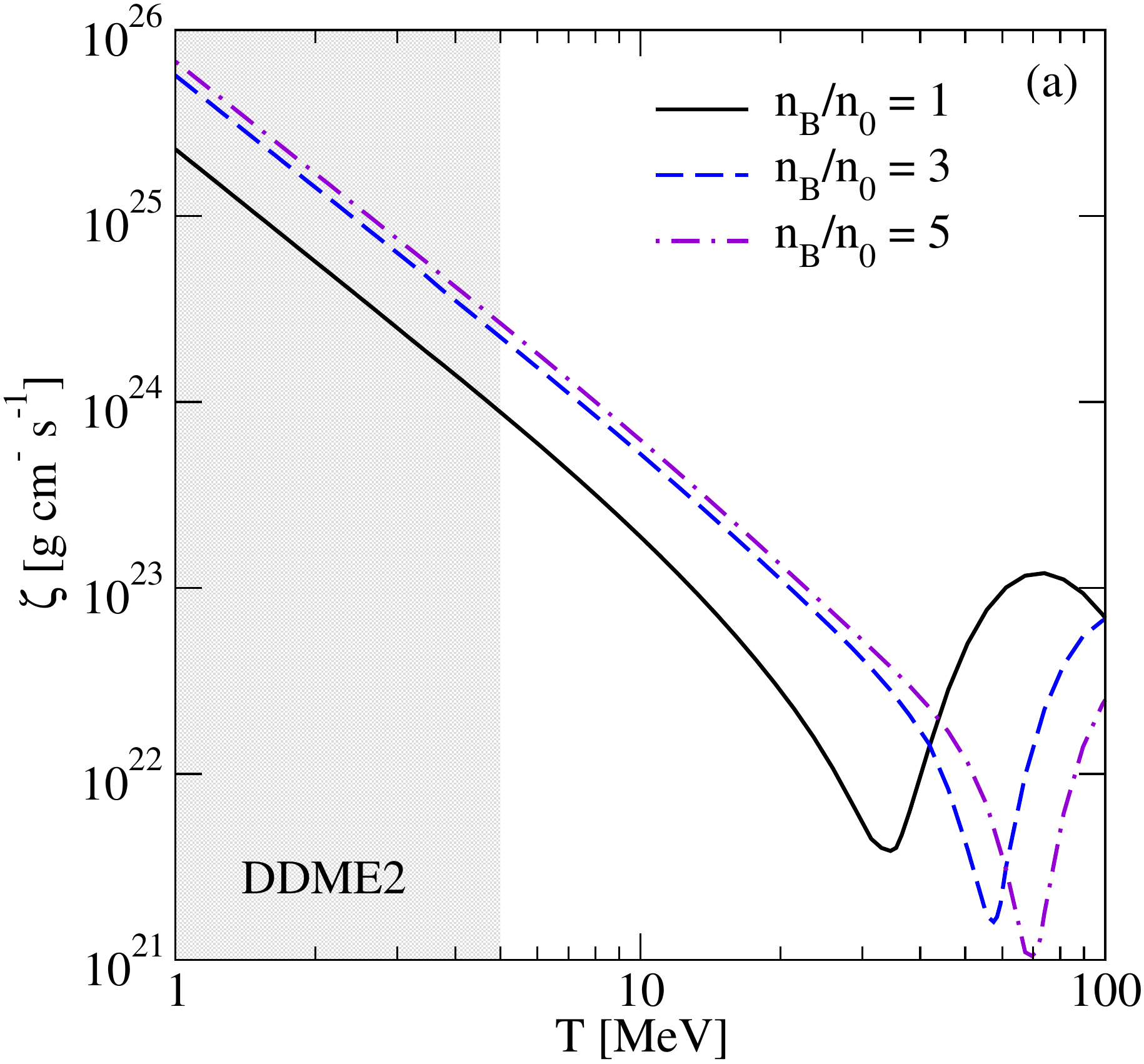}
\hspace{0.5cm} 
\includegraphics[width=0.45\columnwidth, keepaspectratio]{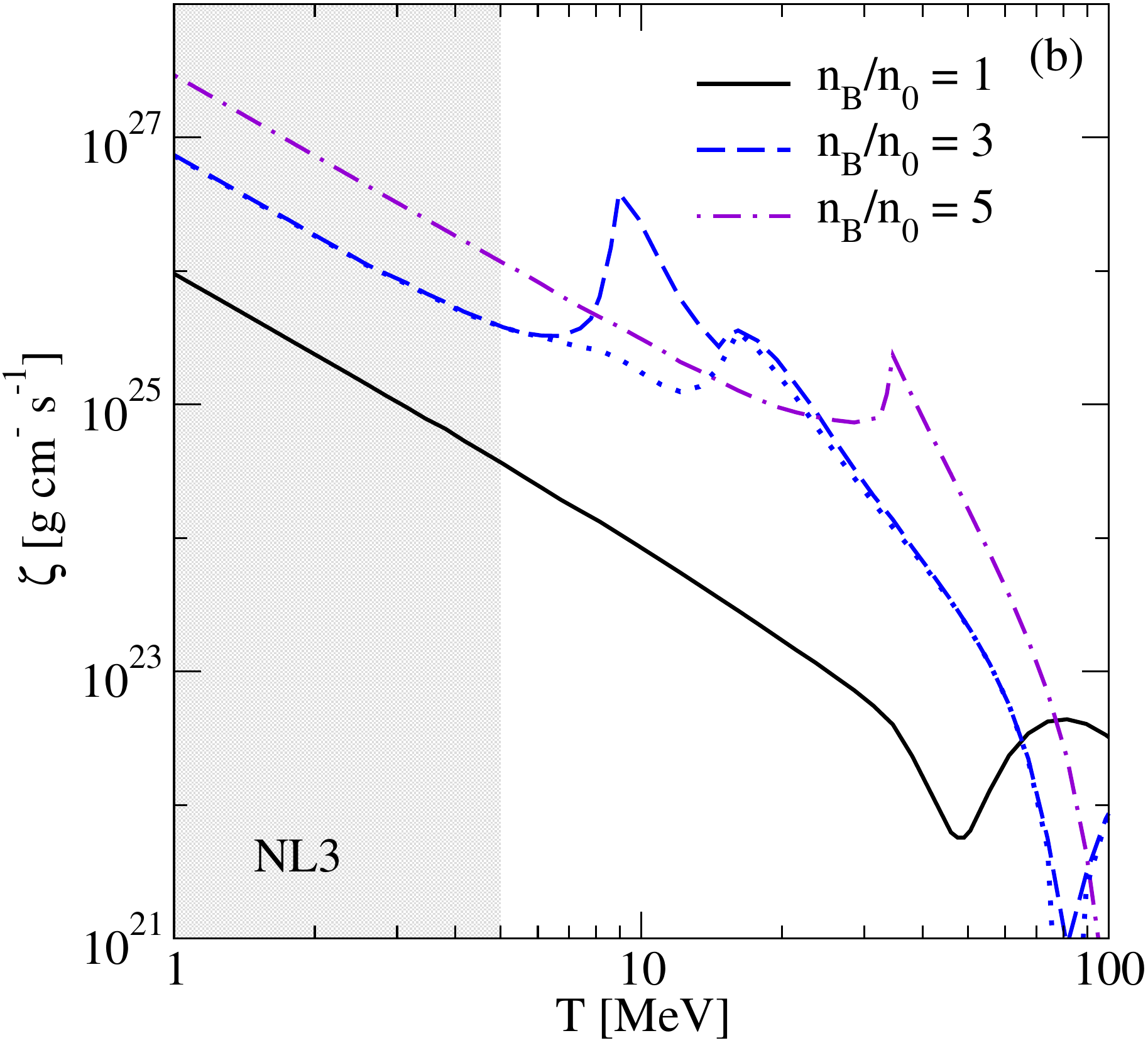}
\caption{ The bulk viscosity of neutrino-trapped $npe\mu$ matter
  as a function of the temperature for (a) model DDME2; (b) model 
  NL3. The region $T\le 5$~MeV is shaded because neutrinos are no longer trapped at those temperatures. All curves assume the slow lepton-equilibration regime except the dotted line in panel (b) which assumes fast lepton equilibration.}
\label{fig:zeta_slow_temp} 
\end{center}
\end{figure}
\begin{figure}[t] 
\begin{center}
\includegraphics[width=0.45\columnwidth, keepaspectratio]{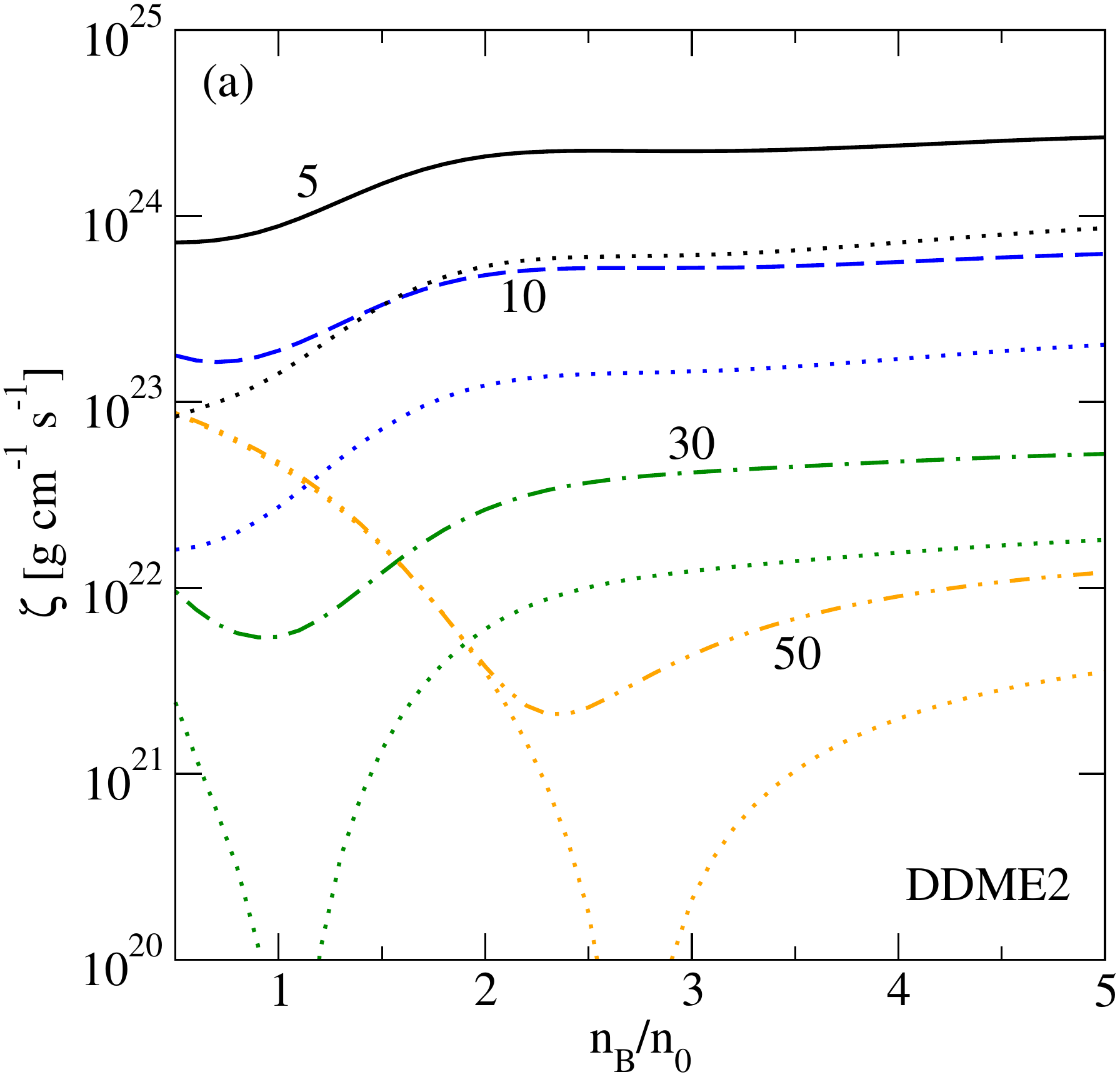}
\hspace{0.5cm} 
\includegraphics[width=0.45\columnwidth, keepaspectratio]{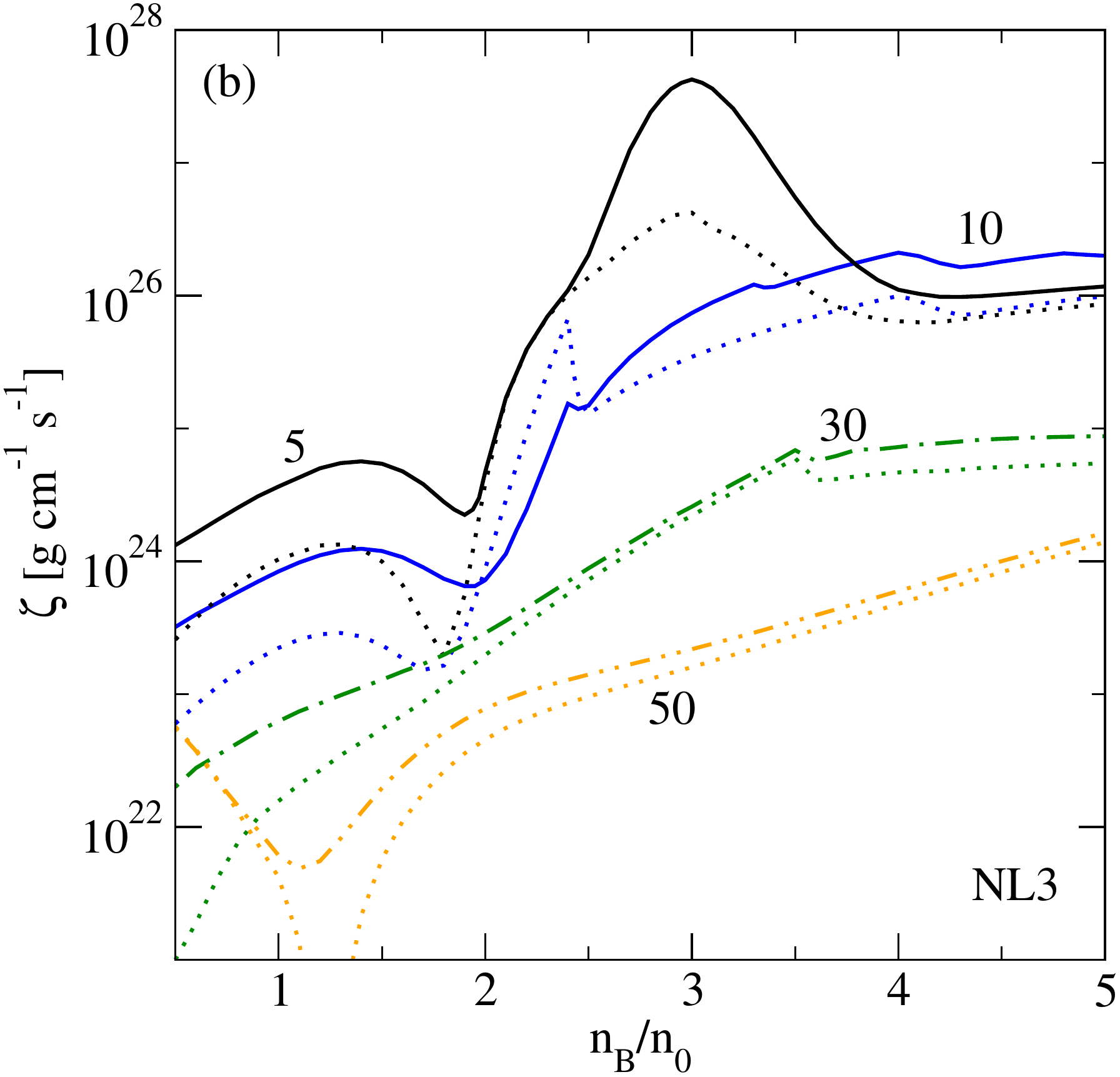}
\caption{ The bulk viscosity of neutrino-trapped $npe\mu$ matter
  as a function of the density for various temperatures indicated
  on the plot for (a) model DDME2; (b) model NL3. The dotted curves 
  show the corresponding bulk viscosities of $npe$ matter. }
\label{fig:zeta_slow_dens} 
\end{center}
\end{figure}

In this section, we present the results of the bulk viscosity
of nuclear matter including the contribution of a muonic component. 
As discussed in Sec.~\ref{sec:rates}, in the case of the DDME2  
model the rates of the leptonic processes are much smaller than 
the rates of the Urca processes, 
see Fig.~\ref{fig:Gamma_lep_DDME2}. Therefore, the bulk 
viscosity of $npe\mu$ matter can be computed  according 
to the slow lepton-equilibration limit, as discussed in Sec.~\ref{sec:bulk}. As the equilibration rates are much larger 
than the oscillation frequency, the bulk viscosity for the 
DDME2 model can be computed from Eq.~\eqref{eq:zeta_slow3}. 
The results are shown in the left panel of
Fig.~\ref{fig:zeta_slow_temp}. 
The generic behavior of the bulk viscosity of 
$npe\mu$ matter is similar to the one of $npe$ matter 
but the former exceeds the latter
by factors from 3 to 10 at the left side of the minimum. Above the
minimum, the bulk viscosity of $npe\mu$ matter is almost the same as the bulk viscosity of $npe$ matter. However, there is an important difference in the high-temperature regime, where the total bulk viscosity has a sharp minimum but does not drop to zero, 
as it was the case of the bulk viscosity of $npe$ matter. This 
behavior is easy to understand by noting that in the relevant 
temperature-density range we have mainly $(A_n+A_p) C_1\ll A_1 C_2$, 
$(A_n+A_p) C_2\ll A_2 C_1$, $(A_n+A_p)^2\ll A_1A_2$, which allows 
to simplify Eq.~\eqref{eq:zeta_slow3} to
\bea\label{eq:zeta_slow4}
\zeta \simeq \frac{\lambda_e (A_1 C_2)^2
+\lambda_\mu (A_2 C_1)^2 }
{\lambda_e \lambda_\mu (A_1A_2)^2}
=\frac{ C_2^2}{\lambda_\mu A_2^2}
+\frac{C_1^2 }{\lambda_e A_1^2}
=\zeta_e +\zeta_\mu,
\eea
where $\zeta_e$ and $\zeta_\mu$ are the partial contributions
of electronic and muonic Urca processes, respectively, to the
bulk viscosity. Both susceptibilities $C_1$ and $C_2$ cross zero 
at high temperatures, but the values of those critical temperatures 
for $C_1$ and $C_2$ are slightly shifted from each other. As a result, 
the summed $\zeta$ has a minimum at a temperature that lies between these two temperatures but does not drop to zero.

Turning to the NL3 model we note that also in this case the matter 
is mainly in the slow lepton-equilibration regime except for the region close to the minimum of equilibration rates, where for $n_B/n_0=3$ and $n_B/n_0=5$ we have the opposite regime of fast lepton-equilibration, see Fig.~\ref{fig:Gamma_lep_NL3}. 
Figure~\ref{fig:zeta_slow_temp}, panel (b) therefore shows the bulk viscosity 
in the slow lepton-equilibration limit by the solid, dashed and the 
dashed-dotted lines. The one exception is the dotted line, which shows the fast lepton equilibration limit [Eq.~\eqref{eq:zeta}] for $n_B/n_0=3$.

At the highest density, $n_B/n_0=5$, the bulk viscosity
has one local maximum as the electronic and muonic Urca process
rates have minima at almost the same  temperature, 
see Fig.~\ref{fig:Gamma_NL3}. For moderate density $n_B/n_0=3$ the minima  of Urca process rates for electrons and muons are at different
temperatures, therefore the bulk viscosity has local maxima at both
temperatures. However, near the maxima we cannot rely on the slow lepton equilibration approximation: in the 
fast lepton-equilibration limit (dotted line) the first maximum is eliminated by leptonic processes, whereas the second maximum remains. At the highest density $n_B/n_0=5$, the numerical results for the bulk viscosity in the 
fast lepton-equilibration limit are found to be very close to those of 
slow lepton-equilibration limit and are not shown on the figure.

The structure of the postmerger object changes	with time from 
initially having double density-peaks, associated with the two
neutron stars, to a single density-peak structure 
corresponding to the
remnant (see, for example,  Refs.~\cite{Perego:2019adq,Hanauske:2019qgs,Hanauske:2017oxo,Kastaun:2016elu,Bernuzzi:2015opx,Foucart:2015gaa,Kiuchi:2012mk,Sekiguchi:2011zd,Ruiz2016,East:2016,Most2019,Bauswein2019}). 
So far, we consider the variations of the bulk viscosity at
fixed density, which corresponds to moving along the constant density
surfaces in such an object. It is also interesting to consider the
isothermal surfaces along which the density is changing.
The
temperature evolution in the postmerger object replicates that	of the	
density, \ie, a double	peak high-temperature structure	evolves	in time	
into a single peak structure.
To account for this type of variation,
we plot the bulk viscosity as a function of the density in
Fig.~\ref{fig:zeta_slow_dens}. 

The density variations of bulk viscosity for each value of temperature represent self-similar curves,  which are shifted with respect to each other by a magnitude 
 which depends on the change in the temperature.
In the case of model NL3 the curves 
$T=5, 30, 50$ MeV correspond to the slow lepton-equilibration limit,
and only the curve $T=10$ MeV shows the results of the fast 
equilibration regime.

\subsection{Damping of density oscillations}
\label{sec:damping}

\begin{figure}[t] 
\centering
\includegraphics[width=0.45\columnwidth, keepaspectratio]{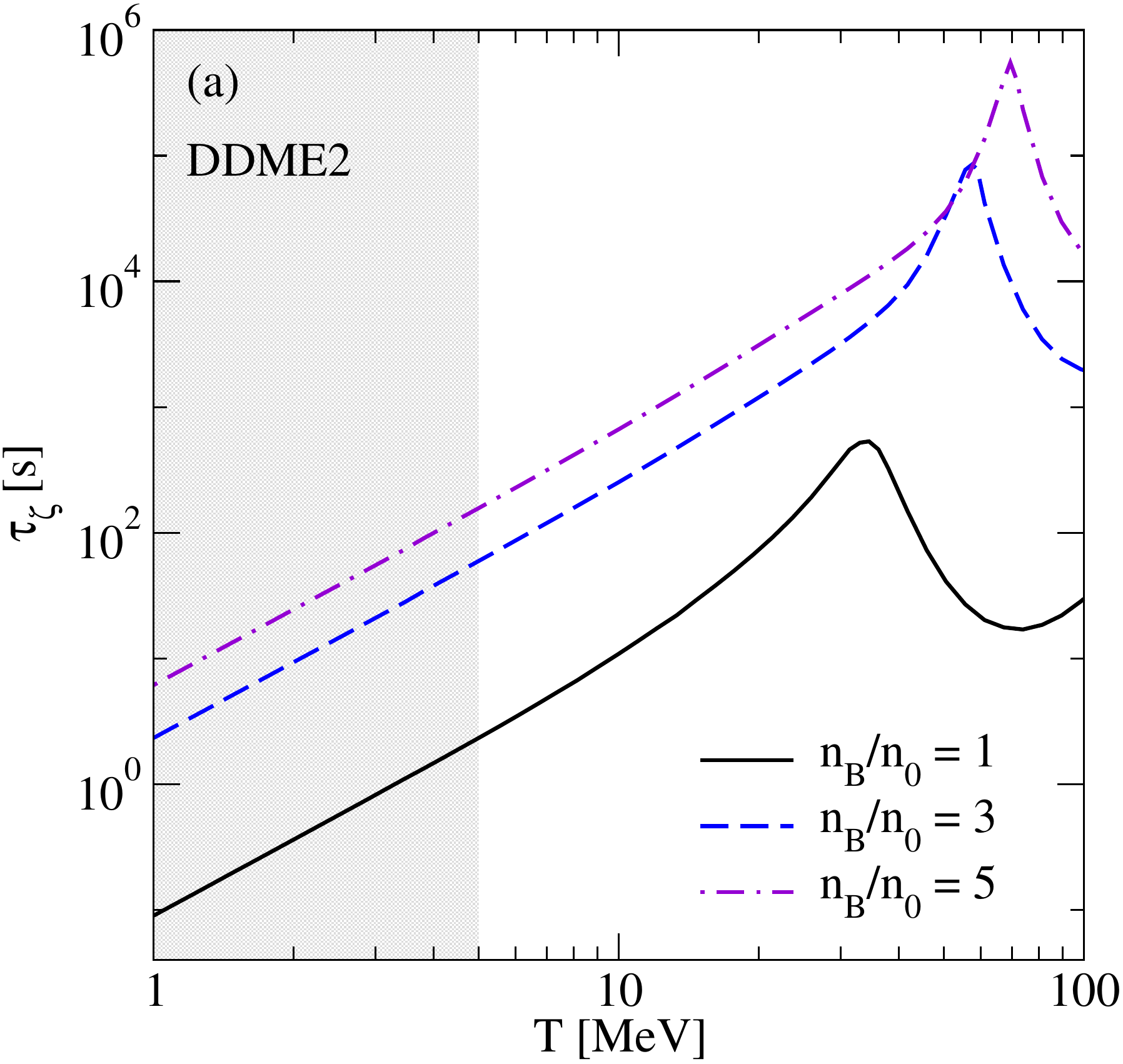}
\hspace{0.5cm}
\includegraphics[width=0.45\columnwidth, keepaspectratio]{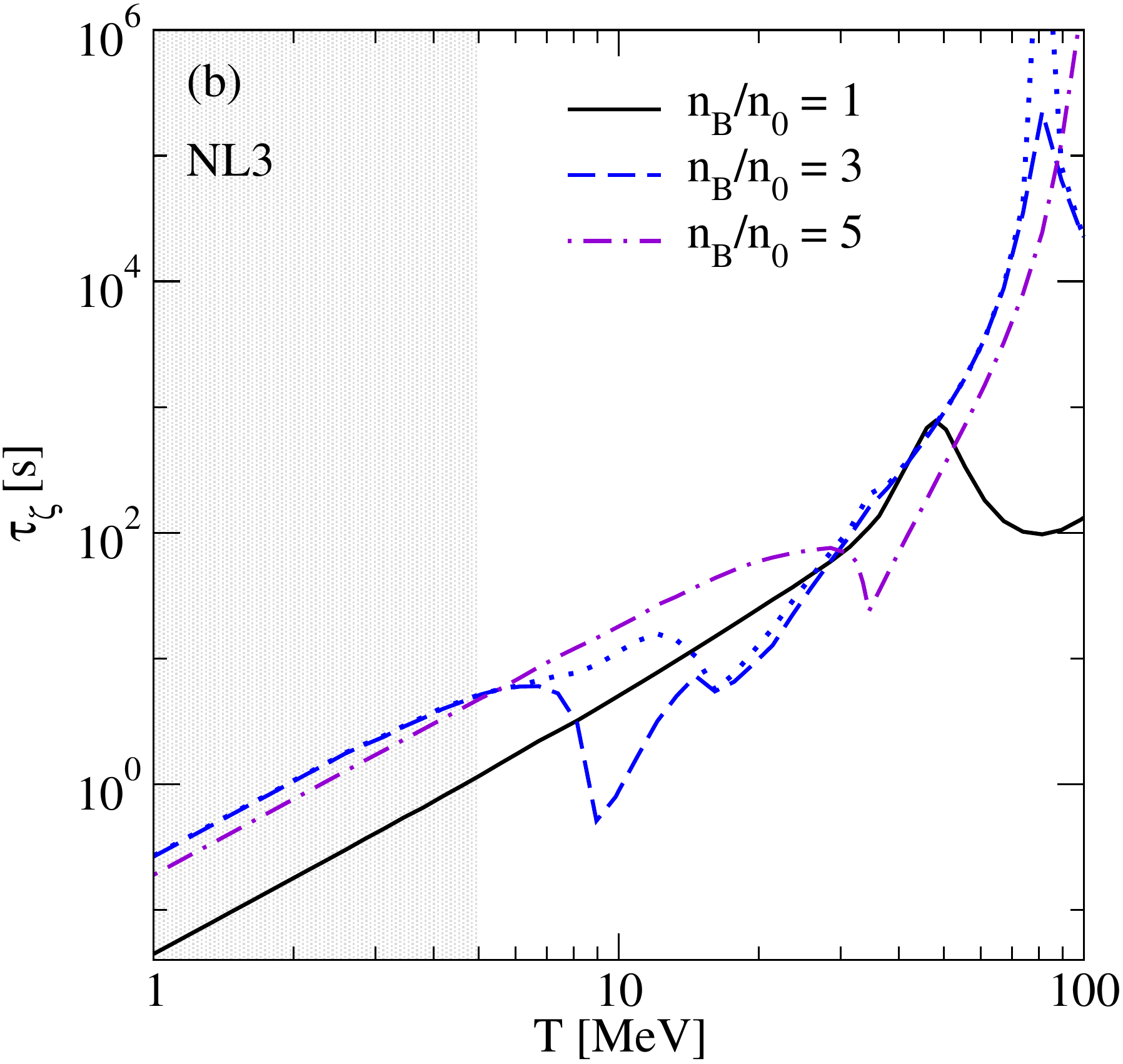}
\caption{ The oscillation damping timescale as a function of 
temperature for various densities and for frequency fixed at 
$f=10$ kHz for (a) model DDME2; (b) model NL3. The result for
the density $n_B/n_0=3$ in panel (b) should be replaced by the
blue dotted line around the minimum.}
\label{fig:tau_damp10} 
\end{figure}

In this last subsection, we estimate the timescales of bulk
viscous damping of density oscillations in neutrino-trapped
$npe\mu$ matter. The characteristic timescale of damping of 
density oscillations is given by~\cite{Alford2018a,Alford2019a,Alford2020}
\bea\label{eq:damping_time}
\tau_{\zeta} =\frac{1}{9}\frac{Kn_B}{\omega^2\zeta},
\eea 
where 
\bea\label{eq:compress}
K=9n_B\frac{\partial^2\varepsilon}{\partial n_B^2}
\eea 
is the (isothermal) incompressibility of nuclear matter. The 
incompressibility of nuclear matter at finite temperatures
is shown in Ref.~\cite{Alford2020}. 

As the bulk viscosity is independent of the oscillation frequency,
the damping timescale is inversely proportional to the square of
$\omega$. We show $\tau_\zeta$ as a function of the temperature
in Fig.~\ref{fig:tau_damp10} for $f=10$\,kHz. The nuclear 
incompressibility is almost independent of the temperature. 
Therefore the damping timescale as a function of the temperature
closely follows the inverse bulk viscosity showing sharp maxima
in the high-temperature regime. In the case of NL3 model there
are local minima resulting from the transition of the matter
from the antineutrino-dominated regime to the neutrino dominated 
regime. However, the damping timescales in the neutrino/antineutrino
trapped regime exceeds the characteristic timescales for the long-term 
postmerger evolution timescale $\lesssim$ 1~s at temperatures above 
5 MeV. At lower frequencies, the damping timescales will be even larger. 
Thus, we conclude that the bulk viscosity of neutrino-trapped $npe\mu$
matter from the Urca processes is not sufficiently large to affect the 
evolution of binary neutron-star mergers in the initial hot regime
and could have an impact close to the neutrino untrapping 
temperature $\sim 5$~MeV.

\end{widetext}

\section{Conclusions}
\label{sec:conclusions}
 
In this work, we studied the bulk viscosity of neutrino-trapped
$npe\mu$ matter from Urca processes under the conditions relevant 
to binary neutron star mergers. We first generalized the computation
of the rates of relevant $\beta$-equilibration processes (\ie, the 
neutron decay and lepton capture) as well as those of relevant 
susceptibilities performed in Ref.~\cite{Alford2019b} to include 
the relativistic corrections to the nucleonic spectra. We find that 
these corrections enhance the equilibration rates by factors from 
1 to 10. The numerical computations were carried out within the 
relativistic density functional theory for two different EoS models 
of nuclear matter.

An interesting feature of full relativistic rates is their strong dependence on the scaled-to-temperature neutrino chemical potential $\alpha_{\nu_l}$. It turns out that if
$\alpha_{\nu_l}\geq -3$ then the neutron decay rate is
Boltzmann-suppressed,
and the only equilibration process is the lepton capture.
This is the case for DDME2 model which has a composition where
the net neutrino densities are mainly positive in the relevant
density-temperature range. The picture is quite different in the case of model NL3 where the net neutrino densities are
positive only in the low temperature and low-density sector, 
and the antineutrino population increases with both density and temperature. At low densities and high temperatures, the lepton capture dominates as in the case of DDME2 model, but
in the low-temperature and high-density domain, we have the opposite limit. Here the antineutrino population is dominant,
and the neutron decay is the main equilibration process as
long as $\alpha_{\nu_l}\leq -6$. For intermediate values 
$-6\leq \alpha_{\nu_l}\leq -3$ both processes are important,
and there is a transition point at around $\alpha_{\nu_l}\simeq 
(-5)$ to $(-4)$ where the rates of neutron decay and lepton capture become equal. Close to this point the net equilibration rate has a sharp minimum.

The relativistic susceptibilities are found to be significantly (up to orders of magnitude)
smaller than their corresponding nonrelativistic counterparts at 
densities $n_B/n_0\geq 2$. Similar to the nonrelativistic case 
we find that the susceptibilities corresponding to the partial 
bulk viscosities from electronic and muonic Urca processes 
vanish at a critical density where the electron/muon fraction has 
a local minimum as a function of density at high temperatures 
$T\gtrsim 30$~MeV. At that point the system becomes scale-invariant: there is no chemical reequilibration induced by
compression which implies zero bulk viscosity on the time scales relevant to mergers.

Neutrino-trapped matter is always in the regime of
fast $\beta$-equilibration, \ie, the relaxation rates are much higher
than the typical frequencies of density oscillations. As a result,
the bulk viscosity is independent of the frequency and decreases
with the temperature. This decrease is followed by sharp drops to
zero at the points where the system becomes scale-invariant. In
the case of model NL3 the bulk viscosity shows also local maxima at
intermediate temperatures where the transition between the antineutrino- 
and neutrino-dominated regimes occurs.  

The proper inclusion of muons in the computation of bulk viscosity 
requires analysis of relative rates of Urca processes and the rates 
of pure leptonic processes, \ie, muon decay, and neutrino/antineutrino 
absorption. We find that the rates of the leptonic reactions are slower 
than the Urca process rates almost in the entire temperature-density range. 
An exception occurs only in the narrow vicinity of the transition point
in the case of NL3 model. We, therefore, conclude that  the bulk viscosity 
of $npe\mu$ matter can typically be computed in the slow lepton-equilibration 
limit. The numerical results show that the bulk viscosity is enhanced by 
factors from 1 to 10 as compared to the viscosity of $npe$ matter. Note 
that our study neglects so far the neutrino flavor conversion, which can 
affect our results. We plan to address this issue in a separate study.

Our estimates of the damping timescales of the density oscillations show 
that  the  bulk  viscosity  of  relativistic $npe\mu$ matter in 
the neutrino-trapped regime is not an important source of damping of 
density oscillations over characteristic timescales of neutron star mergers. 
However, long-lived remnants of mergers, which do not collapse to a black 
hole, can experience bulk viscous dissipation. Young proto-neutron stars 
formed in supernova  explosions offer another setting where the bulk 
viscosity of hot stellar matter could be important for assessing 
their oscillation spectrum and damping time scales.

We finally note that, the methods applied here can be used to obtain 
other microscopic characteristics of dense matter, such as, for example, 
neutrino opacities. The fully relativistic treatment of the rates should 
be of interest in a broader context of radiation and transport in thermal 
quantum field theories with applications to a wide range of relativistic systems.

\section*{Acknowledgments}

  M.~A. is supported by the U. S. Department of Energy, Office of Science, 
  Office of Nuclear Physics under Award No. DE-FG02-05ER41375. The research 
  of A.~H. and A.~S. was funded by the Volkswagen Foundation (Hannover, Germany) 
  grant No. 96 839. They acknowledge the support of the European COST Action 
  ``PHAROS'' (CA16214).  A.~S. acknowledges the support by the Deutsche 
  Forschungsgemeinschaft (DFG) Grant No. SE 1836/5-1. He also  acknowledges 
  the support of the Polish NCN Grant No. 2020/37/B/ST9/01937 at Wroc\l{}aw University.


\appendix

\begin{widetext}

\section{Phase space integrals}
\label{app:rates}

Here we extend the technique of computing the phase-space integrals discussed in \cite{Alford2019a,Alford2019b} to fully relativistic case. Substituting the matrix element of the Urca process~\eqref{eq:matrix_el_full} into the rates~\eqref{eq:Gamma1p_def1} 
and the {\it inverse of} \eqref{eq:Gamma2n_def1} and introducing a 
``dummy'' integration  [we use the same the mapping between the
 particles and their momenta 
$(l) \to k$, $(\nu_l/\bar{\nu}_l) \to k'$, $(p) \to p$,
 and $(n) \to p'$ as before] we obtain
\bea\label{eq:Gamma1p_dummy}
\Gamma_{n\to p l \bar\nu}\, (\mu_{\Delta_l}) &=& 2 G^2\!
\int\! d^4q\! \int\!\! \frac{d^3p}{(2\pi)^3p_0} \int\!\!
\frac{d^3p'}{(2\pi)^3p'_0} \int\!\! \frac{d^3k}{(2\pi)^3k_0}
\int\!\! \frac{d^3k'}{(2 \pi)^3k'_0}(k\cdot p) (k'\cdot p') \nonumber\\
& \times & \bar{f}(k) \bar{f}(p) \bar{f}(k')  f(p') (2\pi)^4
\delta^{(4)}(k+p-q)\delta^{(4)}(k'-p'+q) =2 {G}^2\!\! \int\! d^4q\, I_1(q)\, I_2(q),\\
\label{eq:Gamma2p_dummy}
\Gamma_{n\nu\to p l}\, (\mu_{\Delta_l}) &=& 2 G^2\!
\int\! d^4q\! \int\!\! \frac{d^3p}{(2\pi)^3p_0} \int\!\!
\frac{d^3p'}{(2\pi)^3p'_0} \int\!\! \frac{d^3k}{(2\pi)^3k_0}
\int\!\! \frac{d^3k'}{(2 \pi)^3k'_0}(k\cdot p) (k'\cdot p') \nonumber\\
& \times & \bar{f}(k) \bar{f}(p) f(k') f(p') (2\pi)^4
\delta^{(4)}(k+p-q)\delta^{(4)}(-k'-p'+q) =2 {G}^2\!\! \int\! d^4q\, I_1(q)\, I_3(q),
\eea
where 
\bea
\label{eq:I1}
I_1(q) &=& \int\!\! \frac{d^3p}{(2\pi)^3 p_0}\int\!\!
  \frac{d^3k}{(2\pi)^3 k_0} \bar{f}(k)\bar{f}(p) \,
  (k\cdot p)\, (2\pi)^4\, \delta^{(4)}(k+p-q),\\
\label{eq:I2}
I_2(q) &=& \int\!\! \frac{d^3p'}{(2\pi)^3 p'_0} \int\!\!
  \frac{d^3k'}{(2\pi)^3 k'_0} \bar{f}(k')f(p')\, 
  (k'\cdot p')\,\delta^{(4)}(k'-p'+q) ,\\
\label{eq:I3}
I_3(q) &=& \int\!\! \frac{d^3p'}{(2\pi)^3 p'_0} \int\!\!
  \frac{d^3k'}{(2\pi)^3 k'_0} {f}(k') f(p')\, 
  (k'\cdot p')\, \delta^{(4)}(-k'-p'+q) ,
\eea
with $\delta^{(4)}(k+p-q)=\delta(\veck+\vecp-\vecq)
\delta(\ep_{k}+ \ep_{p}-\omega-\mu_{\Delta_l})$, and
 $\delta^{(4)}(\pm k'-p'+q)=\delta(\pm \veck'-\vecp'+\vecq)
\delta(\pm \ep_{k'}-\ep_{p'}+\omega)$. Here the energy 
conservation $\delta$-function  has been transformed according to 
$\delta(k_0+p_0 \pm k'_0-p'_0)=\delta(\ep_l+\ep_p-\ep_n\pm
\ep_{\bar{\nu}_l/\nu_l}-\mu_{\Delta_l})$,
where we added and subtracted $\mu_{\Delta_l}$ in the argument of the
$\delta$-function, and denoted by $\ep_i$ the energies of the
particles computed from their (effective) chemical potentials, \eg,
$\ep_p=\sqrt{p^2+m^{*2}_p}-\mu_p^*$. 
Since the rates of the 
inverse processes can be obtained by interchanging in Eqs.~\eqref{eq:Gamma1p_dummy}
and \eqref{eq:Gamma2p_dummy} $f(p_i)\leftrightarrow \bar{f}(p_i)$ for all
particles, the problem reduces to the computation of three
$q$-dependent integrals $I_1(q)$, $I_2(q)$ and $I_3(q)$ given by
Eqs.~\eqref{eq:I1}--\eqref{eq:I3}.

To compute the integral $I_1(q)$ we integrate over proton 
momentum and separate the angular part of the remaining 
integral, which gives
\bea 
I_1(q)
&=& (2\pi)^{-1}\! \int_{m_l}^\infty\! \frac{k dk_0}
{p_0}\,\bar{f}(\epsilon_{k})\bar{f}(\bar{\omega}-\epsilon_{k}) 
\int_{-1}^{1} dx\, (\bar{\omega}' k_0 - q k x -m_l^2)\, 
\delta(\epsilon_{k}+\epsilon_{q-k}-\bar{\omega}),\quad
\eea
where $\bar{\omega}= \omega+\mu_{\Delta_l}$, $\bar{\omega}'=
\bar{\omega}+\mu_p^*+\mu_l$, and $x$ is the cosine of the 
angle between $\bm k$ and $\bm q$. The angular integral is done 
by using the $\delta$-function to obtain [recall that $\bar{f}(\ep)=f(-\ep)$]
\bea \label{eq:I1_final}
I_1(q) &=& 
\frac{1}{4\pi q} \left[(\mu_l +\mu_p^* +\bar{\omega})^2
-m_l^2-m_p^{*2}-q^2\right]\int_{m_l-\mu_l}^{\mu_p^* +\bar{\omega}}\! 
d\ep_k\, \bar{f}(\epsilon_{k})f(\epsilon_{k}-\bar{\omega})
\theta(1-\vert x_0\vert),
\eea 
where $x_0$ is the zero of the argument of the $\delta$-function 
\bea\label{eq:x_0}
x_0 = \frac{1}{2kq}\left[-\left(\ep_k -\mu_p^*
-\bar{\omega}\right)^2 +m_p^{*2}+k^2+q^2\right],
\eea
and 
the limits of integration are found from the limits on the lepton energy $\ep_k$ 
\bea 
\label{eq:theta1}
(k-q)^2 + m_p^{*2}\leq \left(\ep_k -\mu_p^* 
-\bar{\omega}\right)^2 \leq (k+q)^2 +m_p^{*2}.
\eea 
The energy integral in Eq.~\eqref{eq:I1_final} could be done analytically, 
but for numerical implementation, the form given above is more suitable.

The computation of the remaining integrals proceeds in full analogy to the 
above. For integral $I_2(q)$ we find 
\bea \label{eq:I2_final}
I_2(q) = \frac{1}{2(2\pi)^5 q} \left[-(\mu_{\nu_l} +\mu_n^*+{\omega})^2 
+m_{\nu_l}^2 +m_n^{*2}+q^2\right]\int_{m_{\nu_l}+\mu_{\nu_l}}^\infty\! 
d\ep_{k'}\,\bar{f}(\epsilon_{k'}) f(\epsilon_{k'}+{\omega})
\theta(1-\vert y_0\vert),
\eea 
where $y_0$ is the zero of the argument of the $\delta$-function, \ie,
\bea\label{eq:y_0}
y_0 = \frac{1}{2k'q}\left[\left(\ep_{k'} +\mu_n^*+
 {\omega}\right)^2-m_n^{*2}-k'^2-q^2\right],
\eea
and the step-function sets the following limits on
the neutrino energy $\ep_{k'}$ 
\bea 
\label{eq:theta2}
(k'-q)^2 +m_n^{*2}\leq \left(\ep_{k'} +\mu_n^*+
 {\omega}\right)^2 \leq (k'+q)^2 +m_n^{*2}.
\eea 
For the integral  $I_3(q)$ we find
\bea \label{eq:I3_final}
I_3(q) =
\frac{1}{2(2\pi)^5q} \left[(\mu_{\nu_l} +\mu_n^*+{\omega})^2 
-m_{\nu_l}^2 -m_n^{*2}-q^2\right]\int_{m_{\nu_l}-\mu_{\nu_l}}^{\omega+\mu_n^*}\! 
d\ep_{k'}\, f(\epsilon_{k'})\bar{f}(\epsilon_{k'}-{\omega})  \theta(1-\vert z_0\vert),
\eea 
where $z_0$ is the zero of the argument of the $\delta$-function, \ie,
\bea\label{eq:z_0}
z_0 = \frac{1}{2k'q}\left[-\left(\ep_{k'} -
\mu_n^* -{\omega}\right)^2 +m_n^{*2}+k'^2+q^2\right],
\eea
and the step-function sets the following limits on
the neutrino energy $\ep_{k'}$ 
\bea 
\label{eq:theta3}
(k'-q)^2 +m_n^{*2}\leq \left(\ep_{k'}-\mu_n^*-
 {\omega}\right)^2 \leq (k'+q)^2 +m_n^{*2}.
\eea 
The expressions for the integrals~\eqref{eq:I1_final}, \eqref{eq:I2_final} 
and \eqref{eq:I3_final} are slightly more general than used in the main body 
of the text because they include the nonzero mass of neutrinos. As we do not 
consider neutrino oscillations they can be neglected hereafter, \ie, 
we put $m_{\nu_l} = 0$. Combining  Eqs.~\eqref{eq:Gamma1p_dummy}, 
\eqref{eq:Gamma2p_dummy}, \eqref{eq:I1_final}, \eqref{eq:I2_final} and 
\eqref{eq:I3_final}, we obtain the final expressions~\eqref{eq:Gamma1p_final1} 
and \eqref{eq:Gamma2n_final1} of the main text.

 Now we are in a position to compute the derivatives of 
 $\Gamma_{n\to p l \bar\nu}$ and $\Gamma_{p l \to n\nu}$ with respect 
to $\mu_{\Delta_l}$. Note that only the integral $I_1$ depends on 
$\mu_{\Delta_l}$, and, exploiting the following identity between 
the Fermi and Bose functions
\bea\label{eq:fermi_bose}
\bar{f}(z) {f}(z-y) = g(-y) [f(z)-f(z-y)],
\eea
 from Eq.~\eqref{eq:I1_final} we obtain
\bea \label{eq:I1_deriv_d}
\frac{\partial I_1}{\partial\mu_{\Delta_l}} &=& \frac{1+g(\bar{\omega})}{4\pi q T}\left[
 g(\bar{\omega})\Lambda_{1}(\bar{\omega})
-   T\frac{\partial }{\partial\bar{\omega}} 
\Lambda_{1}(\bar{\omega})\right],
\eea 
where
\bea 
\Lambda_{1} (\bar{\omega}) =
\left[(\mu_l +\mu_p^* +\bar{\omega})^2 -m_l^2-m_p^{*2}-q^2\right]
\int_{m_l-\mu_l}^{\mu_p^* +\bar{\omega}}\! 
d\ep_k\, \big[{f}(\epsilon_{k})-
f(\epsilon_{k}-\bar{\omega})\big]
\theta(1-\vert x_0\vert).
\eea
The rate derivatives then take the form 
\bea\label{eq:Gamma1p_deriv}
\frac{\partial }{\partial\mu_{\Delta_l}}\Gamma_{n\to p l \bar\nu}\, (\mu_{\Delta_l}) 
&=& -\frac{{G}^2}{(2\pi)^5T} \int_{-\infty}^\infty\!\! d\omega\int_0^{\infty}dq \,
 g(\omega)[1+g(\bar{\omega})]
\left[ g(\bar{\omega}) \Lambda_{1} (\bar{\omega})  
-  T\frac{\partial }{\partial\bar{\omega}}\Lambda_{1} (\bar{\omega})\right]  
\Lambda_{2} ({\omega}), \\
\label{eq:Gamma2p_deriv}
\frac{\partial}{\partial\mu_{\Delta_l}}\Gamma_{n\nu \to p l}\, (\mu_{\Delta_l}) 
&=&  - \frac{{G}^2}{(2\pi)^5T} \int_{-\infty}^\infty\!\! d\omega\int_0^{\infty}dq\, 
 g(\omega)[1+g(\bar{\omega})]
\left[ g(\bar{\omega}) \Lambda_{1} (\bar{\omega})  
-  T\frac{\partial }{\partial\bar{\omega}}\Lambda_{1} (\bar{\omega})\right]  
\Lambda_{3} ({\omega}),
\eea
where
\bea 
\Lambda_{2} ({\omega}) &=&
\left[(\mu_{\nu_l} +\mu_n^*+{\omega})^2 -m_{\nu_l}^2 -m_n^{*2}-q^2\right]\int_{m_{\nu_l}+\mu_{\nu_l}}^\infty\! 
d\ep_{k'}\left[{f}(\epsilon_{k'})-f(\epsilon_{k'}+{\omega})\right]
\theta(1-\vert y_0\vert),\\
\Lambda_{3} ({\omega}) &=&
\left[(\mu_{\nu_l} +\mu_n^*+{\omega})^2 
-m_{\nu_l}^2 -m_n^{*2}-q^2\right]\int_{m_{\nu_l}-\mu_{\nu_l}}^{\omega+\mu_n^*}\! 
d\ep_{k'} \left[f(\epsilon_{k'}) -{f}(\epsilon_{k'}-{\omega}) \right]
\theta(1-\vert z_0\vert).
\eea
The derivatives of the inverse rates can be obtained by replacing $g(\omega)\to 1+g({\omega})$, $g(\bar{\omega})\fromto 1+g(\bar{\omega})$ in Eqs.~\eqref{eq:Gamma1p_deriv} and \eqref{eq:Gamma2p_deriv}.
For the $\lambda$--coefficients we obtain
\bea
\lambda_{n\fromto p l \bar\nu} &=& \frac{{G}^2}{(2\pi)^5T} \int_{-\infty}^\infty\!\!\! 
d\omega\, \!\int_0^\infty\!\! dq\,\Bigg\{
g(\bar{\omega})[1+g(\bar{\omega})] \Lambda_{1}(\bar\omega)
+[g(\omega)-g(\bar{\omega})]  T\frac{\partial }{\partial\bar{\omega}}\Lambda_{1} (\bar{\omega})\Bigg\} \Lambda_{2} ({\omega}),\\
\lambda_{p l \fromto n\nu} &=& \frac{{G}^2}{(2\pi)^5T} \int_{-\infty}^\infty\!\!\! 
d\omega\, \!\int_0^\infty\!\! dq\,\Bigg\{
g(\bar{\omega})[1+g(\bar{\omega})] \Lambda_{1}(\bar\omega)
+[g(\omega)-g(\bar{\omega})]  T\frac{\partial }{\partial\bar{\omega}}\Lambda_{1} (\bar{\omega})\Bigg\} \Lambda_{3} ({\omega}).
\eea

In $\beta$-equilibrium $\bar{\omega}=\omega$ which along with the relations $I_1=-[1+g(\bar{\omega})]\Lambda_{1} (\bar{\omega})$, $I_2=-g({\omega})\Lambda_{2}({\omega})$, $I_3=-g({\omega})\Lambda_{3} ({\omega})$ leads to 
Eqs.~\eqref{eq:lambda1} and \eqref{eq:lambda2} of the main text.


\subsection{Low-$T$ limit of Urca process rates}

In the limit of low temperature the inequalities~\eqref{eq:theta1}, 
\eqref{eq:theta2} and \eqref{eq:theta3} reduce to 
\bea
\theta_x &=& \theta(p_{Fl}+{p}_{Fp}-q)
\theta(q-\vert p_{Fl}- {p}_{Fp}\vert),\\
\theta_y &=& \theta(p_{F{\nu_l}}+p_{Fn}-q)
\theta(q-\vert p_{F{\nu_l}} -p_{Fn}\vert),\\
\theta_z &=& \theta(p_{F{\nu_l}}+p_{Fn}-q)
\theta(q-\vert p_{F{\nu_l}} -p_{Fn}\vert),
\eea 
where we used the notations $\theta_x$, $\theta_y$ and $\theta_z$ introduced in Eqs.~\eqref{eq:Gamma1p_final1} and \eqref{eq:Gamma2n_final1}.
Then the integrals \eqref{eq:I1_final}, \eqref{eq:I2_final} and 
\eqref{eq:I3_final} (in $\beta$-equilibrium) can be approximated as 
\bea\label{eq:I1_lowT}
I_1(q) &\simeq& 
\frac{g(-{\omega})}{4\pi q} \left[(\mu_l +\mu_p^*)^2
-m_l^2-m_p^{*2}-q^2\right]\theta_x \!
\int_{m_l-\mu_l}^{\mu_p^* +{\omega}}\! d\ep_k\,
[f(\epsilon_{k})-f(\epsilon_{k}-{\omega})]\nonumber\\
&\simeq & -\frac{{\omega} g(-{\omega})}{4\pi q} \,\theta_x\,
({p}_{Fp}^2+p_{Fl}^2+2\mu_l\mu_{p}^*-q^2),\\
\label{eq:I2_lowT}
I_2(q) &\simeq& 
\frac{g(\omega)}{2(2\pi)^5 q} \left[-(\mu_{\nu_l} +\mu_n^*)^2 
+m_{\nu_l}^2 +m_n^{*2}+q^2\right]\theta_y\!
\int_{m_{\nu_l}+\mu_{\nu_l}}^\infty\! d\ep_{k'}\,\left[f(\epsilon_{k'})
-f(\epsilon_{k'}+{\omega})\right]\nonumber\\
&\simeq &
-\frac{g(\omega)T}{2(2\pi)^5 q}\,\theta_y\, 
(p_{F{\nu_l}}^2+p_{Fn}^{2} +2\mu_{\nu_l} \mu_n^* -q^2) \ln \Bigg\vert 
\frac{1+\exp\left(-\frac{m_{\nu_l}+\mu_{\nu_l}}{T}\right)} 
{1+\exp\left(-\frac{m_{\nu_l}+\mu_{\nu_l}+{\omega}}{T}\right)}
\Bigg\vert,\\
\label{eq:I3_lowT}
I_3(q) &\simeq& 
\frac{g(\omega)}{2(2\pi)^5q} \left[(\mu_{\nu_l} +\mu_n^*)^2 
-m_{\nu_l}^2 -m_n^{*2}-q^2\right]\theta_z\!
\int_{m_{\nu_l}-\mu_{\nu_l}}^{\omega+\mu_n^*}\! d\ep_{k'}\, \left[f(\epsilon_{k'}-{\omega})-f(\epsilon_{k'})\right]
\nonumber\\
&\simeq &
-\frac{g(\omega)T}{2(2\pi)^5q}\, \theta_z\, 
(p_{Fn}^{2}+p_{F{\nu_l}}^2+2\mu_{\nu_l} \mu_n^*  -q^2)\ln \Bigg\vert 
\frac{1+\exp\left(-\frac{m_{\nu_l}-\mu_{\nu_l}}{T}\right)} 
{1+\exp\left(-\frac{m_{\nu_l}-\mu_{\nu_l}- {\omega}}{T}\right)}
\Bigg\vert.
\eea 
Note that in $I_1(q)$ the integral is approximated as $\omega$ because baryons are highly degenerate; in the remaining integrals, the logarithmic factor should be kept since neutrinos are thermal.  
In the low-temperature neutrino-trapped matter $\mu_{\nu_l}/T\to \infty$, which implies $I_2=0$ and $\Gamma_{n\fromto p l \bar\nu}=0$. In this case 
the logarithm in Eq.~\eqref{eq:I3_lowT} is $-\omega/T$, and
for $\Gamma_{pl \fromto n\nu}$ from Eqs.~\eqref{eq:Gamma2p_dummy} we find (we put again $m_{\nu_l} = 0$)
\bea\label{eq:Gamma2p_lowT_trap}
\Gamma_{pl \fromto n\nu} &=& -2 {G}^2 4\pi\int_{-\infty}^\infty\!\!\!
d\omega\, \omega^2 \int_0^\infty\!\!\! q^2 dq\, 
\frac{g(-{\omega})}{4\pi q} 
\theta(p_{Fl}+{p}_{Fp}-q)\theta(q-\vert p_{Fl}-{p}_{Fp}\vert)
({p}_{Fp}^2+p_{Fl}^2+2\mu_l \mu_{p}^*-q^2)\nonumber\\
&&\times \frac{g(\omega)}{2(2\pi)^5q} \theta(p_{F{\nu_l}}+p_{Fn}-q)
\theta(q-\vert p_{F{\nu_l}} -p_{Fn}\vert)
(p_{Fn}^{2}+p_{F{\nu_l}}^2+2\mu_{\nu_l} \mu_n^* -q^2)\nonumber\\
&=& \frac{{G}^2 T^3}{48\pi^3}  \bigg\{\frac{(p_{Fl}+{p}_{Fp})^5-
( p_{Fn}-p_{F{\nu_l}})^5}{5}-\frac{(p_{Fl}+{p}_{Fp})^3- (p_{Fn}-p_{F{\nu_l}})^3}{3}\nonumber\\
&&\times ({p}_{Fp}^2+
p_{Fl}^2+2\mu_l\mu_{p}^* +p_{Fn}^{2}+
p_{F{\nu_l}}^2+2\mu_{\nu_l} \mu_n^* ) +(p_{Fl}+{p}_{Fp}+p_{F{\nu_l}}- p_{Fn})\nonumber\\
&&\times
({p}_{Fp}^2+p_{Fl}^2+2\mu_l \mu_{p}^*)
(p_{Fn}^{2}+p_{F{\nu_l}}^2+2\mu_{\nu_l} \mu_n^*) \bigg\}
\theta(p_{Fl}+{p}_{Fp}+p_{F{\nu_l}}- p_{Fn}).
\eea
In the nonrelativistic limit for nucleons $\mu_N^*\simeq m^*_N\gg p_{FN}$.
Therefore 
\bea
\Gamma_{pl \fromto n\nu} \simeq  
\frac{{G}^2 T^3}{12\pi^3} m_n^*m_p^*\,
\mu_l\,\mu_{\nu_l} (p_{Fl}+{p}_{Fp}+p_{F\nu_l}-p_{Fn})
\theta(p_{Fl}+{p}_{Fp}+p_{F\nu_l}-p_{Fn}),
\eea
which coincides with our previous calculation if we assume massless leptons $\mu_l=p_{Fl}$ [see Eq. (24) of Ref.~\cite{Alford2019b}].

In the case where the trapped species in the degenerate matter are antineutrinos 
rather than neutrinos we have $\mu_{\nu_l}/T\to -\infty$, therefore $I_3=0$ and $\Gamma_{pl \fromto n\nu}=0$. The logarithm in Eq.~\eqref{eq:I2_lowT} 
in this case is $\omega/T$ and
\bea\label{eq:Gamma1p_lowT_trap}
\Gamma_{n\to pl \bar{\nu}}
&=&- \frac{{G}^2 T^3}{48\pi^3} \bigg\{\frac{(p_{Fl}+{p}_{Fp})^5-
(p_{Fn}-p_{F\bar{\nu}_l})^5}{5}-\frac{(p_{Fl}+{p}_{Fp})^3-
(p_{Fn}-p_{F\bar{\nu}_l})^3}{3}\nonumber\\
&& \times ({p}_{Fp}^2+p_{Fl}^2+2\mu_l\mu_{p}^* +p_{Fn}^{2}+ p_{F\bar{\nu}_l}^2-2|\mu_{\nu_l}| \mu_n^* )
+(p_{Fl}+{p}_{Fp}+p_{F\bar{\nu}_l}- p_{Fn})\nonumber\\
&&\times ({p}_{Fp}^2+p_{Fl}^2+2\mu_l \mu_{p}^*) 
(p_{Fn}^{2}+p_{F\bar{\nu}_l}^2-2|\mu_{\nu_l}| \mu_n^*) \bigg\}
\theta(p_{Fl}+{p}_{Fp}+p_{F\bar{\nu}_l}-p_{Fn}).
\eea

In Fig.~\ref{fig:Gamma_scaled} we show the ratios of summed electron Urca rates $\Gamma_{e}$ to their low-temperature limit given by Eqs.~\eqref{eq:Gamma2p_lowT_trap} and \eqref{eq:Gamma1p_lowT_trap}. We see that the exact rates differ significantly from their low-temperature limit 
typically at $T\geq 10$ MeV, where the deviation
between the exact and the approximate rates reaches up to an order of magnitude. Note that the exact rates are mainly 
larger than their low-temperature limit in neutrino-dominated matter and smaller in the antineutrino-dominated matter.
The analogous ratios for muonic Urca rates are similar and are not shown.

\begin{figure}[t] 
\begin{center}
\includegraphics[width=0.45\columnwidth,keepaspectratio]{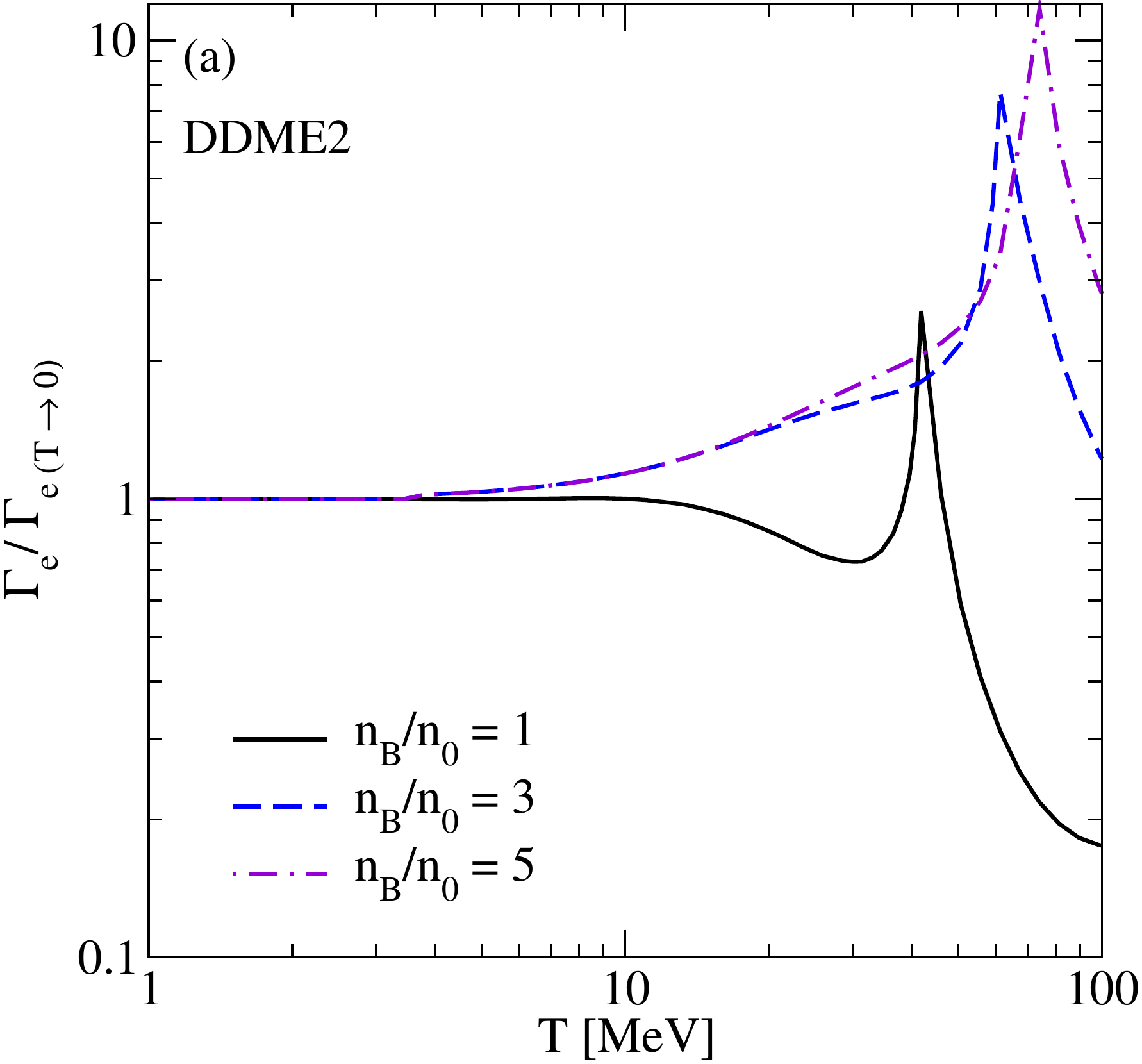}
\hspace{0.5cm}
\includegraphics[width=0.45\columnwidth,keepaspectratio]{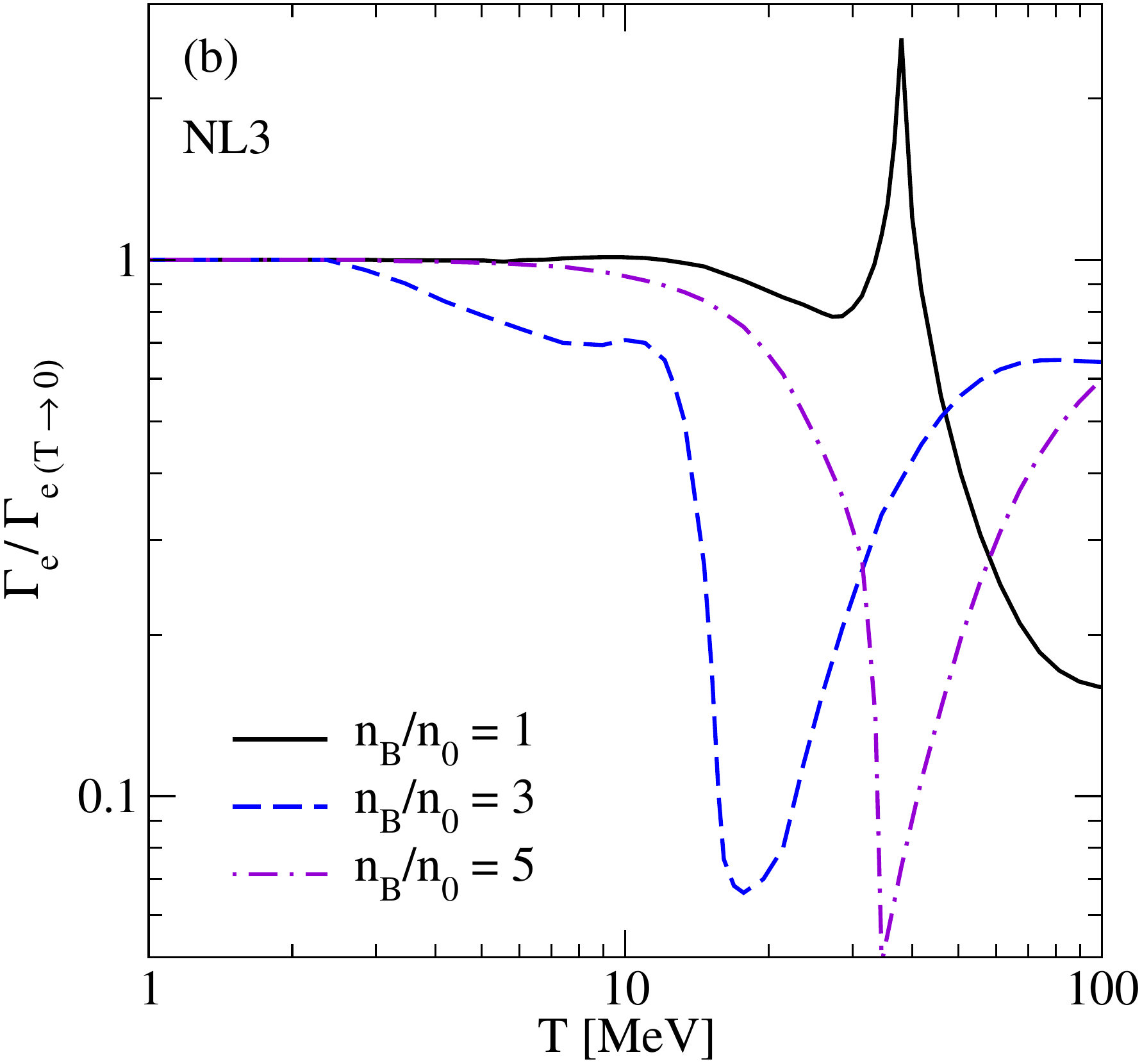}
\caption{ The ratios of summed electron Urca rates $\Gamma_{e}$ to their low-$T$ limit given by (a) Eq.~\eqref{eq:Gamma2p_lowT_trap} for model DDME2 
and (b) Eq.~\eqref{eq:Gamma1p_lowT_trap} model NL3 (b). 
The analogous ratios for muonic Urca rates are similar 
and are not shown.}
\label{fig:Gamma_scaled} 
\end{center}
\end{figure}

\section{Computation of susceptibilities $A_j$}
\label{app:A_coeff}

To compute the susceptibilities $A_{ij}= \left(\frac{\partial \mu_i}
{\partial n_j}\right)_0$ 
we use the following formula for the particle densities
\bea\label{eq:dens}
n_i =\frac{g_i}{2\pi^2}\int_0^\infty p^2dp\, [f_i(p)-\bar{f}_i(p)],
\eea 
where $g_i$ is the spin degeneracy factor, and  $f(p)$ and $\bar{f}(p)$
are the distribution functions for particles and antiparticles, 
respectively. For neutrons, protons, electrons and muons we have 
$g_i=2$, and for neutrinos $g_\nu=1$.

Differentiating the left and right sides of Eq.~\eqref{eq:dens} 
with respect to $n_j$ and exploiting the expressions 
\bea\label{eq:fermi_i}
\frac{\partial f_i}{\partial n_j} =-f_i(1-f_i)\frac{1}{T}
\left(\frac{m^*}{\sqrt{m^{*2}+p^2}}\frac{\partial m^*}{\partial n_j} 
-\frac{\partial \mu^*_i}{\partial n_j}\right),\qquad
\frac{\partial \bar{f}_i}{\partial n_j} =-\bar{f}_i(1-\bar{f}_i)\frac{1}{T}
\left(\frac{m^*}{\sqrt{m^{*2}+p^2}}\frac{\partial m^*}{\partial n_j} 
+\frac{\partial \mu^*_i}{\partial n_j}\right),
\eea
in the case of baryons we obtain
\bea\label{eq:matrix_eq}
\delta_{ij}=-\left(\frac{\partial m^*}{\partial n_j}\right)
{I}_{1i}^- +\left(\frac{\partial
\mu^*_i}{\partial n_j}\right) {I}_{0i}^+,
\eea
where
\bea\label{eq:I_rel}
{I}^{\pm}_{q i}= \frac{1}{\pi^2 T}\int_0^\infty p^2 dp
\left(\frac{m^*}{\sqrt{m^{*2}+p^2}}\right)^q
[f_i(1-f_i)\pm \bar{f}_i(1-\bar{f}_i)],\quad i=\{n,p\}.
\eea 
The average values of the meson fields are given by~\cite{Chatterjee2007}
\bea\label{eq:mean_fields}
g_\omega\omega_0 = \left(\frac{g_\omega}{m_\omega}\right)^2 
(n_n+n_p),\qquad
g_\rho \rho_{03} = \frac{1}{2}\left(\frac{g_\rho}{m_\rho}\right)^2
(n_p -n_n),
\eea
which gives (recall that $\mu^*_i = \mu_i-g_{\omega}\omega_0 
- g_{\rho}\rho_{03}I_{3i}-\Sigma_r$)
\bea\label{eq:b_ij_def}
B_{ij}\equiv \frac{\partial \mu^*_i}{\partial n_j} = 
A_{ij} -\left(\frac{g_\omega}{m_\omega}\right)^2\left[1+
\frac{2n_B}{g_\omega}\frac{\partial g_\omega}{\partial n_B}\right] -
 I_{3i}\left(\frac{g_\rho}{m_\rho}\right)^2 \left[I_{3j}+
 \frac{n_n -n_p}{n_0}a_\rho\right]-\frac{\partial\Sigma_r}{\partial n_j}.
\eea

The scalar field is given by 
\bea\label{eq:sigma}
g_\sigma \sigma =m-m^*=-\frac{g_\sigma}{m_\sigma^2}
\frac{\partial U(\sigma)}{\partial \sigma}+\frac{1}{\pi^2}
\left(\frac{g_\sigma}{m_\sigma}\right)^2\sum_{i=n,p}\int_0^\infty 
p^2dp\frac{m^*}{\sqrt{p^2+m^{*2}}} [f_i(p)+\bar{f}_i(p)],
\eea 
with $U(\sigma)$ being the self-interaction potential of the 
scalar field, therefore up to terms $\partial g_\sigma/\partial n_B$
(which are small and can be neglected) we find
\bea
\frac{\partial m^*}{\partial n_j}=
\frac{g_\sigma}{m_\sigma^2}
\frac{\partial^2 U(\sigma)}{\partial \sigma^2}
\frac{\partial\sigma}{\partial n_j}+
\left(\frac{g_\sigma}{m_\sigma}\right)^2
\left(\frac{\partial m^*}{\partial n_j}\right)
\left({I}_{2n}^+ +{I}_{2p}^+\right) -
\left(\frac{g_\sigma}{m_\sigma}\right)^2
\left(B_{nj}{I}_{1n}^- +B_{pj}{I}_{1p}^-\right)\nonumber\\
-\left(\frac{g_\sigma}{m_\sigma}\right)^2
\left(\frac{\partial m^*}{\partial n_j}\right)
\sum_{i=n,p}\frac{1}{\pi^2}\int_0^\infty\!\!\! p^2dp\frac{p^2}
{(p^2+m^{*2})^{3/2}} [f_i(p)+\bar{f}_i(p)].
\eea 
Denoting 
\bea
\tilde{I}_{2i}^+  = {I}_{2i}^+ -\frac{1}{\pi^2}\int_0^\infty\!\!\! p^2dp
\frac{p^2}{(p^2+m^{*2})^{3/2}} [f_i(p)+\bar{f}_i(p)],\quad {i=n,p},
\eea 
we obtain 
\bea\label{eq:mass_diff}
\frac{\partial m^*}{\partial n_j}=-
\frac{\left(\frac{g_\sigma}{m_\sigma}\right)^2
\left(B_{nj}{I}_{1n}^-+ B_{pj}{I}_{1p}^- \right)}
{1-\left(\frac{g_\sigma}{m_\sigma}\right)^2
\left(\tilde{I}_{2n}^+ +\tilde{I}_{2p}^+\right) +
\frac{1}{m_\sigma^2}\frac{\partial^2 U}{\partial \sigma^2}}.
\eea 
Substituting this into Eq.~\eqref{eq:matrix_eq} we obtain 
the following equations for coefficients $B_{ij}$
\bea\label{eq:matrix_eq1}
B_{ij} {I}_{0i}^+ -\gamma\left(B_{nj}{I}_{1n}^-+ 
B_{pj}{I}_{1p}^-\right){I}_{1i}^- =\delta_{ij},
\eea
where 
\bea\label{eq:gamma_def}
\gamma = \frac{1}{\tilde{I}_{2n}^+ +\tilde{I}_{2p}^+ -\beta},
\qquad \beta =\left(\frac{m_\sigma}{g_\sigma}\right)^2
\left(1+\frac{1}{m_\sigma^2}\frac{\partial^2 U}{\partial \sigma^2}\right).
\eea
In the case of $i\neq j$ we find from Eq.~\eqref{eq:matrix_eq1} 
\bea
B_{np}=\gamma B_{pp}\frac{I_{1p}^-I_{1n}^-}
{I_{0n}^+ -\gamma I_{1n}^{-2}},\qquad
B_{pn}=\gamma B_{nn}\frac{I_{1n}^-I_{1p}^-}
{I_{0p}^+ -\gamma I_{1p}^{-2}}.
\eea
Substituting these expressions into 
Eq.~\eqref{eq:matrix_eq1} for $i=j$ we obtain
\bea\label{eq:b_diag}
B_{nn}=
\frac{I_{0p}^+ -\gamma I_{1p}^{-2}}{I_{0n}^{+}I_{0p}^{+}
-\gamma I_{0p}^+ I_{1n}^{-2}-\gamma I_{0n}^+ I_{1p}^{-2}},\qquad
B_{pp}=
\frac{I_{0n}^+-\gamma I_{1n}^{-2}}{I_{0n}^{+}I_{0p}^{+}
-\gamma I_{0p}^+ I_{1n}^{-2}-\gamma I_{0n}^+ I_{1p}^{-2}},
\eea
and
\bea
\label{eq:b_mix}
B_{np}=B_{pn}=
\frac{\gamma I_{1p}^-I_{1n}^-}{I_{0n}^{+}I_{0p}^{+}
-\gamma I_{0p}^+ I_{1n}^{-2}-\gamma I_{0n}^+ I_{1p}^{-2}}.
\eea

Substituting Eqs.~\eqref{eq:b_diag} and \eqref{eq:b_mix} in 
Eq.~\eqref{eq:b_ij_def} and recalling the definitions 
$A_n=A_{nn}-A_{pn}$, $A_p=A_{pp}-A_{np}$ we obtain 
\bea\label{eq:A_n_final}
A_{n}=\frac{I_{0p}^+ -\gamma I_{1p}^{-}(I_{1p}^{-}+ I_{1n}^-)}
{I_{0n}^{+}I_{0p}^{+} -\gamma I_{0p}^+ 
I_{1n}^{-2}-\gamma I_{0n}^+ I_{1p}^{-2}}
+\left(\frac{g_\rho}{m_\rho}\right)^2\left(\frac{1}{2}
-\frac{n_n -n_p}{n_0}a_\rho \right),\\
\label{eq:A_p_final}
A_{p}=\frac{I_{0n}^+-\gamma I_{1n}^{-}(I_{1p}^{-}+ I_{1n}^-)}
{I_{0n}^{+}I_{0p}^{+}-\gamma I_{0p}^+ 
I_{1n}^{-2}-\gamma I_{0n}^+ I_{1p}^{-2}}
+\left(\frac{g_\rho}{m_\rho}\right)^2\left(\frac{1}{2}
+\frac{n_n -n_p}{n_0}a_\rho \right).
\eea
For leptons we have simply
\bea\label{A_j_int_lep}
A_{l} =\frac{1}{{I}_{0l}^+},\quad 
A_{\nu_l} =\frac{2}{{I}_{0\nu_l}^+},\quad l=\{e,\mu\}.
\eea 


\end{widetext}

\bibliographystyle{JHEP}                                
\bibliography{urca_bulk_light.bib}

\end{document}